\documentclass[twocolumn,preprintnumbers,superscriptaddress,amsmath,amssymb,prb]{revtex4-2}

\usepackage{graphicx}
\usepackage{dcolumn}
\usepackage{subfigure}
\usepackage{bm}
\usepackage[utf8]{inputenc}  
\usepackage{comment}
\usepackage{units}
\usepackage[unicode=true, bookmarks=false, breaklinks=false, pdfborder={0 0 1},colorlinks=false]{hyperref}
\usepackage{breakurl}
\usepackage{url}
\usepackage{natbib}
\usepackage{multirow}
\usepackage{float}
\usepackage{xcolor}
\renewenvironment{widetext@grid}{%
  \par\ignorespaces
  \setbox\widetext@top\vbox{%
   \vskip15\p@
   \hb@xt@\hsize{%
    \leaders\hrule\hfil
    \vrule\@height6\p@
   }%
   \vskip6\p@
  }%
  \setbox\widetext@bot\hb@xt@\hsize{%
    \vrule\@depth6\p@
    \leaders\hrule\hfil
  }%
  \onecolumngrid
  \let\set@footnotewidth\set@footnotewidth@ii
}{%
  \par
  \twocolumngrid\global\@ignoretrue
  \@endpetrue
}%
\renewcommand{\vec}[1]{\boldsymbol{#1}}

\begin{document} 


\title{Influence of non-local damping on magnon properties of ferromagnets}

\newcommand{\KTH}{Department of Applied Physics, School of Engineering Sciences, KTH Royal Institute of Technology, 
AlbaNova University Center, SE-10691 Stockholm, Sweden}
\newcommand{\SeRC}{SeRC (Swedish e-Science Research Center), KTH Royal Institute of Technology, SE-10044 Stockholm, Sweden}
\newcommand{\Uppsala}{Department of Physics and Astronomy, Uppsala University, Box 516, SE-75120 Uppsala, Sweden}
\newcommand{\Orebro}{School of Science and Technology, \"Orebro University, SE-701 82, \"Orebro, Sweden}
\newcommand{\Stockholm}{Department of Materials and Environmental Chemistry, Stockholm University, SE-10691 Stockholm, Sweden}

\author{Zhiwei Lu}
\thanks{These two authors contributed equally}
    \affiliation{\KTH}
    \email[Corresponding author:\ ]{zhiweil@kth.se}
\author{I. P. Miranda}
\thanks{These two authors contributed equally}
    \affiliation{\Uppsala}
\author{Simon Streib}
    \affiliation{\Uppsala}

\author{Manuel Pereiro}
    \affiliation{\Uppsala}

\author{Erik Sjöqvist}
    \affiliation{\Uppsala}
    
\author{Olle Eriksson}
    \affiliation{\Uppsala}
    \affiliation{\Orebro}

\author{Anders Bergman}
    \affiliation{\Uppsala}
    
\author{Danny Thonig}
    \affiliation{\Orebro}
    \affiliation{\Uppsala}
    
\author{Anna Delin}
    \affiliation{\KTH}
    \affiliation{\SeRC}

\date{\today}

\begin{abstract}
We study the influence of non-local damping on magnon properties of Fe, Co, Ni and Fe$_{1-x}$Co$_{x}$ ($x=30\%,50\%$) alloys. The Gilbert damping parameter is typically considered as a local scalar both in experiment and in theoretical modelling. However, recent works have revealed that Gilbert damping is a non-local quantity that allows for energy dissipation between atomic sites. With the Gilbert damping parameters calculated from a state-of-the-art real-space electronic structure method, magnon lifetimes are evaluated from spin dynamics and linear response, where a good agreement is found between these two methods. It is found that non-local damping affects the magnon lifetimes in different ways depending on the system. Specifically, we find that in Fe, Co, and Ni the non-local damping decreases the magnon lifetimes, while in $\rm Fe_{70}Co_{30}$ and Fe$_{50}$Co$_{50}$ an opposite, non-local damping effect is observed, and our data show that it is much stronger in the former. 
\end{abstract}

\maketitle 

\section*{Introduction}

In recent years, there has been a growing interest in magnonics, which uses quasi-particle excitations in magnetically ordered materials to perform information transport and processing on the nanoscale. Comparing to the conventional information device, the magnonics device exhibits lower energy consumption, easier integrability with complementary metal-oxide semiconductor (CMOS) structure, anisotropic properties, and efficient tunability by various external stimuli to name a few \cite{barman20212021,pirro2021advances,rana2019towards,mahmoud2020introduction,serga2010yig,lendinez2019magnetization,rana2019towards,zakeri2020magnonic,awschalom2021quantum,chen2021skyrmion,lenk2011building}.
Yttrium iron garnet (YIG) \cite{liu2018long} as well as other iron garnets with rare-earth elements (Tm, Tb, Dy, Ho, Er) \cite{sheng2021magnonics} are very promising candidates for magnonics device applications due to their low energy dissipation properties and, thus, long spin wave propagation distances up to tens of $\unit{\mu m}$. Contrary, the damping of other materials for magnonics, like CoFeB, is typically two orders of magnitude higher compared to YIG \cite{sheng2021magnonics}, leading to much shorter spin wave propagation distances. 
A clear distinction can be made between materials with an ultra-low damping parameter, like in YIG, and those with a signficiantly larger, but still small, damping parameter. Materials like YIG are insulating, which hinders many of the microscopic mechanisms for damping, resulting in the low observed damping parameter. In contrast, materials like CoFeB are metallic. In research projects that utilize low damping materials, YIG and similar non-metallic low damping systems are typically favored. However, metallic systems have an advantage, since magnetic textures can easily by influenced by electrical currents. Hence, there is good reason to consider metallic systems for low damping applications, even though their damping typically is larger than in YIG. One can conclude that
 Gilbert damping is one of the major bottlenecks for the choice of material in magnonics applications and a detailed experimental as well as theoretical characterisation is fundamental for this field of research, especially for metallic systems. Thus, a more advanced and detailed understanding of Gilbert damping is called for, in order to overcome this obstacle for further development of magnonics-based technology.

Whereas most studies consider chemical modifications of the materials in order to tune damping \cite{burrowes2012enhanced,correa2019exploring}, only a few focus on the fundamental physical properties as well as dependencies of the Gilbert damping. Often Gilbert damping is considered as a phenomenological scalar parameter in the equation of motion of localized atomistic magnetic moments, i.e. the Landau-Lifshitz-Gilbert (LLG) equation \cite{eriksson2017atomistic}. However, from using the general Rayleigh dissipation function in the derivation proposed by Gilbert \cite{gilbert2004phenomenological}, it was theoretically found that the Gilbert damping should be anisotropic, a tensor, and non-local. Furthermore, it depends on the temperature and, thus, on underlying magnon as well as phonon configurations \cite{gilmore2010anisotropic,fahnle2006breathing,bhattacharjee2012atomistic,thonig2018nonlocal}. This is naturally built into the multiple theoretical methods developed to predict the damping parameter, including breathing Fermi surface model \cite{kambersky1976ferromagnetic}, torque correlation model \cite{gilmore2007identification}, and linear response formulation \cite{ebert2011ab}. For instance, the general Gilbert damping tensor as a function of the non-collinear spin configuration has been proposed in Ref. \onlinecite{brinker2022generalization}. 

Nonetheless, an experimental verification is still missing due to lacking insights into the impact of the generalised damping on experimental observables. In a recent experiment, however, the anisotropic behavior of the damping has been confirmed for $\rm Co_{50}Fe_{50}$ thin films and was measured to be of the order of $\rm 400\% $ \cite{li2019giant}, with respect to changing the magnetization direction. Changes of Gilbert damping in a magnetic domain wall and, thus, its dependency on the magnetic configuration was measured in Ref.  \cite{weindler2014magnetic} and fitted to the Landau-Lifshitz-Baryakhtar (LLBar) equation, which includes non-locality of the damping by an additional dissipation term proportional to the gradient of the magnetisation \cite{bar1984phenomenological,dvornik2013micromagnetic,wang2015phenomenological}. However, the pair-wise non-local damping $\alpha_{ij}$ has not yet been measured.

The most common experimental techniques of evaluating damping are ferromagnetic resonance (FMR) \cite{ma2017metrology} and time-resolved magneto-optical Kerr effect (TR-MOKE) \cite{zhu2019magnetization}. In these experiments, Gilbert damping is related to the relaxation rate when \textit{(i)} slightly perturbing the coherent magnetic moment out of equilibrium by an external magnetic field \cite{urban2001gilbert} or \textit{(ii)} when disordered magnetic moments remagnetise after pumping by an ultrafast laser pulse \cite{Schoen2016}. Normally, in case \textit{(i)} the non-locality is suppressed due to the coherent precession of the atomic magnetic moments.  However, this coherence can be perturbed by temperature, making non-locality in principle measurable. One possible other path to link non-local damping with experiment is magnon lifetimes. Theoretically, the magnon properties as well as the impact of damping on these properties can be assessed from the dynamical structure factor, and atomistic spin-dynamics simulations have been demonstrated to yield magnon dispersion relations that are in good agreement with experiment \cite{etz2015atomistic}.
In experiment, neutron scattering \cite{nambu2020observation} and electron scattering \cite{balashov2008inelastic} are the most common methods for probing magnon excitations, where the linewidth broadening of magnon excitations is related to damping and provides a way to evaluate the magnon lifetimes \cite{balashov2014magnon}. It is found in ferromagnets that the magnon lifetimes is wave vector (magnon energy) dependent \cite{costa2010spin,qin2013magnons,chakraborty2015lifetimes}. It has been reported that the magnon energy in Co films is nearly twice as large as in Fe films, but they have similar magnon lifetimes, which is related to the intrinsic damping mechanism of materials \cite{zhang2012relaxation}. However, this collective effect of damping and magnon energy on magnon lifetimes is still an open question. The study of this collective effect is of great interest for both theory and device applications.

 Here, we report an implementation for solving the stochastic Landau-Lifshitz-Gilbert (SLLG) equation incorporating the non-local damping. With the dynamical structure factor extracted from the spin dynamics simulations, we investigate the collective effect of non-local damping and magnon energy on the magnon lifetimes. We propose an efficient method to evaluate magnon lifetimes from linear response theory and verify its validity.

The paper is organized as follows. In Sec. \ref{sec:methods}, we give the simulation details of the spin dynamics, the adiabatic magnon spectra and dynamical structure factor, and the methodology of DFT calculations and linear response. Sec. \ref{sec:results} presents the non-local damping in real-space, non-local damping effects on the spin dynamics and magnon properties including magnon lifetimes of pure ferromagnets (Fe, Co, Ni), and Fe$_{1-x}$Co$_{x}$ ($x=30\%,50\%$) alloys.  In Sec. \ref{sec:conclusion}, we give a summary and an outlook. 
\section{Theory}
\label{sec:methods}
\subsection{Non-local damping in atomistic spin dynamics}
The dynamical properties of magnetic materials at finite temperature have been so far simulated from atomistic spin dynamics by means of the stochastic Landau-Lifshitz-Gilbert equation with scalar local energy dissipation. Here, the time evolution of the magnetic moments $ \vec{m}_i=m_i \vec{e}_i$ at atom site $i$ is well described by:
\begin{equation}\label{eq:sllg}
\frac{\partial \vec{m}_i}{\partial t}=\vec{m}_i\times\left(-\gamma \left[\vec{B}_i+\vec{b}_i(t)\right]+\frac{\alpha}{m_i}\frac{\partial \vec{m}_i}{\partial t}\right),
\end{equation}
where $\gamma$ is the gyromagnetic ratio. The effective field $\vec{B}_i$ acting on each magnetic moment is obtained from:
\begin{equation}
 \vec{B}_i=-\frac{\partial \mathcal{H}}{\partial \vec{m}_i}.
\end{equation}
The here considered spin-Hamiltonian $\mathcal{H}$ consists of a
Heisenberg spin-spin exchange:
 \begin{equation}
 \mathcal{H}=-\sum_{i\neq j}J_{ij}\vec{e}_i\cdot \vec{e}_j.
 \end{equation}
Here, $J_{ij}$ -- the Heisenberg exchange parameter -- couples the spin at site $i$ with the spin at site $j$ and is calculated from first principles (see Section \ref{sec:details-dft}). Furthermore, 
$\alpha$ is the scalar phenomenological Gilbert damping parameter. Finite temperature $T$ is included in Eq.~\eqref{eq:sllg} via the fluctuating field $\bm{b}_i(t)$, which is modeled by uncorrelated Gaussian white noise: $\left \langle \bm{b}_i(t) \right \rangle=0 $ and $\left \langle b_i^\mu(t)b_j^\nu(t') \right \rangle=2D\delta_{ij}\delta_{\mu\nu}\delta(t-t') $, where $\delta$ is the Kronecker delta, $i,j$ are site and $\mu,\nu=\{x,y,z\}$ Cartesian indices. Furthermore, the fluctuation-dissipation theorem gives $D=\alpha\frac{k_BT}{\gamma m_i} $  \cite{mentink2010stable}, with the Boltzman constant $k_B$. 

A more generalized form of the SLLG equation that includes non-local tensorial damping has been reported in previous studies \cite{brataas2011mag,vittoria2010relaxation, thonig2018nonlocal} and is:
\begin{equation}\label{eq:sllg_nonlocal}
\frac{\partial \vec{m}_i}{\partial t}=\vec{m}_i\times\left(-\gamma \left[\vec{B}_i+\vec{b}_i(t)\right]+\sum_j\frac{\alpha_{ij}}{m_j}\frac{\partial \vec{m}_j}{\partial t}\right),
\end{equation}
which can be derived from Rayleigh dissipation functional in the Lagrange formalism used by Gilbert \cite{gilbert2004phenomenological}.  In the presence of non-local damping, the Gaussian fluctuating field fulfills \cite{Rossi2005,brataas2011mag,Rueckriegel2015}
\begin{equation}\label{eq:flucdispnonlocal}
    \left \langle b_i^\mu(t)b_j^\nu(t') \right \rangle=2D_{ij}^{\mu\nu}\delta(t-t'),
\end{equation}
with $D_{ij}^{\mu\nu}=\alpha_{ij}^{\mu\nu}\frac{k_BT}{\gamma m_i}$. The damping tensor $\alpha_{ij}^{\mu\nu}$ must be positive definite in order to be physically-defined. Along with spatial non-locality, the damping can also be non-local in time, as discussed in Ref. \cite{thonig2015gilbert}. To prove the fluctuation-dissipation theorem in Eq.~\eqref{eq:flucdispnonlocal}, the Fokker-Planck equation has to be analysed in the presence of non-local damping, similar to Ref. \cite{eriksson2017atomistic}. This is, however, not the purpose of this paper. Instead, we will use the approximation $\alpha_{ij}^{\mu\nu}=\frac{1}{3}\textnormal{Tr}\{\alpha_{ii}\}\delta_{ij}\delta_{\mu\nu}$ within the diffusion constant $D$. Such an approximation is strictly valid only in the low temperature limit. 
 
To solve this SLLG equation incorporating the non-local damping, we have implemented an implicit midpoint solver in the UppASD code \cite{UppASD}. This iterative fix-point scheme converges within an error of $10^{-10}\,\mu_{B}$, which is typically equivalent to 6 iteration steps. More details of this solver are provided in Appendix \ref{appendix:solver}. The initial spin configuration in the typical $N=\rm 20\times20\times20 $ supercell with periodic boundary conditions starts from totally random state. The spin-spin exchange interactions and non-local damping parameters are included up to at least  30  shells of neighbors, in order to guarantee the convergence with respect to the spatial expansion of these parameters (a discussion about the convergence is given in Section \ref{subsec:onsite-nonlocal}).
Observables from our simulations are typically the average magnetisation $\vec{M}=\frac{1}{N}\sum_i^N\vec{m}_i$ as well as the magnon dispersion.

\subsection{Magnon dispersion}\label{sec:magnon-dispersion}
Two methods to simulate the magnon spectrum are applied in this paper: \textit{i)} the dynamical structure factor and \textit{ii)} frozen magnon approach.\\

For the dynamical structure factor  $ S(\vec{q},\omega) $ at finite temperature and damping \cite{bergman2010magnon,etz2015atomistic}, the spatial and time correlation function between two magnetic moments $i$ at position $\vec{r}$ and $j$ at position $\vec{r}^\prime$ as well as different time $0$ and $t$ is expressed as:
\begin{equation}
C^\mu(\vec{r}-\vec{r'},t)= \left\langle m_{\vec{r}}^\mu(t)m_{\vec{r'}}^\mu(0)\right \rangle-\left \langle m_{\vec{r}}^\mu(t)\right \rangle\left \langle m_{\vec{r'}}^\mu(0) \right \rangle.
\end{equation}
Here $\left\langle \cdot  \right \rangle $ denotes the ensemble average and $\mu$ are Cartesian components. The dynamical structure factor can be obtained from the time and space Fourier transform of the correlation function, namely:
\begin{equation}\label{eq:sqw}
S^\mu(\vec{q},\omega)=\frac{1}{\sqrt{2\pi}N}\sum_{\vec{r},\vec{r'}}e^{i\vec{q}\cdot(\vec{r}-\vec{r'})}\int_{-\infty}^{\infty}e^{i\omega t}C^\mu(\vec{r}-\vec{r'},t)\mathrm{d}t.
\end{equation}
The magnon dispersion is obtained from the peak positions of $S(\vec{q},\omega)$ along different magnon wave vectors $ \bm{q} $ in the Brillouin zone and magnon energies $\omega$. It should be noted that $S(\vec{q},\omega)$ is related to the scattering intensity in inelastic neutron scattering experiments \cite{mourigal2010field}. The broadening of the magnon spectrum correlates to the lifetime of spin waves mediated by Gilbert damping as well as intrinsic magnon-magnon scattering processes. Good agreement between $S(\vec{q},\omega)$ and experiment has been found previously \cite{etz2015atomistic}.\\

The second method -- the frozen magnon approach -- determines the magnon spectrum directly from the Fourier transform of the spin-spin exchange parameters $J_{ij}$ \cite{kubler2017theory,halilov1998adiabatic} and non-local damping $\alpha_{ij}$. At zero temperature, a time-dependent external magnetic field is considered,
\begin{equation}
B_{i}^{\pm}(t)=\frac{1}{N}\sum_{{\vec{q}}}B_{{\vec{q}}}^{\pm}e^{i{\vec{q}}\cdot\vec{R}_{i}-i\omega t},
\end{equation}
where $N$ is the total number of lattice sites and $B_{\vec{q}}^\pm=B_{\vec{q}}^x\pm i B_{\vec{q}}^y$. The linear response to this field is then given by
\begin{equation}
M_{{\vec{q}}}^{\pm}=\chi^{\pm}({\vec{q}},\omega)B_{{\vec{q}}}^{\pm}.
\end{equation}
We obtain for the transverse dynamic magnetic susceptibility \cite{Skadsem2007,Mankovsky2018}
\begin{equation}
\chi^{\pm}({\vec{q}},\omega)=\frac{\pm\gamma M_{s}}{\omega\pm\omega_{{\vec{q}}}\mp i\omega\alpha_{{\vec{q}}}},
\end{equation}
with saturation magnetization $M_s$, spin-wave frequency $\omega_{{\vec{q}}}=E({\vec{q}})/\hbar$ and damping
\begin{equation}\label{eq:fourier-transform-damping}
\alpha_{{\vec{q}}}=\sum_{j}\alpha_{0j}e^{-i{\vec{q}}\cdot(\vec{R}_{0}-\vec{R}_{j})}.
\end{equation}
We can extract the spin-wave spectrum from the imaginary part of the susceptibility,
\begin{equation}\label{eq:susceptibility}
\text{Im}\chi^{\pm}({\vec{q}},\omega)=\frac{\gamma M_{s}\alpha_{{\vec{q}}}\omega}{\left[\omega\pm\omega_{{\vec{q}}}\right]^{2}+\alpha_{{\vec{q}}}^{2}\omega^{2}},
\end{equation}
which is equivalent to the correlation function $S^\pm({\vec{q}},\omega)$ due to the fluctuation-dissipation theorem \cite{Marshall1968}. We find that the spin-wave lifetime $\tau_{\vec{q}}$ is determined by the Fourier transform of the non-local damping (for $\alpha_{\vec{q}}\ll 1$),
\begin{equation}\label{eq:magnon lifetime}
    \tau_{\vec{q}}=\frac{\pi}{\alpha_{\vec{q}}\omega_{\vec{q}}}.
\end{equation}
The requirement of positive definiteness of the damping matrix $\alpha_{ij}$ directly implies $\alpha_{\vec{q}}>0$, since $\alpha_{ij}$ is diagonalized by Fourier transformation due to translational invariance. Hence, $\alpha_{\vec{q}}>0$ is a criterion to evaluate whether the damping quantity in real-space is physically consistent and whether first-principles calculations are well converged. If $\alpha_{\vec{q}}<0$ for some wave vector $\vec{q}$, energy is pumped into the spin system through the correspondent magnon mode, preventing the system to fully reach the saturation magnetization at sufficiently low temperatures. 

The effective damping $\alpha_{0}$ of the FMR mode at ${\vec{q}}=0$ is determined by the sum over all components of the damping matrix, following Eqn.\ref{eq:fourier-transform-damping}, 
\begin{equation}\label{eq:damp-effective}
\alpha_\text{tot}\equiv\alpha_{0}=\sum_{j}\alpha_{0j}.
\end{equation}
Therefore, an effective local damping should be based on $\alpha_\text{tot}$ if the full non-local damping is not taken into account.

\subsection{Details of the DFT calculations}
\label{sec:details-dft}
The electronic structure calculations, in the framework of density functional theory (DFT), were performed using the fully self-consistent real-space linear muffin-tin orbital in the atomic sphere approximation (RS-LMTO-ASA) \cite{Peduto1991,FrotaPessoa1992}. The RS-LMTO-ASA uses the Haydock recursion method \cite{Haydock1980} to solve the eigenvalue problem based on a Green's functions methodology directly in real-space. In the recursion method, the continued fractions have been truncated using the Beer-Pettifor terminator \cite{Beer1984}, after a number $LL$ of recursion levels. The LMTO-ASA \cite{Andersen1975} is a linear method which gives precise results around an energy $E_{\nu}$, usually taken as the center of the $s$, $p$ and $d$ bands. Therefore, as we calculate fine quantities as the non-local damping parameters, we here consider an expression accurate to $(E-E_{\nu})^{2}$ starting from the orthogonal representation of the LMTO-ASA formalism \cite{FrotaPessoa2000}. 

For bcc FeCo alloys and bcc Fe we considered $LL=31$, while for fcc Co and fcc Ni much higher $LL$ values (51 and 47, respectively), needed to better describe the density of states and Green's functions at the Fermi level.

The spin-orbit coupling (SOC) is included as a $l\cdot s$ \cite{Andersen1975} term computed in each variational step \cite{FrotaPessoa2004}. All calculations were performed within the local spin density approximation (LSDA) exchange-functional (XC) by von Barth and Hedin \cite{Barth1972}, as it gives general magnetic information with equal or better quality as, \textit{e.g.}, the generalized gradient approximation (GGA). 
Indeed, the choice of XC between LSDA and GGA \cite{Perdew1996} have a minor impact on the onsite damping and the shape of the $\alpha_{\vec{q}}$ curves, 
when considering the same lattice parameters (data not shown). No orbital polarization \cite{Eriksson1990} was considered here. Each bulk system was modelled by a big cluster containing $\sim55000$ (bcc) and $\sim696000$ (fcc) atoms located in the perfect crystal positions with the respective lattice parameters of $a=2.87$\AA$\,$(bcc Fe and bcc Fe$_{1-x}$Co$_{x}$, sufficiently close to experimental observations \cite{Ohnuma2002}), $a=3.54$\AA$\,$(fcc Co \cite{thonig2018nonlocal,American1972}), and $a=3.52$\AA$\,$(fcc Ni \cite{Wijn1991}). To account for the chemical disorder in the Fe$_{70}$Co$_{30}$ and Fe$_{50}$Co$_{50}$ bulks, the electronic structure calculated within the simple virtual crystal approximation (VCA), which has shown to work well for the ferromagnetic transition metals alloys (particularly for elements next to each other in the Periodic Table, such as FeCo and CoNi) \cite{Burkert2004,Masin2013,Soderlind1992,Bergman2006,Fahnle2006,Trinastic2013,Seemann2011,Lourembam2021}, and also describe in a good agreement the damping trends in both FeCo and CoNi (see Appendix \ref{sec:vca-comparison}).

As reported in Ref. \cite{Miranda2021}, the total damping of site $i$, influenced by the interaction with neighbors $j$, 
can be decomposed in two main contributions: the onsite (for $i=j$), and the non-local (for $i\neq j$). Both can be calculated, in the collinear framework, by the following expression,
\begin{equation}\label{eq:damping-dft}
\begin{split}
\alpha_{ij}^{\mu\nu}=\alpha_{C}\int_{-\infty}^{\infty}\eta(\epsilon)\textnormal{Tr}\left(\hat{T}_{i}^{\mu}\hat{A}_{ij}(\hat{T}_{j}^{\nu})^{\dagger}\hat{A}_{ji}\right)d\epsilon\xrightarrow{T\to 0 K}\\
\alpha_{C}\textnormal{Tr}\left(\hat{T}_{i}^{\mu}\hat{A}_{ij}(\epsilon_F+i\delta)(\hat{T}_{j}^{\nu})^{\dagger}\hat{A}_{ji}(\epsilon_F+i\delta)\right),
\end{split}
\end{equation}

\noindent where we define $\hat{A}_{ij}(\epsilon+i\delta)=\frac{1}{2i}(\hat{G}_{ij}(\epsilon+i\delta)-\hat{G}_{ji}^{\dagger}(\epsilon+i\delta))$ the anti-Hermitian part of the retarded physical Green's functions in the LMTO formalism, and $\alpha_{C}=\frac{g}{m_{t_i}\pi}$ a pre-factor related to the $i$-th site magnetization. The imaginary part, $\delta$, is obtained from the terminated continued fractions. Also in Eq. \ref{eq:damping-dft}, $\hat{T}^{\mu}_i=\left[\sigma^{\mu}_i,\mathcal{H}_{so}\right]$ is the so-called torque operator \cite{thonig2018nonlocal} evaluated in each Cartesian direction $\mu,\nu=\{x,y,z\}$ and at site $i$, $\eta(\epsilon)=-\frac{\partial f(\epsilon)}{\partial\epsilon}$ is the derivative of the Fermi-Dirac distribution $f(\epsilon)$ with respect to the energy $\epsilon$, $g=2\left(1+\frac{m_{orb}}{m_{spin}}\right)$ the $g$-factor (not considering here the spin-mixing parameter \cite{Shaw2021}), $\sigma^{\mu}$ are the Pauli matrices, and $m_{t_i}$ is the total magnetic moment of site $i$ ($m_{t_i}=m_{orb_i}+m_{spin_i}$). This results in a $3\times3$ tensor with terms $\alpha_{ij}^{\mu\nu}$. In the real-space bulk calculations performed in the present work, the $\alpha_{ij}$ (with $i\neq j$) matrices contain off-diagonal terms which are cancelled by the summation of the contributions of all neighbors within a given shell, resulting in a purely diagonal damping tensor, as expected for symmetry reasons \cite{eriksson2017atomistic}. Therefore, as in the DFT calculations the spin quantization axis is considered to be in the $z$ ($[001]$) direction (collinear model), we can ascribe a scalar damping value $\alpha_{ij}$ as the average $\alpha_{ij}=\frac{1}{2}(\alpha_{ij}^{xx}+\alpha_{ij}^{yy})=\alpha_{ij}^{xx}$ for the systems investigated here. This scalar $\alpha_{ij}$ is, then, used in the SLLG equation (Eq. \ref{eq:sllg}).

The exchange parameters $J_{ij}$ in the Heisenberg model were calculated by the Liechtenstein-Katsnelson-Antropov-Gubanov (LKAG) formalism \cite{Liechtenstein1987}, according to the implementation in the RS-LMTO-ASA method \cite{FrotaPessoa2000}. Hence all parameters needed for the atomistic LLG equation have been evaluated from ab-initio electronic structure theory.
\section{RESULTS}
\label{sec:results}
\subsection{Onsite and non-local dampings}
\label{subsec:onsite-nonlocal}
Table \ref{tab:properties} shows the relevant \textit{ab-initio} magnetic properties of each material; the $T_C$ values refer to the Curie temperature calculated within the random-phase approximation (RPA) \cite{Pajda2001}, based on the computed $J_{ij}$ set. Despite the systematic $\alpha_{\textnormal{tot}}$ values found in the lower limit of available experimental results (in similar case with, \textit{e.g.}, Ref. \cite{Starikov2010}), in part explained by the fact that we analyze only the intrinsic damping, a good agreement between theory and experiment can be seen. When the whole VCA Fe$_{1-x}$Co$_{x}$ series is considered (from $x=0\%$ to $x=60\%$), the expected Slater-Pauling behavior of the total magnetic moment \cite{Bardos1969,Fahnle2006} is obtained (data not shown).

\begin{table*}[t]
\caption{\label{tab:properties}%
Spin ($m_{spin}$) and orbital ($m_{orb}$) magnetic moments, onsite ($\alpha_{ii}$) damping, total ($\alpha_{\textnormal{tot}}$) damping, and Curie temperature ($T_C$) of the investigated systems. The theoretical $T_C$ value is calculated within the RPA. In turn, $m_t$ denotes the total moments for experimental results of Ref. \cite{Bardos1969}.}
\begin{ruledtabular}
\begin{tabular}{cccccc}
\multicolumn{1}{c}{}&
\multicolumn{1}{c}{$m_{spin}$ ($\mu_B$)}& $m_{orb}$ ($\mu_B$) &
$\alpha_{ii}$ $(\times10^{-3})$ & $\alpha_{\textnormal{tot}}$ $(\times10^{-3})$ & $T_C$ (K)\\
\colrule \\
bcc Fe (theory) & 2.23 & 0.05 & 2.4 & 2.1 & 919 \\
bcc Fe (expt.) & 2.13 \cite{Wijn1991} & $0.08$ \cite{Wijn1991} & $-$ & $1.9-7.2$ \cite{Oogane2006,Schoen2016,Mankovsky2013,Khodadadi2020,Scheck2007,Bhagat1974,Hsu1978,Schoen2017} & 1044 \\ \\
bcc Fe$_{70}$Co$_{30}$ (theory) & 2.33 & 0.07 & 0.5 & 0.9 & 1667 \\
bcc Fe$_{70}$Co$_{30}$ (expt.) & \multicolumn{2}{c}{$m_t=2.457$ \cite{Bardos1969}} & $-$ & $0.5-1.7$\footnote{The lower limit refers to polycrystalline Fe$_{75}$Co$_{25}$ 10 nm-thick films from Ref. \cite{Schoen2016}. Lee \textit{et al.} \cite{Lee2017} also found a low Gilbert damping in an analogous system, where $\alpha_{\textnormal{tot}}<1.4\times10^{-3}$. For the exact $30\%$ of Co concentration, however, previous results \cite{Schoen2016,Mankovsky2013,Zhao2018} indicate that we should expect a slightly higher damping than in Fe$_{75}$Co$_{25}$.} \cite{Schoen2016,Oogane2006,Lee2017} & 1258~\cite{karipoth2013synthesis}\\ \\
bcc Fe$_{50}$Co$_{50}$ (theory) & 2.23 & 0.08 & 1.5 & 1.6 & 1782\\
bcc Fe$_{50}$Co$_{50}$ (expt.) & \multicolumn{2}{c}{$m_t=2.355$ \cite{Bardos1969}} & $-$ & $2.0-3.2$\footnote{The upper limit refers to the approximate minimum intrinsic value for a 10 nm-thick film of Fe$_{50}$Co$_{50}|$Pt (easy magnetization axis).} \cite{Oogane2006,Schoen2016,li2019giant} & 1242~\cite{karipoth2016magnetic}\\ \\
fcc Co (theory) & 1.62 & $0.08$ & 7.4 & 1.4 & 1273\\
fcc Co  (expt.) & $1.68(6)$ \cite{Liu1996} & $-$ & 
 $-$ & $2.8(5)$ \cite{Schoen2016,Schoen2017} & $1392$ \\ \\
fcc Ni (theory) & $0.61$ & $0.05$ & 160.1 & 21.6 & 368 \\
fcc Ni (expt.) & $0.57$ \cite{Wijn1991} & $0.05$ \cite{Wijn1991} & $-$ & $23.6-64$ \cite{Bhagat1974,Oogane2006,gilmore2007identification,Walowski2008,Heinrich1979,Hsu1978,Schoen2017} & 631 \\
\end{tabular}
\end{ruledtabular}
\end{table*}

\begin{figure}[h]
\centering
\includegraphics[width=\columnwidth]{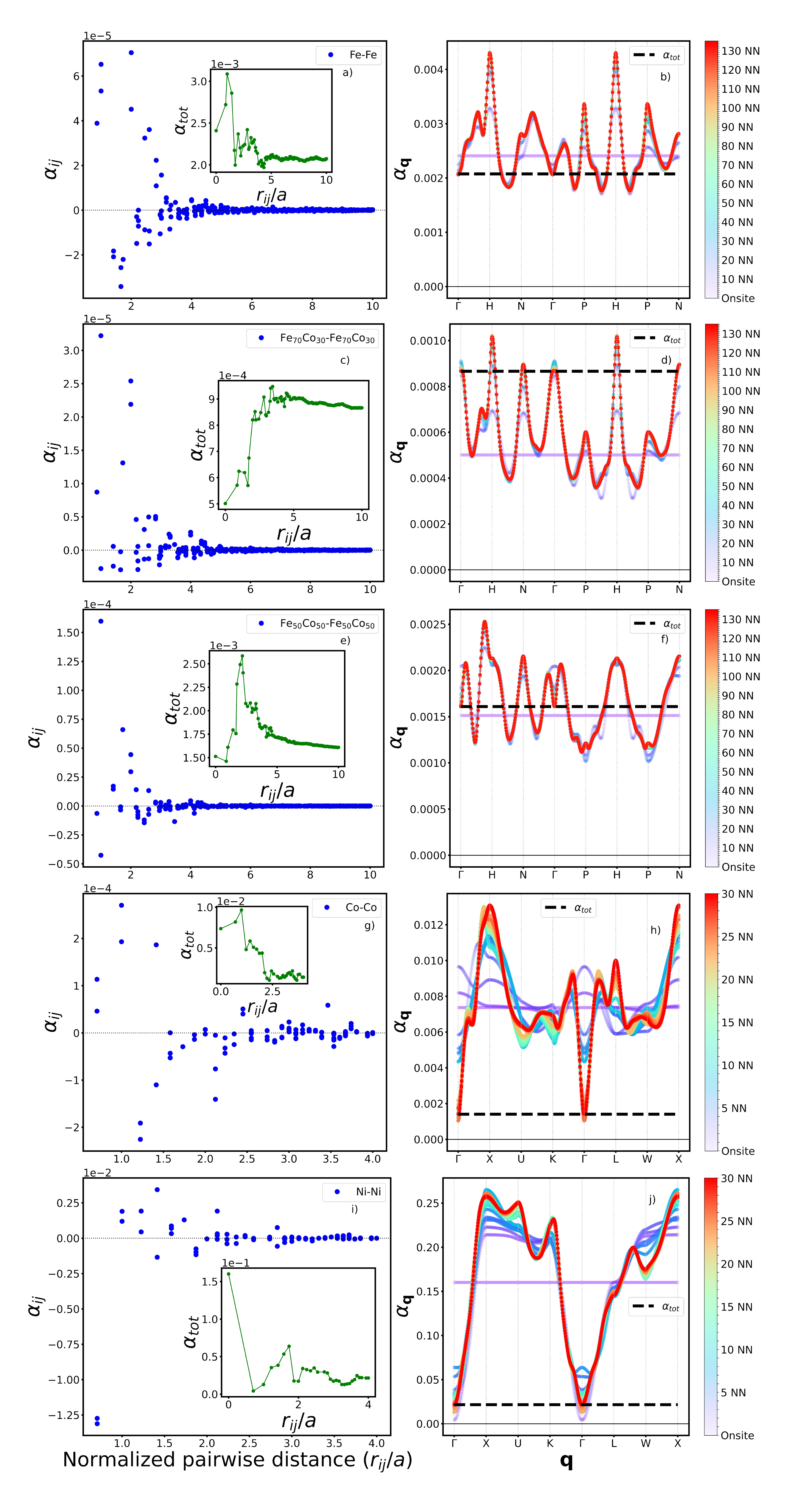} 
\caption{Non-local damping ($\alpha_{ij}$) as a function of the normalized real-space pairwise $(i,j)$ distance computed for each neighboring shell, and corresponding Fourier transform $\alpha_{\vec{q}}$ (see Eq. \ref{eq:fourier-transform-damping}) from the onsite value ($\alpha_{ii}$) up to 136 shells of neighbors (136 NN) for: (a,b) bcc Fe; (c,d) bcc Fe$_{70}$Co$_{30}$; (e,f) bcc Fe$_{50}$Co$_{50}$ in the virtual-crystal approximation; and up to 30 shells of neighbors (30 NN) for: (g,h) fcc Co; (i,j) fcc Ni. The \textit{insets} in subfigures (a,c,e,g,i) show the convergence of $\alpha_{\textnormal{tot}}$ in real-space. 
The obtained onside damping values are shown in Table \ref{tab:properties}. In the \textit{insets} of the left panel, green full lines are guides for the eyes.
}
\label{fig:non-local-damping-systems} 
\end{figure}

For all systems studied here, the dissipation is dominated by the onsite ($\alpha_{ii}$) term, while the non-local parameters ($\alpha_{ij}$, $i\neq j$) exhibit values at least one order of magnitude lower; however, as it will be demonstrated in the next sections, these smaller terms still cause a non-negligible impact on the relaxation of the average magnetization as well as magnon lifetimes. Figure \ref{fig:non-local-damping-systems} shows the non-local damping parameters for the investigated ferromagnets as a function of the $(i,j)$ pairwise distance $r_{ij}/a$, 
together with the correspondent Fourier transforms $\alpha_{\vec{q}}$ over the first Brillouin Zone (BZ). The first point to notice is the overall strong dependence of $\alpha$ on the wave vector $\vec{q}$. The second point is the fact that, as also reported in Ref. \cite{thonig2018nonlocal}, $\alpha_{ij}$ can be an anisotropic quantity with respect to the same shell of neighbors, due to the broken symmetry imposed by a preferred spin quantization axis. This means that, in the collinear model and for a given neighboring shell, $\alpha_{ij}$ is isotropic only for equivalent sites around the magnetization as a symmetry axis.

Another important feature that can be seen in Fig. \ref{fig:non-local-damping-systems} is the presence of negative $\alpha_{ij}$ values. Real-space negative non-local damping parameters have been reported previously \cite{Miranda2021,thonig2018nonlocal,Umetsu2012}. They are related to the decrease of damping at the $\Gamma$-point, but may also increase $\alpha_{\vec{q}}$ from the onsite value in specific $\vec{q}$ points inside the BZ; therefore, they cannot be seen as  
\textit{ad hoc} anti-dissipative contributions. In the ground-state, these negative non-local dampings originate from the overlap between the anti-Hermitian parts of the two Green's functions at the Fermi level, each associated with a spin-dependent phase factor $\Phi^{\sigma}$ ($\sigma=\uparrow,\downarrow$) \cite{thonig2018nonlocal,Pajda2001}. 

Finally, as shown in the \textit{insets} of Fig. \ref{fig:non-local-damping-systems}, a long-range convergence can be seen for all cases investigated. An illustrative example is the bcc Fe$_{50}$Co$_{50}$ bulk, for which the effective damping can be $\sim60\%$ higher than the converged $\alpha_{\textnormal{tot}}$ if only the first 7 shells of neighbors are considered in Eq. \ref{eq:damp-effective}. The non-local damping of each neighboring shell is found to follow a $\frac{1}{r_{ij}^{2}}$ trend, as previously argued by Thonig \textit{et al.} \cite{thonig2018nonlocal} and Umetsu \textit{et al.} \cite{Umetsu2012}. Explicitly,  
\begin{align}
\label{eq:alpha-long-range}
 \alpha_{ij}\propto\frac{\sin(\vec{k}^{\uparrow}\cdot\vec{r}_{ij}+\Phi^{\uparrow})\sin(\vec{k}^{\downarrow}\cdot\vec{r}_{ij}+\Phi^{\downarrow})}{\left|\vec{r}_{ij}\right|^{2}},   
\end{align}
\noindent which also qualitatively justifies the existence of negative $\alpha_{ij}$'s. Thus, the convergence in real-space is typically slower than other magnetic quantities, such as exchange interactions ($J_{ij}\propto\frac{1}{\left|\vec{r}_{ij}\right|^{3}}$) \cite{Pajda2001}, and also depends on the imaginary part $\delta$ (see Eq. \ref{eq:damping-dft}) \cite{thonig2018nonlocal}. The difference in the asymptotic behaviour of the damping and the Heisenberg exchange is distinctive; the first scales with the inverse of the square of the distance while the latter as the inverse of the cube of the distance. Although this asymptotic behaviour can be derived from similar arguments, both using the Greens function of the free electron gas, the results are different. The reason for this difference is simply that the damping parameter is governed by states close to the Fermi surface, while the exchange parameter involves an integral over all occupied states \cite{thonig2018nonlocal, Liechtenstein1987}.

From bcc Fe to bcc Fe$_{50}$Co$_{50}$ (Fig. \ref{fig:non-local-damping-systems}(a-f)), with increasing Co content, the average first neighbors $\alpha_{ij}$ decreases to a negative value, while the next-nearest neighbors contributions reach a minimum, and then increase again. Similar oscillations can be found in further shells.
Among the interesting features in the Fe$_{1-x}$Co$_{x}$ systems ($x=0\%,30\%,50\%$), we highlight the low $\alpha_{\vec{q}}$ around the high-symmetry point $H$, along the $H-P$ and $H-N$ directions, consistently lower than the FMR damping. Both $\alpha$ values are strongly influenced by non-local contributions $\gtrsim5$ NN. Also consistent is the high $\alpha_{\vec{q}}$ obtained for $\vec{q}=\vec{H}$. For long wavelengths in bcc Fe, some $\alpha_{\vec{q}}$ anisotropy is observed around $\Gamma$, which resembles the same trait obtained for the corresponding magnon dispersion curves \cite{Pajda2001}. This anisotropy changes to a more isotropic behavior by FeCo alloying.

Far from the more noticeable high-symmetry points, $\alpha_{\vec{q}}$ presents an oscillatory behavior along BZ, around the onsite value. It is noteworthy, however, that these oscillatory $\alpha_{\vec{q}}$ parameters exhibit variations up to $\sim2$ times $\alpha_{ii}$, thus showing a pronounced non-local influence in specific $\vec{q}$ points. 

In turn, for fcc Co (Fig. \ref{fig:non-local-damping-systems}(g,h)) the first values are characterized by an oscillatory behavior around zero, which also reflects on the damping of the FMR mode, $\alpha_{\vec{q}=0}$. In full agreement with Ref. \cite{thonig2018nonlocal}, we compute a peak of $\alpha_{ij}$ contribution at $r_{ij}\sim3.46a$, which shows the long-range character that non-local damping can exhibit for specific materials. Despite the relatively small magnitude of $\alpha_{ij}$, the multiplicity of the nearest neighbors shells drives a converged $\alpha_{\vec{q}}$ dispersion with non-negligible variations from the onsite value along the BZ, specially driven by the negative third neighbors. The maximum damping is found to be in the region around the high-symmetry point \textit{X}, where thus the lifetime of magnon excitations are expected to be reduced. Similar situation is found for fcc Ni (Fig. \ref{fig:non-local-damping-systems}(i,j)), where the first neighbors $\alpha_{ij}$ are found to be highly negative, consequently resulting in a spectrum in which $\alpha_{\vec{q}}>\alpha_{\vec{q}=0}$ for every $\vec{q}\neq0$. In contrast with fcc Co, however, no notable peak contributions are found.

\subsection{Remagnetization}\label{sec:remagnetization}
Gilbert damping in magnetic materials determines the rate of energy that dissipates from the magnetic to other reservoirs, like phonons or electron correlations. To explore what impact non-local damping has on the energy dissipation process, we performed atomistic spin dynamics (ASD) simulations for the aforementioned ferromagnets: bcc Fe$_{1-x}$Co$_{x}$ ($x=0\%,30\%,50\%$), fcc Co, and fcc Ni, for the \textit{(i)} fully non-local  $\alpha_{ij}$ and \textit{(ii)} effective $\alpha_{\textnormal{tot}}$(defined in \ref{eq:damp-effective})  dissipative case. We note that, although widely considered in ASD calculations, the adoption of a constant $\alpha_{\textnormal{tot}}$ value (case \textit{(ii)}) is only a good approximation for long wavelength magnons close to $\vec{q}=0$.

\begin{figure}[!htb]
       
        \includegraphics[width=\columnwidth]{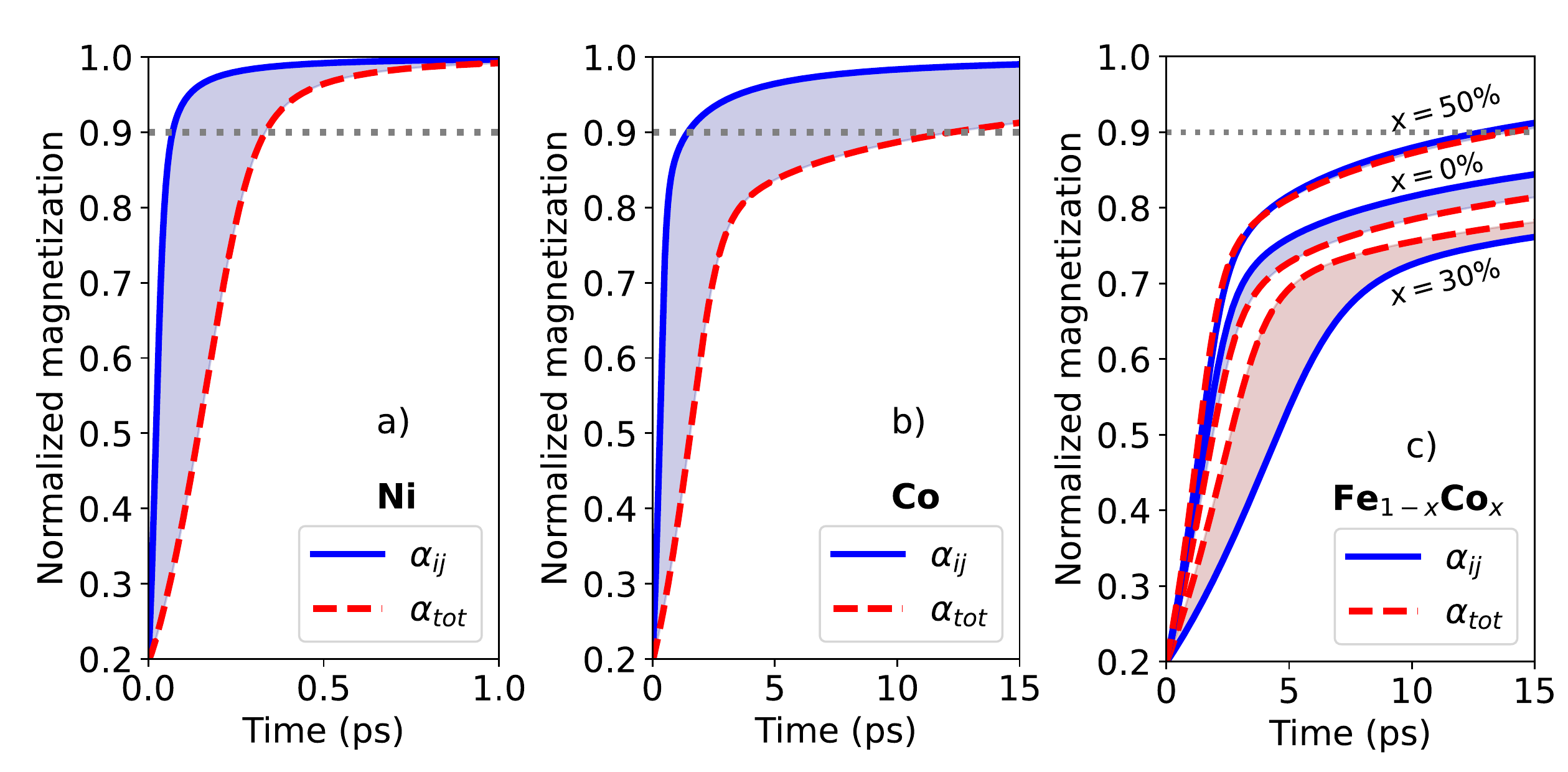} 
        \caption{Remagnetization process simulated with ASD, considering fully non-local Gilbert damping ($\alpha_{ij}$, blue sold lines), and the effective damping ($\alpha_{\textnormal{tot}}$, red dashed lines), for: (a) fcc Ni; (b) fcc Co; and (c) bcc Fe$_{1-x}$Co$_{x}$ ($x=0\%,30\%,50\%$). The dashed gray lines indicate the stage of $90\%$ of the saturation magnetization.} 
        \label{fig:spin dynamics} 
\end{figure}

First, we are interested on the role of non-local damping in the remagnetization processes as it was already discussed by Thonig \textit{et al.} \cite{thonig2018nonlocal} and as it is important for, \textit{e.g.}, ultrafast pump-probe experiments as well as all-optical switching. In the simulations presented here, the relaxation starts from a totally
random magnetic configuration. The results of re-magnetization simulations  are shown in Figure \ref{fig:spin dynamics}. The fully non-local damping \textit{(i)} in the equation of motion enhances the energy dissipation process compared to the case when only the effective damping \textit{(ii)} is used. This effect is found to be more pronounced in fcc Co and fcc Ni compared to bcc Fe and bcc Fe$_{50}$Co$_{50}$. Thus, the remagnetization time to $90\%$ of the saturation magnetisation becomes $\sim5-8$ times faster for case \textit{(i)} compared to the case \textit{(ii)}. This is due to the increase of $\alpha_{\vec{q}}$ away from the $\Gamma$ point in the whole spectrum for Co and Ni (see Fig. \ref{fig:non-local-damping-systems}), where in Fe and Fe$_{50}$Co$_{50}$ it typically oscillates around $\alpha_{\textnormal{tot}}$. 

For bcc Fe$_{70}$Co$_{30}$, the effect of non-local damping on the dynamics is opposite to the data in Fig. \ref{fig:spin dynamics}; the relaxation process is decelerated. In this case, almost the entire $\alpha_{\vec{q}}$ spectrum is below $\alpha_{\vec{q}=0}$, which is an interesting result given the fact that FMR measurements of the damping parameter in this system is already considered an ultra-low value, when compared to other metallic ferromagnets \cite{Schoen2016}. Thus, in the remagnetization process of Fe$_{70}$Co$_{30}$, the majority of magnon modes lifetimes is underestimated when a constant $\alpha_{\textnormal{tot}}$ is considered in the spin dynamics simulations, which leads to a faster overall relaxation rate. 

Although bcc Fe presents the highest Gilbert damping obtained in the series of the Fe-Co alloys (see Table \ref{tab:properties}) the remagnetization rate is found to be faster in bcc Fe$_{50}$Co$_{50}$. 
This can be explained by the fact that the exchange interactions for this particular alloy are stronger ($\sim80\%$ higher for nearest-neighbors) than in pure bcc Fe, leading to an enhanced Curie temperature (see Table \ref{tab:properties}). In view of Eq. \ref{eq:magnon lifetime} and Fig. \ref{fig:non-local-damping-systems}, the difference in the remagnetization time between bcc Fe$_{50}$Co$_{50}$ and elemental bcc Fe arises from $\alpha_{\vec{q}}$ values that are rather close, but where the magnon spectrum of Fe$_{50}$Co$_{50}$  has much higher frequencies, with corresponding faster dynamics and hence shorter remagnetization times. 

From our calculations we find that the sum of non-local damping $\left( \sum_{i \neq j} \alpha_{ij} \right)$ contributes with $-13\%$, $-81\%$, $-87\%$, $+80\%$, and $+7\%$ to the local damping in bcc Fe, fcc Co, fcc Ni, bcc Fe$_{70}$Co$_{30}$, and bcc Fe$_{50}$Co$_{50}$, respectively. 
The high positive ratio found in Fe$_{70}$Co$_{30}$ indicates that, in contrast to the other systems analyzed, the non-local contributions act like an anti-damping torque, diminishing the local damping torque. 
A similar anti-damping effect in antiferromagnetic (AFM) materials have been reported in theoretical and experimental investigations (\textit{e.g.}, \cite{chen2018antidamping,mahfouzi2018damping}), induced by electrical current. Here we find that an anti-damping torque effect can have an intrinsic origin.

To provide a deeper understanding of the anti-damping effect caused by a positive non-local contribution, we analytically solved the equation of motion for a two spin model system, e.g. a dimer. In the particular case when the onsite damping $\alpha_{11}$ is equal to the non-local contribution $\alpha_{12}$, we observed that the system becomes undamped (see Appendix \ref{undamped dimer}). As demonstrated in Appendix \ref{undamped dimer}, ASD simulations of such a dimer corroborate the result of undamped dynamics. It should be further noticed that this proposed model system was used to analyse the stability of the ASD solver, verifying whether it can preserve both the spin length and total energy. Full detail of the analytical solution and ASD simulation of a spin-dimer and the anti-damping effect are provided in Appendix \ref{undamped dimer}.
 
\subsection{Magnon spectra} \label{sec:magnon-spectra}
\begin{figure*}[!ht]
        
        \includegraphics[width=2\columnwidth]{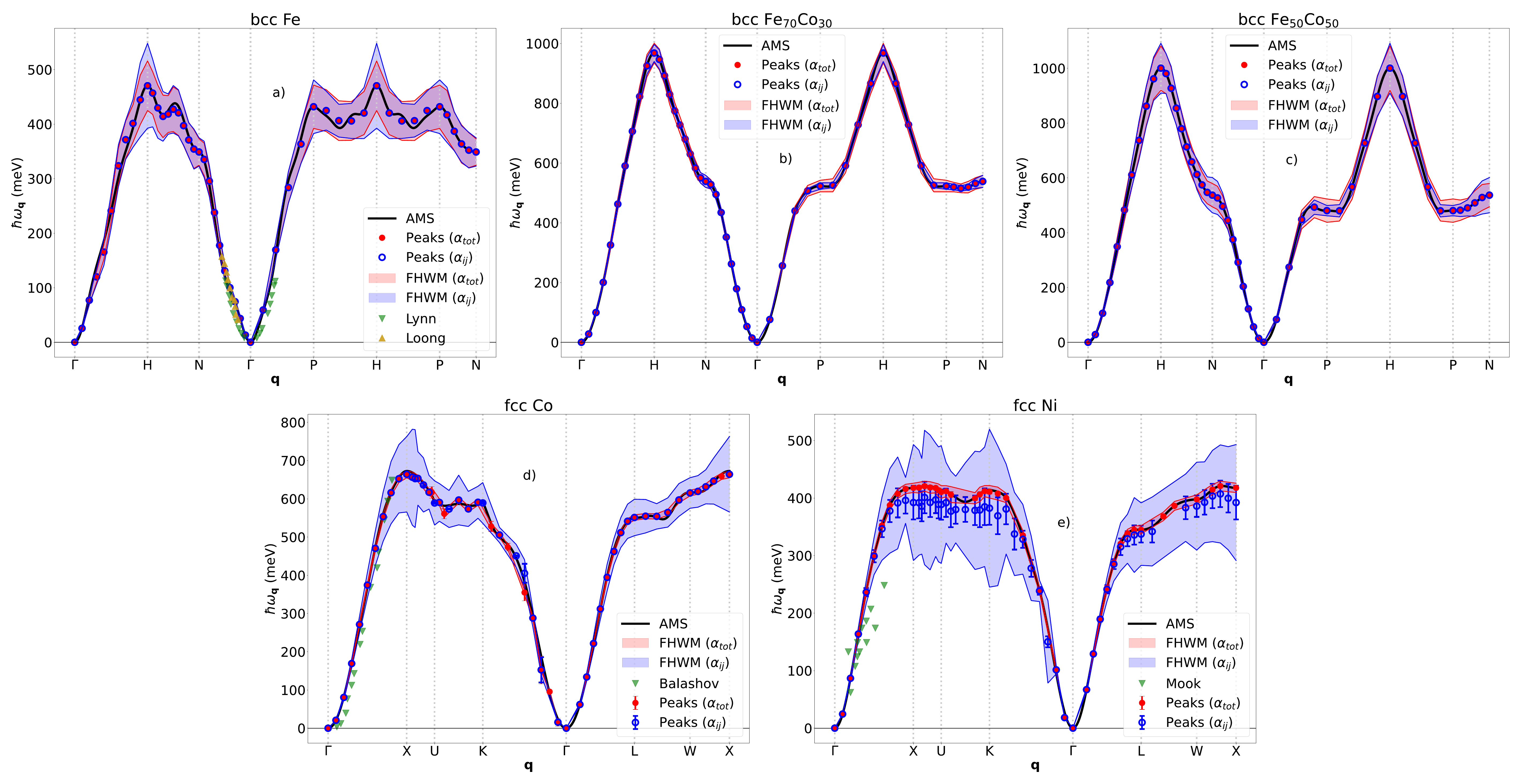} 
        \caption{Magnon spectra calculated with non-local Gilbert damping and effective Gilbert damping in: (a) bcc Fe; (b) bcc Fe$_{70}$Co$_{30}$; (c) bcc Fe$_{50}$Co$_{50}$; (d) fcc Co; and (e) fcc Ni. The black lines denote the adiabatic magnon spectra calculated from Eq. \ref{eq:sqw}. Full red and open blue points denote the peak positions of $S(\vec{q},\omega)$ at each $\vec{q}$ vector for $\alpha_{\textnormal{tot}}$ and $\alpha_{ij}$ calculations, respectively,  at $T=0.1$ K. The width of transparent red and blue areas corresponds to the full width half maximum (FWHM) on the energy axis fitted from a Lorentzian curve, following the same color scheme. To highlight the difference of FWHM between the two damping modes, the FWHMs shown in the magnon spectrum of Fe$_{1-x}$Co$_{x}$, Co, and Ni are multiplied by  20, 5, and $\frac{1}{2}$ times, in this order. The triangles represent experimental results: in (a), Fe at 10 K \cite{Loong1984} (yellow up) and Fe with $12\%$ Si at room-temperature \cite{Lynn1975} (green down); in (d), Co(9 ML)/Cu(100) at room-temperature \cite{Balashov2009} (green down); in (e) Ni at room-temperature (green down) \cite{Mook1985}. The standard deviation of the peaks are represented as error bars.} 
        \label{fig:magnon spectra} 
\end{figure*}

In order to demonstrate the influence of damping on magnon properties at finite temperatures, we have performed ASD simulations to obtain the excitation spectra from the dynamical structure factor introduced in Section \ref{sec:methods}. Here, we consider 16 NN shells for $S(\bm{q},\omega)$ calculations both from simulations that include non-local damping as well as the effective total damping (see Appendix \ref{sec:effect-further} for a focused discussion). In Fig. \ref{fig:magnon spectra}, the simulated magnon spectra of the here investigated ferromagnets are shown. 
We note that a general good agreement can be observed between our computed magnon spectra (both from the the frozen magnon approach as well as from the dynamical structure factor) and previous theoretical as well as experimental results \cite{etz2015atomistic,Pajda2001,halilov1998adiabatic,Mook1985,Lynn1975,Loong1984,Balashov2009}, where deviations from experiments is largest for fcc Ni. This exception, however, is well known and has already been discussed elsewhere \cite{katsnelson2004magnetic}.

The main feature that the non-local damping causes to the magnon spectra in all systems investigated here, is in changes of the full width at half maximum (FWHM) ${\triangle_{\vec{q}}}$ of $S(\vec{q},\omega)$. Usually, $\triangle_{\vec{q}}$ is determined from the superposition of thermal fluctuations and damping processes. More specifically, the non-local damping broadens the FWHM compared to simulations based solely on an effective damping, for most of the high-symmetry paths in all of the here analyzed ferromagnets, with the exception of Fe$_{70}$Co$_{30}$. The most extreme case is for fcc Ni, as $\alpha_{\vec{q}}$ exceeds the $0.25$ threshold for $\vec{q}=\vec{X}$, which is comparable to the damping of ultrathin magnetic films on high-SOC metallic hosts \cite{Barati2014}. As a comparison, the largest difference of FWHM between the non-local damping process and effective damping process in bcc Fe is $\sim2$ meV, while in fcc Ni the largest difference can reach $\sim258$ meV. 
In contrast, the difference is $\sim-1$ meV in $\rm Fe_{70}Co_{30}$ and the largest non-local damping effect occurs around $\vec{q}=\vec{N}$ and in the $H-P$ direction, corroborating with the discussion in Section \ref{subsec:onsite-nonlocal}. 
At the $\Gamma$ point, which corresponds to the mode measured in FMR experiments, all spins in the system have a coherent precession. This implies that $\frac{\partial \bm{m}_j}{\partial t} $ in Eq. \ref{eq:sllg_nonlocal} is the same for all moments and, thus, both damping scenarios discussed here (effecive local and the one that also takes into account non-local contributions) make no difference to the spin dynamics. As a consequence, only a tiny (negligible) difference of the FWHM is found between effective and non-local damping for the FMR mode at low temperatures.

The broadening of the FWHM on the magnon spectrum is temperature dependent. 
Thus, the effect of non-local damping to the width near $\Gamma$ can be of great interest for experiments. More specifically, taking bcc Fe as an example, the difference between width in effective damping and non-local damping process increases with temperature, where the difference can be enhanced up to one order of magnitude from $T=0.1$ K to $T=25$ K. Note that this enhancement might be misleading due to the limits of finite temperature assumption made here. This temperature dependent damping effect on FWHM suggests a path for the measurement of non-local damping in FMR experiments. 

We have also compared the difference in the imaginary part of the transverse dynamical magnetic susceptibility computed from non-local and effective damping. Defined by Eq. \ref{eq:susceptibility}, the imaginary part of susceptibility is related to the FWHM \cite{eriksson2017atomistic}. Similar to the magnon spectra shown in Fig. \ref{fig:magnon spectra}, the susceptibility difference is significant at the BZ boundaries. Taking the example of fcc Co, $\text{Im}\chi^{\pm}(\mathbf{q},\omega)$ for effective damping processes can be $11.8$ times larger than in simulations that include non-local damping processes, which is consistent to the lifetime peak that occurs at high the symmetry point, $X$, depicted in Fig. \ref{fig:magnon lifetime}. In the Fe$_{1-x}$Co$_{x}$ alloy, and $\rm Fe_{70}Co_{30}$, the largest ratio is $1.7$ and $2.7$ respectively. The intensity at $\Gamma$ point is zero since $\alpha_{\mathbf{q}}$ is independent on the coupling vector and equivalent in both damping modes. The effect of non-local damping on susceptibility coincides well with the magnon spectra from spin dynamics. Thus, this method allows us to evaluate the magnon properties in a more efficient way.

\subsection{Magnon lifetimes}

By fitting the 
$S(\vec{q},\omega)$ curve at each wave vector with a Lorentzian curve, the FWHF and hence the magnon lifetimes, $\tau_{\vec{q}}$, can be obtained from the simple relation \cite{eriksson2017atomistic}

\begin{equation}\label{eq:fwhm-magnon-lifetimes}
\tau_{\vec{q}}=\frac{2\pi}{\triangle_{\vec{q}}}.
\end{equation} 

\begin{figure*}[htpb]
        \flushleft
        \includegraphics[width=2\columnwidth]{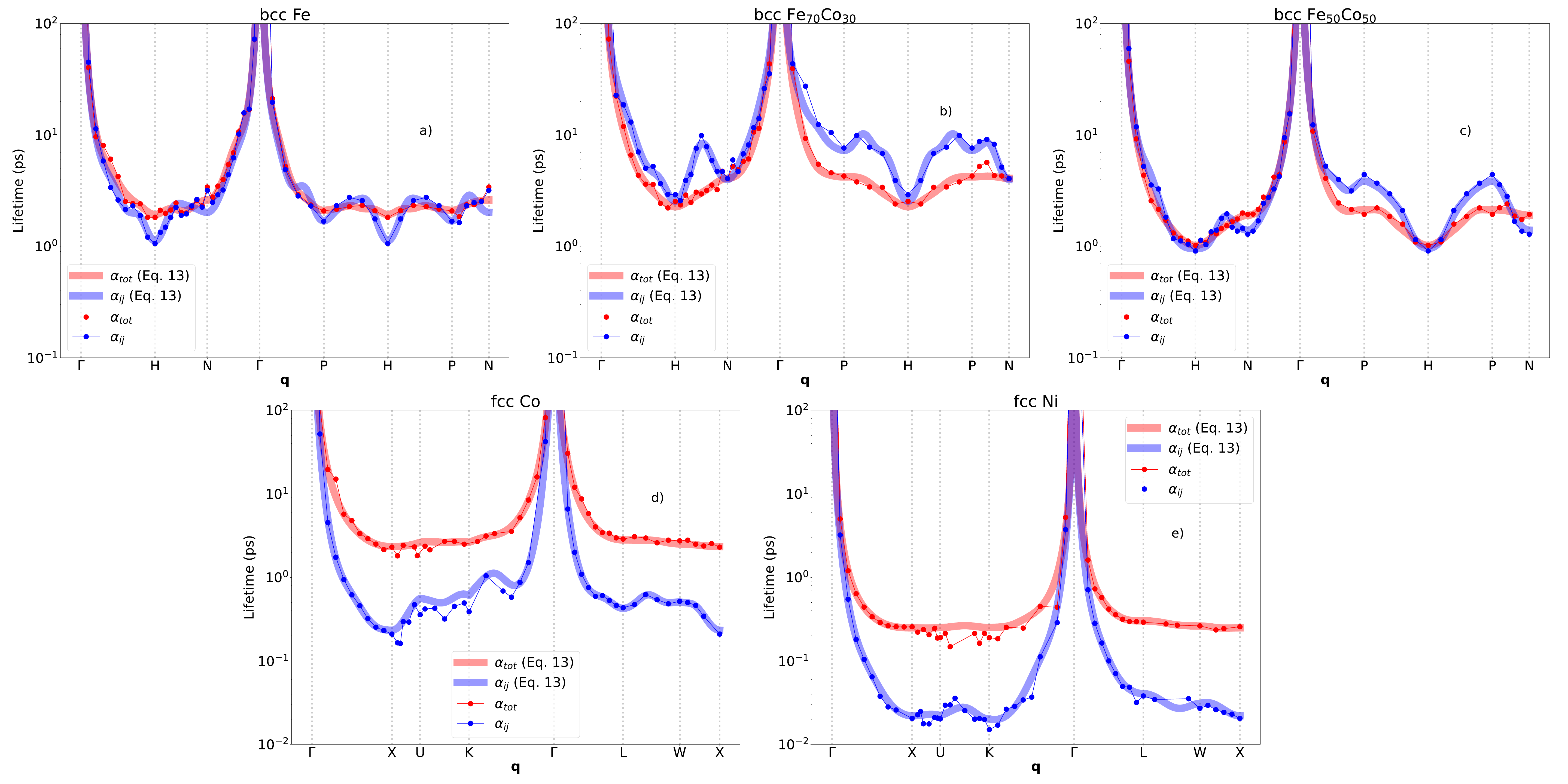} 
        \caption{Magnon lifetimes $\tau_{\vec{q}}$ of: (a) bcc Fe; (b) bcc Fe$_{70}$Co$_{30}$; (c) bcc Fe$_{50}$Co$_{50}$; (d) fcc Co;  and (e) fcc Ni as function of $\vec{q}$, shown in logarithmic scale. 
        The color scheme is the same of Fig. \ref{fig:magnon spectra}, where blue and red represents $\tau_{\vec{q}}$ computed in the effective and non-local damping models. The transparent lines and opaque points depict the lifetimes calculated with Eq. \ref{eq:magnon lifetime} and by the FWHM of $S(\vec{q},\omega)$ at $T=0.1$ K (see Eq. \ref{eq:fwhm-magnon-lifetimes}). The lifetime asymptotically diverges around the $\Gamma$-point due to the absence of anisotropy effects or external magnetic field in the spin-Hamiltonian.}
       
        \label{fig:magnon lifetime} 
\end{figure*}

Figure \ref{fig:magnon lifetime} shows the lifetimes computed in the high-symmetry lines in the BZ for all ferromagnets here investigated. As expected, $\tau_{\vec{q}}$ is much lower at the $\rm \mathbf{q}$ vectors far away from the zone center, being of the order of 1 ps for the Fe$_{1-x}$Co$_{x}$ alloys ($x=0\%,30\%,50\%$), and from $\sim0.01-1$ ps in fcc Co and Ni. In view of Eq. \ref{eq:magnon lifetime}, the magnon lifetime is inversely proportional to both damping and magnon frequency. In the effective damping process, $\rm \alpha_\mathbf{q}$ is a constant and independent of $\vec{q}$; thus,
the lifetime in the entire BZ is dictated only by $\omega_{\vec{q}}$. The situation becomes more complex in the non-local damping process, where the $\tau_{\vec{q}}$ is influenced by the combined effect of changing damping and magnon frequency. Taking $\rm Fe_{70}Co_{30}$ as an example, 
even though the $\alpha_{\vec{q}}$ is higher around the $\rm \Gamma$, the low magnon frequency compensates the damping effect, leading to an asymptotically divergent magnon lifetime as $\omega_{\vec{q}}\rightarrow0$. However, this divergence becomes finite when including e.g. magnetocrystalline anisotropy or an external magnetic field to the spin-Hamiltonian. In the $H-N$ path, the magnon energy of $\rm Fe_{70}Co_{30}$ is large, but $\alpha_{\vec{q}}$ reaches $\sim 4\times10^{-4}$ at $\vec{q}=\left(\frac{1}{4},\frac{1}{4},\frac{1}{2}\right)$, 
resulting in a magnon lifetime peak of $\sim 10$ ps. This value is not found for the effective damping model. 

In the elemental ferromagnets, as well as for Fe$_{50}$Co$_{50}$, it is found that non-local damping decreases the magnon lifetimes. This non-local damping effect is significant in both Co and Ni, where the magnon lifetimes from the $\alpha_{ij}$ model 
differ by an order of magnitude from the effective model (see Fig. \ref{fig:magnon lifetime}). In fact, considering $\tau_{\vec{q}}$ obtained from Eq. \ref{eq:magnon lifetime}, the effective model predicts a lifetime already higher by more than 50\% when the magnon frequencies are $\sim33$ meV and $\sim14$ meV in the $K-\Gamma$ path (\textit{i.e.}, near $\Gamma$) of Ni and Co, respectively. This difference mainly arises, in real-space, from the strong negative contriutions of $\alpha_{ij}$ in the close neighborhood around the reference site, namely the NN in Ni and third neighbors in Co.
In contrast, due to the $\alpha_{\vec{q}}$ spectrum composed of almost all dampings lower than $\alpha_{\textnormal{tot}}$, already discussed in Section \ref{subsec:onsite-nonlocal}, the opposite trend on $\tau_{\vec{q}}$ is observed for Fe$_{70}$Co$_{30}$: the positive overall non-local contribution guide an anti-damping effect, and the lifetimes are enhanced in the non-local model. 

Another way to evaluate the magnon lifetimes is from the linear response theory. As introduced in Section \ref{sec:magnon-dispersion}, we have access to magnon lifetimes at low temperatures from the imaginary part of the susceptibility. The $\tau_{\vec{q}}$ calculated from Eq. \ref{eq:magnon lifetime} is also displayed in Fig. \ref{fig:magnon lifetime}. Here the spin-wave frequency $\omega_{\vec{q}} $ is from the frozen magnon method. The magnon lifetimes from linear response have a very good agreement with the results from the dynamical structure factor, showing the equivalence between both methods. 
Part of the small discrepancies are related to magnon-magnon scattering induced by the temperature effect in the dynamical structure factor method. We also find a good agreement on the magnon lifetimes of effective damping in pure Fe with previous studies \cite{wu2018magnon}. They are in the similar order and decrease with the increasing magnon energy. However, their results are more diffused since the simulations are performed at room-temperature.

\section{Conclusion}
\label{sec:conclusion}

We have presented the influence of non-local damping on spin dynamics and magnon properties of elemental ferromagnets (bcc Fe, fcc Co, fcc Ni) and the bcc Fe$_{70}$Co$_{30}$ and bcc Fe$_{50}$Co$_{50}$ alloys in the virtual-crystal approximation. It is found that the non-local damping has important effects on relaxation processes and magnon properties. Regarding the relaxation process, the non-local damping in Fe, Co, and Ni has a negative contribution to the local (onsite) part, which accelerates the remagnetization. Contrarily, influenced by the positive contribution of $\alpha_{ij}$ ($i\neq j$), the magnon lifetimes of $\rm Fe_{70}Co_{30}$ and Fe$_{50}$Co$_{50}$ are increased in the non-local model, typically at the boundaries of the BZ, decelerating the remagnetization. 

Concerning the magnon properties, the non-local damping has a significant effect in Co and Ni. More specifically, the magnon lifetimes can be overestimated by an order of magnitude in the effective model for these two materials. In real-space, this difference arises as a result of strong negative non-local contributions in the close neighborhood around the reference atom, namely the NN in Ni and the third neighbors in Co.

Although the effect of non-local damping to the stochastic thermal field in spin dynamics is not included in this work, we still obtain coherent magnon lifetimes comparing to the analytical solution from linear response theory. Notably, it is predicted that the magnon lifetimes at certain wave vectors are higher for the non-local damping model in some materials. An example is $\rm Fe_{70}Co_{30}$, in which the lifetime can be $\sim3$ times higher in the $H-N$ path for the non-local model. On the other hand, we have proposed a fast method based on linear response to evaluate these lifetimes, which can be used to high-throughput computations of magnonic materials. 

Finally, our study provides a link on how non-local damping can be measured in FMR and neutron scattering experiments. Even further, it gives insight into optimising excitation of magnon modes with possible long lifetimes. This optimisation is important for any spintronics applications. As a natural consequence of any real-space \textit{ab-initio} formalism, our methodology and findings also open routes for the investigation of other materials with preferably longer lifetimes caused by non-local energy dissipation at low excitation modes. Such materials research could also include tuning the local chemical environments by doping or defects.

\section{Acknowledgments}

Financial support from Vetenskapsrådet (grant numbers VR 2016-05980 and VR 2019-05304), and the Knut and Alice Wallenberg foundation (grant number 2018.0060) is acknowledged. Support from the Swedish Research Council (VR), the Foundation for Strategic Research (SSF), the Swedish Energy Agency (Energimyndigheten), the European Research Council (854843-FASTCORR), eSSENCE and STandUP is acknowledged by O.E.~. Support from the Swedish Research Council (VR) is acknowledged by D.T.~and A.D.~. The China Scholarship Council (CSC) is acknowledged by Z.L.. The computations/data handling were enabled by resources provided by the Swedish National Infrastructure for Computing (SNIC) at the National Supercomputing Centre (NSC, Tetralith cluster), partially funded by the Swedish Research Council through grant agreement No.\,2016-07213. 
\bibliography{references.bib}

\begin{thebibliography}{116}%
\makeatletter
\providecommand \@ifxundefined [1]{%
 \@ifx{#1\undefined}
}%
\providecommand \@ifnum [1]{%
 \ifnum #1\expandafter \@firstoftwo
 \else \expandafter \@secondoftwo
 \fi
}%
\providecommand \@ifx [1]{%
 \ifx #1\expandafter \@firstoftwo
 \else \expandafter \@secondoftwo
 \fi
}%
\providecommand \natexlab [1]{#1}%
\providecommand \enquote  [1]{``#1''}%
\providecommand \bibnamefont  [1]{#1}%
\providecommand \bibfnamefont [1]{#1}%
\providecommand \citenamefont [1]{#1}%
\providecommand \href@noop [0]{\@secondoftwo}%
\providecommand \href [0]{\begingroup \@sanitize@url \@href}%
\providecommand \@href[1]{\@@startlink{#1}\@@href}%
\providecommand \@@href[1]{\endgroup#1\@@endlink}%
\providecommand \@sanitize@url [0]{\catcode `\\12\catcode `\$12\catcode
  `\&12\catcode `\#12\catcode `\^12\catcode `\_12\catcode `\%12\relax}%
\providecommand \@@startlink[1]{}%
\providecommand \@@endlink[0]{}%
\providecommand \url  [0]{\begingroup\@sanitize@url \@url }%
\providecommand \@url [1]{\endgroup\@href {#1}{\urlprefix }}%
\providecommand \urlprefix  [0]{URL }%
\providecommand \Eprint [0]{\href }%
\providecommand \doibase [0]{https://doi.org/}%
\providecommand \selectlanguage [0]{\@gobble}%
\providecommand \bibinfo  [0]{\@secondoftwo}%
\providecommand \bibfield  [0]{\@secondoftwo}%
\providecommand \translation [1]{[#1]}%
\providecommand \BibitemOpen [0]{}%
\providecommand \bibitemStop [0]{}%
\providecommand \bibitemNoStop [0]{.\EOS\space}%
\providecommand \EOS [0]{\spacefactor3000\relax}%
\providecommand \BibitemShut  [1]{\csname bibitem#1\endcsname}%
\let\auto@bib@innerbib\@empty
\bibitem [{\citenamefont {Barman}\ \emph {et~al.}(2021)\citenamefont {Barman},
  \citenamefont {Gubbiotti}, \citenamefont {Ladak}, \citenamefont {Adeyeye},
  \citenamefont {Krawczyk}, \citenamefont {Gr{\"a}fe}, \citenamefont
  {Adelmann}, \citenamefont {Cotofana}, \citenamefont {Naeemi}, \citenamefont
  {Vasyuchka} \emph {et~al.}}]{barman20212021}%
  \BibitemOpen
  \bibfield  {author} {\bibinfo {author} {\bibfnamefont {A.}~\bibnamefont
  {Barman}}, \bibinfo {author} {\bibfnamefont {G.}~\bibnamefont {Gubbiotti}},
  \bibinfo {author} {\bibfnamefont {S.}~\bibnamefont {Ladak}}, \bibinfo
  {author} {\bibfnamefont {A.~O.}\ \bibnamefont {Adeyeye}}, \bibinfo {author}
  {\bibfnamefont {M.}~\bibnamefont {Krawczyk}}, \bibinfo {author}
  {\bibfnamefont {J.}~\bibnamefont {Gr{\"a}fe}}, \bibinfo {author}
  {\bibfnamefont {C.}~\bibnamefont {Adelmann}}, \bibinfo {author}
  {\bibfnamefont {S.}~\bibnamefont {Cotofana}}, \bibinfo {author}
  {\bibfnamefont {A.}~\bibnamefont {Naeemi}}, \bibinfo {author} {\bibfnamefont
  {V.~I.}\ \bibnamefont {Vasyuchka}}, \emph {et~al.},\ }\href
  {https://doi.org/10.1088/1361-648x/abec1a} {\bibfield  {journal} {\bibinfo
  {journal} {J. Phys. Condens. Matter}\ }\textbf {\bibinfo {volume} {33}},\
  \bibinfo {pages} {413001} (\bibinfo {year} {2021})}\BibitemShut {NoStop}%
\bibitem [{\citenamefont {Pirro}\ \emph {et~al.}(2021)\citenamefont {Pirro},
  \citenamefont {Vasyuchka}, \citenamefont {Serga},\ and\ \citenamefont
  {Hillebrands}}]{pirro2021advances}%
  \BibitemOpen
  \bibfield  {author} {\bibinfo {author} {\bibfnamefont {P.}~\bibnamefont
  {Pirro}}, \bibinfo {author} {\bibfnamefont {V.~I.}\ \bibnamefont
  {Vasyuchka}}, \bibinfo {author} {\bibfnamefont {A.~A.}\ \bibnamefont
  {Serga}},\ and\ \bibinfo {author} {\bibfnamefont {B.}~\bibnamefont
  {Hillebrands}},\ }\href {https://doi.org/10.1038/s41578-021-00332-w}
  {\bibfield  {journal} {\bibinfo  {journal} {Nat. Rev. Mater}\ }\textbf
  {\bibinfo {volume} {6}},\ \bibinfo {pages} {1114} (\bibinfo {year}
  {2021})}\BibitemShut {NoStop}%
\bibitem [{\citenamefont {Rana}\ and\ \citenamefont
  {Otani}(2019)}]{rana2019towards}%
  \BibitemOpen
  \bibfield  {author} {\bibinfo {author} {\bibfnamefont {B.}~\bibnamefont
  {Rana}}\ and\ \bibinfo {author} {\bibfnamefont {Y.}~\bibnamefont {Otani}},\
  }\href {https://doi.org/10.1038/s42005-019-0189-6} {\bibfield  {journal}
  {\bibinfo  {journal} {Commun. Phys}\ }\textbf {\bibinfo {volume} {2}},\
  \bibinfo {pages} {1} (\bibinfo {year} {2019})}\BibitemShut {NoStop}%
\bibitem [{\citenamefont {Mahmoud}\ \emph {et~al.}(2020)\citenamefont
  {Mahmoud}, \citenamefont {Ciubotaru}, \citenamefont {Vanderveken},
  \citenamefont {Chumak}, \citenamefont {Hamdioui}, \citenamefont {Adelmann},\
  and\ \citenamefont {Cotofana}}]{mahmoud2020introduction}%
  \BibitemOpen
  \bibfield  {author} {\bibinfo {author} {\bibfnamefont {A.}~\bibnamefont
  {Mahmoud}}, \bibinfo {author} {\bibfnamefont {F.}~\bibnamefont {Ciubotaru}},
  \bibinfo {author} {\bibfnamefont {F.}~\bibnamefont {Vanderveken}}, \bibinfo
  {author} {\bibfnamefont {A.~V.}\ \bibnamefont {Chumak}}, \bibinfo {author}
  {\bibfnamefont {S.}~\bibnamefont {Hamdioui}}, \bibinfo {author}
  {\bibfnamefont {C.}~\bibnamefont {Adelmann}},\ and\ \bibinfo {author}
  {\bibfnamefont {S.}~\bibnamefont {Cotofana}},\ }\href
  {https://doi.org/10.1063/5.0019328} {\bibfield  {journal} {\bibinfo
  {journal} {J. Appl. Phys.}\ }\textbf {\bibinfo {volume} {128}},\ \bibinfo
  {pages} {161101} (\bibinfo {year} {2020})}\BibitemShut {NoStop}%
\bibitem [{\citenamefont {Serga}\ \emph {et~al.}(2010)\citenamefont {Serga},
  \citenamefont {Chumak},\ and\ \citenamefont {Hillebrands}}]{serga2010yig}%
  \BibitemOpen
  \bibfield  {author} {\bibinfo {author} {\bibfnamefont {A.}~\bibnamefont
  {Serga}}, \bibinfo {author} {\bibfnamefont {A.}~\bibnamefont {Chumak}},\ and\
  \bibinfo {author} {\bibfnamefont {B.}~\bibnamefont {Hillebrands}},\ }\href
  {https://doi.org/10.1088/0022-3727/43/26/264002} {\bibfield  {journal}
  {\bibinfo  {journal} {J. Phys. D: Appl. Phys.}\ }\textbf {\bibinfo {volume}
  {43}},\ \bibinfo {pages} {264002} (\bibinfo {year} {2010})}\BibitemShut
  {NoStop}%
\bibitem [{\citenamefont {Lendinez}\ and\ \citenamefont
  {Jungfleisch}(2019)}]{lendinez2019magnetization}%
  \BibitemOpen
  \bibfield  {author} {\bibinfo {author} {\bibfnamefont {S.}~\bibnamefont
  {Lendinez}}\ and\ \bibinfo {author} {\bibfnamefont {M.}~\bibnamefont
  {Jungfleisch}},\ }\href {https://doi.org/10.1088/1361-648x/ab3e78} {\bibfield
   {journal} {\bibinfo  {journal} {J. Phys. Condens. Matter}\ }\textbf
  {\bibinfo {volume} {32}},\ \bibinfo {pages} {013001} (\bibinfo {year}
  {2019})}\BibitemShut {NoStop}%
\bibitem [{\citenamefont {Zakeri}(2020)}]{zakeri2020magnonic}%
  \BibitemOpen
  \bibfield  {author} {\bibinfo {author} {\bibfnamefont {K.}~\bibnamefont
  {Zakeri}},\ }\href {https://doi.org/10.1088/1361-648x/ab88f2} {\bibfield
  {journal} {\bibinfo  {journal} {J. Phys. Condens. Matter}\ }\textbf {\bibinfo
  {volume} {32}},\ \bibinfo {pages} {363001} (\bibinfo {year}
  {2020})}\BibitemShut {NoStop}%
\bibitem [{\citenamefont {Awschalom}\ \emph {et~al.}(2021)\citenamefont
  {Awschalom}, \citenamefont {Du}, \citenamefont {He}, \citenamefont
  {Heremans}, \citenamefont {Hoffmann}, \citenamefont {Hou}, \citenamefont
  {Kurebayashi}, \citenamefont {Li}, \citenamefont {Liu}, \citenamefont
  {Novosad} \emph {et~al.}}]{awschalom2021quantum}%
  \BibitemOpen
  \bibfield  {author} {\bibinfo {author} {\bibfnamefont {D.~D.}\ \bibnamefont
  {Awschalom}}, \bibinfo {author} {\bibfnamefont {C.}~\bibnamefont {Du}},
  \bibinfo {author} {\bibfnamefont {R.}~\bibnamefont {He}}, \bibinfo {author}
  {\bibfnamefont {J.}~\bibnamefont {Heremans}}, \bibinfo {author}
  {\bibfnamefont {A.}~\bibnamefont {Hoffmann}}, \bibinfo {author}
  {\bibfnamefont {J.}~\bibnamefont {Hou}}, \bibinfo {author} {\bibfnamefont
  {H.}~\bibnamefont {Kurebayashi}}, \bibinfo {author} {\bibfnamefont
  {Y.}~\bibnamefont {Li}}, \bibinfo {author} {\bibfnamefont {L.}~\bibnamefont
  {Liu}}, \bibinfo {author} {\bibfnamefont {V.}~\bibnamefont {Novosad}}, \emph
  {et~al.},\ }\bibfield  {journal} {\bibinfo  {journal} {IEEE Trans. Quantum
  Eng}\ }\href {https://doi.org/10.1109/TQE.2021.3057799}
  {10.1109/TQE.2021.3057799} (\bibinfo {year} {2021})\BibitemShut {NoStop}%
\bibitem [{\citenamefont {Chen}\ and\ \citenamefont
  {Ma}(2021)}]{chen2021skyrmion}%
  \BibitemOpen
  \bibfield  {author} {\bibinfo {author} {\bibfnamefont {Z.}~\bibnamefont
  {Chen}}\ and\ \bibinfo {author} {\bibfnamefont {F.}~\bibnamefont {Ma}},\
  }\href {https://doi.org/10.1063/5.0061832} {\bibfield  {journal} {\bibinfo
  {journal} {J. Appl. Phys.}\ }\textbf {\bibinfo {volume} {130}},\ \bibinfo
  {pages} {090901} (\bibinfo {year} {2021})}\BibitemShut {NoStop}%
\bibitem [{\citenamefont {Lenk}\ \emph {et~al.}(2011)\citenamefont {Lenk},
  \citenamefont {Ulrichs}, \citenamefont {Garbs},\ and\ \citenamefont
  {M{\"u}nzenberg}}]{lenk2011building}%
  \BibitemOpen
  \bibfield  {author} {\bibinfo {author} {\bibfnamefont {B.}~\bibnamefont
  {Lenk}}, \bibinfo {author} {\bibfnamefont {H.}~\bibnamefont {Ulrichs}},
  \bibinfo {author} {\bibfnamefont {F.}~\bibnamefont {Garbs}},\ and\ \bibinfo
  {author} {\bibfnamefont {M.}~\bibnamefont {M{\"u}nzenberg}},\ }\href
  {https://doi.org/https://doi.org/10.1016/j.physrep.2011.06.003} {\bibfield
  {journal} {\bibinfo  {journal} {Phys. Rep.}\ }\textbf {\bibinfo {volume}
  {507}},\ \bibinfo {pages} {107} (\bibinfo {year} {2011})}\BibitemShut
  {NoStop}%
\bibitem [{\citenamefont {Liu}\ \emph {et~al.}(2018)\citenamefont {Liu},
  \citenamefont {Chen}, \citenamefont {Liu}, \citenamefont {Heimbach},
  \citenamefont {Yu}, \citenamefont {Xiao}, \citenamefont {Hu}, \citenamefont
  {Liu}, \citenamefont {Chang}, \citenamefont {Stueckler} \emph
  {et~al.}}]{liu2018long}%
  \BibitemOpen
  \bibfield  {author} {\bibinfo {author} {\bibfnamefont {C.}~\bibnamefont
  {Liu}}, \bibinfo {author} {\bibfnamefont {J.}~\bibnamefont {Chen}}, \bibinfo
  {author} {\bibfnamefont {T.}~\bibnamefont {Liu}}, \bibinfo {author}
  {\bibfnamefont {F.}~\bibnamefont {Heimbach}}, \bibinfo {author}
  {\bibfnamefont {H.}~\bibnamefont {Yu}}, \bibinfo {author} {\bibfnamefont
  {Y.}~\bibnamefont {Xiao}}, \bibinfo {author} {\bibfnamefont {J.}~\bibnamefont
  {Hu}}, \bibinfo {author} {\bibfnamefont {M.}~\bibnamefont {Liu}}, \bibinfo
  {author} {\bibfnamefont {H.}~\bibnamefont {Chang}}, \bibinfo {author}
  {\bibfnamefont {T.}~\bibnamefont {Stueckler}}, \emph {et~al.},\ }\href
  {https://doi.org/10.1038/s41467-018-03199-8} {\bibfield  {journal} {\bibinfo
  {journal} {Nat. Commun.}\ }\textbf {\bibinfo {volume} {9}},\ \bibinfo {pages}
  {1} (\bibinfo {year} {2018})}\BibitemShut {NoStop}%
\bibitem [{\citenamefont {Sheng}\ \emph {et~al.}(2021)\citenamefont {Sheng},
  \citenamefont {Chen}, \citenamefont {Wang},\ and\ \citenamefont
  {Yu}}]{sheng2021magnonics}%
  \BibitemOpen
  \bibfield  {author} {\bibinfo {author} {\bibfnamefont {L.}~\bibnamefont
  {Sheng}}, \bibinfo {author} {\bibfnamefont {J.}~\bibnamefont {Chen}},
  \bibinfo {author} {\bibfnamefont {H.}~\bibnamefont {Wang}},\ and\ \bibinfo
  {author} {\bibfnamefont {H.}~\bibnamefont {Yu}},\ }\href
  {https://doi.org/10.7566/JPSJ.90.081005} {\bibfield  {journal} {\bibinfo
  {journal} {J. Phys. Soc. Jpn.}\ }\textbf {\bibinfo {volume} {90}},\ \bibinfo
  {pages} {081005} (\bibinfo {year} {2021})}\BibitemShut {NoStop}%
\bibitem [{\citenamefont {Burrowes}\ \emph {et~al.}(2012)\citenamefont
  {Burrowes}, \citenamefont {Heinrich}, \citenamefont {Kardasz}, \citenamefont
  {Montoya}, \citenamefont {Girt}, \citenamefont {Sun}, \citenamefont {Song},\
  and\ \citenamefont {Wu}}]{burrowes2012enhanced}%
  \BibitemOpen
  \bibfield  {author} {\bibinfo {author} {\bibfnamefont {C.}~\bibnamefont
  {Burrowes}}, \bibinfo {author} {\bibfnamefont {B.}~\bibnamefont {Heinrich}},
  \bibinfo {author} {\bibfnamefont {B.}~\bibnamefont {Kardasz}}, \bibinfo
  {author} {\bibfnamefont {E.}~\bibnamefont {Montoya}}, \bibinfo {author}
  {\bibfnamefont {E.}~\bibnamefont {Girt}}, \bibinfo {author} {\bibfnamefont
  {Y.}~\bibnamefont {Sun}}, \bibinfo {author} {\bibfnamefont {Y.-Y.}\
  \bibnamefont {Song}},\ and\ \bibinfo {author} {\bibfnamefont
  {M.}~\bibnamefont {Wu}},\ }\href {https://doi.org/10.1063/1.3690918}
  {\bibfield  {journal} {\bibinfo  {journal} {Appl. Phys. Lett.}\ }\textbf
  {\bibinfo {volume} {100}},\ \bibinfo {pages} {092403} (\bibinfo {year}
  {2012})}\BibitemShut {NoStop}%
\bibitem [{\citenamefont {Correa}\ \emph {et~al.}(2019)\citenamefont {Correa},
  \citenamefont {Santos}, \citenamefont {Silva}, \citenamefont {Raza},
  \citenamefont {Della~Pace}, \citenamefont {Chesman}, \citenamefont {Sommer},\
  and\ \citenamefont {Bohn}}]{correa2019exploring}%
  \BibitemOpen
  \bibfield  {author} {\bibinfo {author} {\bibfnamefont {M.}~\bibnamefont
  {Correa}}, \bibinfo {author} {\bibfnamefont {J.}~\bibnamefont {Santos}},
  \bibinfo {author} {\bibfnamefont {B.}~\bibnamefont {Silva}}, \bibinfo
  {author} {\bibfnamefont {S.}~\bibnamefont {Raza}}, \bibinfo {author}
  {\bibfnamefont {R.}~\bibnamefont {Della~Pace}}, \bibinfo {author}
  {\bibfnamefont {C.}~\bibnamefont {Chesman}}, \bibinfo {author} {\bibfnamefont
  {R.}~\bibnamefont {Sommer}},\ and\ \bibinfo {author} {\bibfnamefont
  {F.}~\bibnamefont {Bohn}},\ }\href
  {https://doi.org/https://doi.org/10.1016/j.jmmm.2019.04.072} {\bibfield
  {journal} {\bibinfo  {journal} {J. Magn. Magn. Mater}\ }\textbf {\bibinfo
  {volume} {485}},\ \bibinfo {pages} {75} (\bibinfo {year} {2019})}\BibitemShut
  {NoStop}%
\bibitem [{\citenamefont {Eriksson}\ \emph {et~al.}(2017)\citenamefont
  {Eriksson}, \citenamefont {Bergman}, \citenamefont {Bergqvist},\ and\
  \citenamefont {Hellsvik}}]{eriksson2017atomistic}%
  \BibitemOpen
  \bibfield  {author} {\bibinfo {author} {\bibfnamefont {O.}~\bibnamefont
  {Eriksson}}, \bibinfo {author} {\bibfnamefont {A.}~\bibnamefont {Bergman}},
  \bibinfo {author} {\bibfnamefont {L.}~\bibnamefont {Bergqvist}},\ and\
  \bibinfo {author} {\bibfnamefont {J.}~\bibnamefont {Hellsvik}},\ }\href@noop
  {} {\emph {\bibinfo {title} {Atomistic spin dynamics: Foundations and
  applications}}}\ (\bibinfo  {publisher} {Oxford university press},\ \bibinfo
  {year} {2017})\BibitemShut {NoStop}%
\bibitem [{\citenamefont {Gilbert}(2004)}]{gilbert2004phenomenological}%
  \BibitemOpen
  \bibfield  {author} {\bibinfo {author} {\bibfnamefont {T.~L.}\ \bibnamefont
  {Gilbert}},\ }\href {https://doi.org/10.1109/TMAG.2004.836740} {\bibfield
  {journal} {\bibinfo  {journal} {IEEE Trans. Magn.}\ }\textbf {\bibinfo
  {volume} {40}},\ \bibinfo {pages} {3443} (\bibinfo {year}
  {2004})}\BibitemShut {NoStop}%
\bibitem [{\citenamefont {Gilmore}\ \emph {et~al.}(2010)\citenamefont
  {Gilmore}, \citenamefont {Stiles}, \citenamefont {Seib}, \citenamefont
  {Steiauf},\ and\ \citenamefont {F{\"a}hnle}}]{gilmore2010anisotropic}%
  \BibitemOpen
  \bibfield  {author} {\bibinfo {author} {\bibfnamefont {K.}~\bibnamefont
  {Gilmore}}, \bibinfo {author} {\bibfnamefont {M.}~\bibnamefont {Stiles}},
  \bibinfo {author} {\bibfnamefont {J.}~\bibnamefont {Seib}}, \bibinfo {author}
  {\bibfnamefont {D.}~\bibnamefont {Steiauf}},\ and\ \bibinfo {author}
  {\bibfnamefont {M.}~\bibnamefont {F{\"a}hnle}},\ }\href
  {https://doi.org/10.1103/PhysRevB.81.174414} {\bibfield  {journal} {\bibinfo
  {journal} {Phys. Rev. B}\ }\textbf {\bibinfo {volume} {81}},\ \bibinfo
  {pages} {174414} (\bibinfo {year} {2010})}\BibitemShut {NoStop}%
\bibitem [{\citenamefont {F{\"a}hnle}\ and\ \citenamefont
  {Steiauf}(2006)}]{fahnle2006breathing}%
  \BibitemOpen
  \bibfield  {author} {\bibinfo {author} {\bibfnamefont {M.}~\bibnamefont
  {F{\"a}hnle}}\ and\ \bibinfo {author} {\bibfnamefont {D.}~\bibnamefont
  {Steiauf}},\ }\href {https://doi.org/10.1103/PhysRevB.73.184427} {\bibfield
  {journal} {\bibinfo  {journal} {Phys. Rev. B}\ }\textbf {\bibinfo {volume}
  {73}},\ \bibinfo {pages} {184427} (\bibinfo {year} {2006})}\BibitemShut
  {NoStop}%
\bibitem [{\citenamefont {Bhattacharjee}\ \emph {et~al.}(2012)\citenamefont
  {Bhattacharjee}, \citenamefont {Nordstr{\"o}m},\ and\ \citenamefont
  {Fransson}}]{bhattacharjee2012atomistic}%
  \BibitemOpen
  \bibfield  {author} {\bibinfo {author} {\bibfnamefont {S.}~\bibnamefont
  {Bhattacharjee}}, \bibinfo {author} {\bibfnamefont {L.}~\bibnamefont
  {Nordstr{\"o}m}},\ and\ \bibinfo {author} {\bibfnamefont {J.}~\bibnamefont
  {Fransson}},\ }\href {https://doi.org/10.1103/PhysRevLett.108.057204}
  {\bibfield  {journal} {\bibinfo  {journal} {Phys. Rev. Lett.}\ }\textbf
  {\bibinfo {volume} {108}},\ \bibinfo {pages} {057204} (\bibinfo {year}
  {2012})}\BibitemShut {NoStop}%
\bibitem [{\citenamefont {Thonig}\ \emph {et~al.}(2018)\citenamefont {Thonig},
  \citenamefont {Kvashnin}, \citenamefont {Eriksson},\ and\ \citenamefont
  {Pereiro}}]{thonig2018nonlocal}%
  \BibitemOpen
  \bibfield  {author} {\bibinfo {author} {\bibfnamefont {D.}~\bibnamefont
  {Thonig}}, \bibinfo {author} {\bibfnamefont {Y.}~\bibnamefont {Kvashnin}},
  \bibinfo {author} {\bibfnamefont {O.}~\bibnamefont {Eriksson}},\ and\
  \bibinfo {author} {\bibfnamefont {M.}~\bibnamefont {Pereiro}},\ }\href
  {https://doi.org/10.1103/PhysRevMaterials.2.013801} {\bibfield  {journal}
  {\bibinfo  {journal} {Phys. Rev. Mater.}\ }\textbf {\bibinfo {volume} {2}},\
  \bibinfo {pages} {013801} (\bibinfo {year} {2018})}\BibitemShut {NoStop}%
\bibitem [{\citenamefont {Kambersk{\`y}}(1976)}]{kambersky1976ferromagnetic}%
  \BibitemOpen
  \bibfield  {author} {\bibinfo {author} {\bibfnamefont {V.}~\bibnamefont
  {Kambersk{\`y}}},\ }\href {https://doi.org/10.1007/BF01587621} {\bibfield
  {journal} {\bibinfo  {journal} {Czechoslovak Journal of Physics}\ }\textbf
  {\bibinfo {volume} {26}},\ \bibinfo {pages} {1366} (\bibinfo {year}
  {1976})}\BibitemShut {NoStop}%
\bibitem [{\citenamefont {Gilmore}\ \emph {et~al.}(2007)\citenamefont
  {Gilmore}, \citenamefont {Idzerda},\ and\ \citenamefont
  {Stiles}}]{gilmore2007identification}%
  \BibitemOpen
  \bibfield  {author} {\bibinfo {author} {\bibfnamefont {K.}~\bibnamefont
  {Gilmore}}, \bibinfo {author} {\bibfnamefont {Y.}~\bibnamefont {Idzerda}},\
  and\ \bibinfo {author} {\bibfnamefont {M.~D.}\ \bibnamefont {Stiles}},\
  }\href {https://doi.org/10.1103/PhysRevLett.99.027204} {\bibfield  {journal}
  {\bibinfo  {journal} {Phys. Rev. Lett.}\ }\textbf {\bibinfo {volume} {99}},\
  \bibinfo {pages} {027204} (\bibinfo {year} {2007})}\BibitemShut {NoStop}%
\bibitem [{\citenamefont {Ebert}\ \emph {et~al.}(2011)\citenamefont {Ebert},
  \citenamefont {Mankovsky}, \citenamefont {K{\"o}dderitzsch},\ and\
  \citenamefont {Kelly}}]{ebert2011ab}%
  \BibitemOpen
  \bibfield  {author} {\bibinfo {author} {\bibfnamefont {H.}~\bibnamefont
  {Ebert}}, \bibinfo {author} {\bibfnamefont {S.}~\bibnamefont {Mankovsky}},
  \bibinfo {author} {\bibfnamefont {D.}~\bibnamefont {K{\"o}dderitzsch}},\ and\
  \bibinfo {author} {\bibfnamefont {P.~J.}\ \bibnamefont {Kelly}},\ }\href
  {https://doi.org/10.1103/PhysRevLett.107.066603} {\bibfield  {journal}
  {\bibinfo  {journal} {Phys. Rev. Lett.}\ }\textbf {\bibinfo {volume} {107}},\
  \bibinfo {pages} {066603} (\bibinfo {year} {2011})}\BibitemShut {NoStop}%
\bibitem [{\citenamefont {Brinker}\ \emph {et~al.}(2022)\citenamefont
  {Brinker}, \citenamefont {dos Santos~Dias},\ and\ \citenamefont
  {Lounis}}]{brinker2022generalization}%
  \BibitemOpen
  \bibfield  {author} {\bibinfo {author} {\bibfnamefont {S.}~\bibnamefont
  {Brinker}}, \bibinfo {author} {\bibfnamefont {M.}~\bibnamefont {dos
  Santos~Dias}},\ and\ \bibinfo {author} {\bibfnamefont {S.}~\bibnamefont
  {Lounis}},\ }\href {https://doi.org/10.1088/1361-648x/ac699d} {\bibfield
  {journal} {\bibinfo  {journal} {J. Phys. Condens. Matter}\ }\textbf {\bibinfo
  {volume} {34}},\ \bibinfo {pages} {285802} (\bibinfo {year}
  {2022})}\BibitemShut {NoStop}%
\bibitem [{\citenamefont {Li}\ \emph {et~al.}(2019)\citenamefont {Li},
  \citenamefont {Zeng}, \citenamefont {Zhang}, \citenamefont {Shin},
  \citenamefont {Saglam}, \citenamefont {Karakas}, \citenamefont {Ozatay},
  \citenamefont {Pearson}, \citenamefont {Heinonen}, \citenamefont {Wu} \emph
  {et~al.}}]{li2019giant}%
  \BibitemOpen
  \bibfield  {author} {\bibinfo {author} {\bibfnamefont {Y.}~\bibnamefont
  {Li}}, \bibinfo {author} {\bibfnamefont {F.}~\bibnamefont {Zeng}}, \bibinfo
  {author} {\bibfnamefont {S.~S.-L.}\ \bibnamefont {Zhang}}, \bibinfo {author}
  {\bibfnamefont {H.}~\bibnamefont {Shin}}, \bibinfo {author} {\bibfnamefont
  {H.}~\bibnamefont {Saglam}}, \bibinfo {author} {\bibfnamefont
  {V.}~\bibnamefont {Karakas}}, \bibinfo {author} {\bibfnamefont
  {O.}~\bibnamefont {Ozatay}}, \bibinfo {author} {\bibfnamefont {J.~E.}\
  \bibnamefont {Pearson}}, \bibinfo {author} {\bibfnamefont {O.~G.}\
  \bibnamefont {Heinonen}}, \bibinfo {author} {\bibfnamefont {Y.}~\bibnamefont
  {Wu}}, \emph {et~al.},\ }\href
  {https://doi.org/10.1103/PhysRevLett.122.117203} {\bibfield  {journal}
  {\bibinfo  {journal} {Phys. Rev. Lett.}\ }\textbf {\bibinfo {volume} {122}},\
  \bibinfo {pages} {117203} (\bibinfo {year} {2019})}\BibitemShut {NoStop}%
\bibitem [{\citenamefont {Weindler}\ \emph {et~al.}(2014)\citenamefont
  {Weindler}, \citenamefont {Bauer}, \citenamefont {Islinger}, \citenamefont
  {Boehm}, \citenamefont {Chauleau},\ and\ \citenamefont
  {Back}}]{weindler2014magnetic}%
  \BibitemOpen
  \bibfield  {author} {\bibinfo {author} {\bibfnamefont {T.}~\bibnamefont
  {Weindler}}, \bibinfo {author} {\bibfnamefont {H.}~\bibnamefont {Bauer}},
  \bibinfo {author} {\bibfnamefont {R.}~\bibnamefont {Islinger}}, \bibinfo
  {author} {\bibfnamefont {B.}~\bibnamefont {Boehm}}, \bibinfo {author}
  {\bibfnamefont {J.-Y.}\ \bibnamefont {Chauleau}},\ and\ \bibinfo {author}
  {\bibfnamefont {C.}~\bibnamefont {Back}},\ }\href
  {https://doi.org/10.1103/PhysRevLett.113.237204} {\bibfield  {journal}
  {\bibinfo  {journal} {Phys. Rev. Lett.}\ }\textbf {\bibinfo {volume} {113}},\
  \bibinfo {pages} {237204} (\bibinfo {year} {2014})}\BibitemShut {NoStop}%
\bibitem [{\citenamefont {Bar'yakhtar}(1984)}]{bar1984phenomenological}%
  \BibitemOpen
  \bibfield  {author} {\bibinfo {author} {\bibfnamefont {V.}~\bibnamefont
  {Bar'yakhtar}},\ }\href@noop {} {\bibfield  {journal} {\bibinfo  {journal}
  {Sov. Phys. JETP}\ }\textbf {\bibinfo {volume} {60}},\ \bibinfo {pages} {863}
  (\bibinfo {year} {1984})}\BibitemShut {NoStop}%
\bibitem [{\citenamefont {Dvornik}\ \emph {et~al.}(2013)\citenamefont
  {Dvornik}, \citenamefont {Vansteenkiste},\ and\ \citenamefont
  {Van~Waeyenberge}}]{dvornik2013micromagnetic}%
  \BibitemOpen
  \bibfield  {author} {\bibinfo {author} {\bibfnamefont {M.}~\bibnamefont
  {Dvornik}}, \bibinfo {author} {\bibfnamefont {A.}~\bibnamefont
  {Vansteenkiste}},\ and\ \bibinfo {author} {\bibfnamefont {B.}~\bibnamefont
  {Van~Waeyenberge}},\ }\href {https://doi.org/10.1103/PhysRevB.88.054427}
  {\bibfield  {journal} {\bibinfo  {journal} {Phys. Rev. B}\ }\textbf {\bibinfo
  {volume} {88}},\ \bibinfo {pages} {054427} (\bibinfo {year}
  {2013})}\BibitemShut {NoStop}%
\bibitem [{\citenamefont {Wang}\ \emph {et~al.}(2015)\citenamefont {Wang},
  \citenamefont {Dvornik}, \citenamefont {Bisotti}, \citenamefont
  {Chernyshenko}, \citenamefont {Beg}, \citenamefont {Albert}, \citenamefont
  {Vansteenkiste}, \citenamefont {Waeyenberge}, \citenamefont {Kuchko},
  \citenamefont {Kruglyak} \emph {et~al.}}]{wang2015phenomenological}%
  \BibitemOpen
  \bibfield  {author} {\bibinfo {author} {\bibfnamefont {W.}~\bibnamefont
  {Wang}}, \bibinfo {author} {\bibfnamefont {M.}~\bibnamefont {Dvornik}},
  \bibinfo {author} {\bibfnamefont {M.-A.}\ \bibnamefont {Bisotti}}, \bibinfo
  {author} {\bibfnamefont {D.}~\bibnamefont {Chernyshenko}}, \bibinfo {author}
  {\bibfnamefont {M.}~\bibnamefont {Beg}}, \bibinfo {author} {\bibfnamefont
  {M.}~\bibnamefont {Albert}}, \bibinfo {author} {\bibfnamefont
  {A.}~\bibnamefont {Vansteenkiste}}, \bibinfo {author} {\bibfnamefont {B.~V.}\
  \bibnamefont {Waeyenberge}}, \bibinfo {author} {\bibfnamefont {A.~N.}\
  \bibnamefont {Kuchko}}, \bibinfo {author} {\bibfnamefont {V.~V.}\
  \bibnamefont {Kruglyak}}, \emph {et~al.},\ }\href
  {https://doi.org/10.1103/PhysRevB.92.054430} {\bibfield  {journal} {\bibinfo
  {journal} {Phys. Rev. B}\ }\textbf {\bibinfo {volume} {92}},\ \bibinfo
  {pages} {054430} (\bibinfo {year} {2015})}\BibitemShut {NoStop}%
\bibitem [{\citenamefont {Ma}\ and\ \citenamefont
  {Seiler}(2017)}]{ma2017metrology}%
  \BibitemOpen
  \bibfield  {author} {\bibinfo {author} {\bibfnamefont {Z.}~\bibnamefont
  {Ma}}\ and\ \bibinfo {author} {\bibfnamefont {D.~G.}\ \bibnamefont
  {Seiler}},\ }\href@noop {} {\emph {\bibinfo {title} {Metrology and Diagnostic
  Techniques for Nanoelectronics}}}\ (\bibinfo  {publisher} {Jenny Stanford
  Publishing},\ \bibinfo {year} {2017})\BibitemShut {NoStop}%
\bibitem [{\citenamefont {Zhu}\ \emph {et~al.}(2019)\citenamefont {Zhu},
  \citenamefont {Zhu}, \citenamefont {Li}, \citenamefont {Wu}, \citenamefont
  {Xi}, \citenamefont {Jin},\ and\ \citenamefont
  {Zhang}}]{zhu2019magnetization}%
  \BibitemOpen
  \bibfield  {author} {\bibinfo {author} {\bibfnamefont {W.}~\bibnamefont
  {Zhu}}, \bibinfo {author} {\bibfnamefont {Z.}~\bibnamefont {Zhu}}, \bibinfo
  {author} {\bibfnamefont {D.}~\bibnamefont {Li}}, \bibinfo {author}
  {\bibfnamefont {G.}~\bibnamefont {Wu}}, \bibinfo {author} {\bibfnamefont
  {L.}~\bibnamefont {Xi}}, \bibinfo {author} {\bibfnamefont {Q.}~\bibnamefont
  {Jin}},\ and\ \bibinfo {author} {\bibfnamefont {Z.}~\bibnamefont {Zhang}},\
  }\href {https://doi.org/https://doi.org/10.1016/j.jmmm.2019.01.087}
  {\bibfield  {journal} {\bibinfo  {journal} {J. Magn. Magn. Mater}\ }\textbf
  {\bibinfo {volume} {479}},\ \bibinfo {pages} {179} (\bibinfo {year}
  {2019})}\BibitemShut {NoStop}%
\bibitem [{\citenamefont {Urban}\ \emph {et~al.}(2001)\citenamefont {Urban},
  \citenamefont {Woltersdorf},\ and\ \citenamefont
  {Heinrich}}]{urban2001gilbert}%
  \BibitemOpen
  \bibfield  {author} {\bibinfo {author} {\bibfnamefont {R.}~\bibnamefont
  {Urban}}, \bibinfo {author} {\bibfnamefont {G.}~\bibnamefont {Woltersdorf}},\
  and\ \bibinfo {author} {\bibfnamefont {B.}~\bibnamefont {Heinrich}},\ }\href
  {https://doi.org/10.1103/PhysRevLett.87.217204} {\bibfield  {journal}
  {\bibinfo  {journal} {Phys. Rev. Lett.}\ }\textbf {\bibinfo {volume} {87}},\
  \bibinfo {pages} {217204} (\bibinfo {year} {2001})}\BibitemShut {NoStop}%
\bibitem [{\citenamefont {Schoen}\ \emph {et~al.}(2016)\citenamefont {Schoen},
  \citenamefont {Thonig}, \citenamefont {Schneider}, \citenamefont {Silva},
  \citenamefont {Nembach}, \citenamefont {Eriksson}, \citenamefont {Karis},\
  and\ \citenamefont {Shaw}}]{Schoen2016}%
  \BibitemOpen
  \bibfield  {author} {\bibinfo {author} {\bibfnamefont {M.~A.}\ \bibnamefont
  {Schoen}}, \bibinfo {author} {\bibfnamefont {D.}~\bibnamefont {Thonig}},
  \bibinfo {author} {\bibfnamefont {M.~L.}\ \bibnamefont {Schneider}}, \bibinfo
  {author} {\bibfnamefont {T.}~\bibnamefont {Silva}}, \bibinfo {author}
  {\bibfnamefont {H.~T.}\ \bibnamefont {Nembach}}, \bibinfo {author}
  {\bibfnamefont {O.}~\bibnamefont {Eriksson}}, \bibinfo {author}
  {\bibfnamefont {O.}~\bibnamefont {Karis}},\ and\ \bibinfo {author}
  {\bibfnamefont {J.~M.}\ \bibnamefont {Shaw}},\ }\href
  {https://doi.org/10.1038/nphys3770} {\bibfield  {journal} {\bibinfo
  {journal} {Nat. Phys.}\ }\textbf {\bibinfo {volume} {12}},\ \bibinfo {pages}
  {839} (\bibinfo {year} {2016})}\BibitemShut {NoStop}%
\bibitem [{\citenamefont {Etz}\ \emph {et~al.}(2015)\citenamefont {Etz},
  \citenamefont {Bergqvist}, \citenamefont {Bergman}, \citenamefont {Taroni},\
  and\ \citenamefont {Eriksson}}]{etz2015atomistic}%
  \BibitemOpen
  \bibfield  {author} {\bibinfo {author} {\bibfnamefont {C.}~\bibnamefont
  {Etz}}, \bibinfo {author} {\bibfnamefont {L.}~\bibnamefont {Bergqvist}},
  \bibinfo {author} {\bibfnamefont {A.}~\bibnamefont {Bergman}}, \bibinfo
  {author} {\bibfnamefont {A.}~\bibnamefont {Taroni}},\ and\ \bibinfo {author}
  {\bibfnamefont {O.}~\bibnamefont {Eriksson}},\ }\href
  {https://doi.org/10.1088/0953-8984/27/24/243202} {\bibfield  {journal}
  {\bibinfo  {journal} {J. Phys. Condens. Matter}\ }\textbf {\bibinfo {volume}
  {27}},\ \bibinfo {pages} {243202} (\bibinfo {year} {2015})}\BibitemShut
  {NoStop}%
\bibitem [{\citenamefont {Nambu}\ \emph {et~al.}(2020)\citenamefont {Nambu},
  \citenamefont {Barker}, \citenamefont {Okino}, \citenamefont {Kikkawa},
  \citenamefont {Shiomi}, \citenamefont {Enderle}, \citenamefont {Weber},
  \citenamefont {Winn}, \citenamefont {Graves-Brook}, \citenamefont {Tranquada}
  \emph {et~al.}}]{nambu2020observation}%
  \BibitemOpen
  \bibfield  {author} {\bibinfo {author} {\bibfnamefont {Y.}~\bibnamefont
  {Nambu}}, \bibinfo {author} {\bibfnamefont {J.}~\bibnamefont {Barker}},
  \bibinfo {author} {\bibfnamefont {Y.}~\bibnamefont {Okino}}, \bibinfo
  {author} {\bibfnamefont {T.}~\bibnamefont {Kikkawa}}, \bibinfo {author}
  {\bibfnamefont {Y.}~\bibnamefont {Shiomi}}, \bibinfo {author} {\bibfnamefont
  {M.}~\bibnamefont {Enderle}}, \bibinfo {author} {\bibfnamefont
  {T.}~\bibnamefont {Weber}}, \bibinfo {author} {\bibfnamefont
  {B.}~\bibnamefont {Winn}}, \bibinfo {author} {\bibfnamefont {M.}~\bibnamefont
  {Graves-Brook}}, \bibinfo {author} {\bibfnamefont {J.}~\bibnamefont
  {Tranquada}}, \emph {et~al.},\ }\href
  {https://doi.org/10.1103/PhysRevLett.125.027201} {\bibfield  {journal}
  {\bibinfo  {journal} {Phys. Rev. Lett.}\ }\textbf {\bibinfo {volume} {125}},\
  \bibinfo {pages} {027201} (\bibinfo {year} {2020})}\BibitemShut {NoStop}%
\bibitem [{\citenamefont {Balashov}\ \emph {et~al.}(2008)\citenamefont
  {Balashov}, \citenamefont {Tak{\'a}cs}, \citenamefont {D{\"a}ne},
  \citenamefont {Ernst}, \citenamefont {Bruno},\ and\ \citenamefont
  {Wulfhekel}}]{balashov2008inelastic}%
  \BibitemOpen
  \bibfield  {author} {\bibinfo {author} {\bibfnamefont {T.}~\bibnamefont
  {Balashov}}, \bibinfo {author} {\bibfnamefont {A.}~\bibnamefont
  {Tak{\'a}cs}}, \bibinfo {author} {\bibfnamefont {M.}~\bibnamefont
  {D{\"a}ne}}, \bibinfo {author} {\bibfnamefont {A.}~\bibnamefont {Ernst}},
  \bibinfo {author} {\bibfnamefont {P.}~\bibnamefont {Bruno}},\ and\ \bibinfo
  {author} {\bibfnamefont {W.}~\bibnamefont {Wulfhekel}},\ }\href
  {https://doi.org/10.1103/PhysRevB.78.174404} {\bibfield  {journal} {\bibinfo
  {journal} {Phys. Rev. B}\ }\textbf {\bibinfo {volume} {78}},\ \bibinfo
  {pages} {174404} (\bibinfo {year} {2008})}\BibitemShut {NoStop}%
\bibitem [{\citenamefont {Balashov}\ \emph {et~al.}(2014)\citenamefont
  {Balashov}, \citenamefont {Buczek}, \citenamefont {Sandratskii},
  \citenamefont {Ernst},\ and\ \citenamefont {Wulfhekel}}]{balashov2014magnon}%
  \BibitemOpen
  \bibfield  {author} {\bibinfo {author} {\bibfnamefont {T.}~\bibnamefont
  {Balashov}}, \bibinfo {author} {\bibfnamefont {P.}~\bibnamefont {Buczek}},
  \bibinfo {author} {\bibfnamefont {L.}~\bibnamefont {Sandratskii}}, \bibinfo
  {author} {\bibfnamefont {A.}~\bibnamefont {Ernst}},\ and\ \bibinfo {author}
  {\bibfnamefont {W.}~\bibnamefont {Wulfhekel}},\ }\href
  {https://doi.org/10.1088/0953-8984/26/39/394007} {\bibfield  {journal}
  {\bibinfo  {journal} {J. Phys. Condens. Matter}\ }\textbf {\bibinfo {volume}
  {26}},\ \bibinfo {pages} {394007} (\bibinfo {year} {2014})}\BibitemShut
  {NoStop}%
\bibitem [{\citenamefont {Costa}\ \emph {et~al.}(2010)\citenamefont {Costa},
  \citenamefont {Muniz}, \citenamefont {Lounis}, \citenamefont {Klautau},\ and\
  \citenamefont {Mills}}]{costa2010spin}%
  \BibitemOpen
  \bibfield  {author} {\bibinfo {author} {\bibfnamefont {A.}~\bibnamefont
  {Costa}}, \bibinfo {author} {\bibfnamefont {R.}~\bibnamefont {Muniz}},
  \bibinfo {author} {\bibfnamefont {S.}~\bibnamefont {Lounis}}, \bibinfo
  {author} {\bibfnamefont {A.}~\bibnamefont {Klautau}},\ and\ \bibinfo {author}
  {\bibfnamefont {D.}~\bibnamefont {Mills}},\ }\href
  {https://doi.org/10.1103/PhysRevB.82.014428} {\bibfield  {journal} {\bibinfo
  {journal} {Phys. Rev. B}\ }\textbf {\bibinfo {volume} {82}},\ \bibinfo
  {pages} {014428} (\bibinfo {year} {2010})}\BibitemShut {NoStop}%
\bibitem [{\citenamefont {Qin}\ \emph {et~al.}(2013)\citenamefont {Qin},
  \citenamefont {Zakeri}, \citenamefont {Ernst}, \citenamefont {Chuang},
  \citenamefont {Chen}, \citenamefont {Meng},\ and\ \citenamefont
  {Kirschner}}]{qin2013magnons}%
  \BibitemOpen
  \bibfield  {author} {\bibinfo {author} {\bibfnamefont {H.}~\bibnamefont
  {Qin}}, \bibinfo {author} {\bibfnamefont {K.}~\bibnamefont {Zakeri}},
  \bibinfo {author} {\bibfnamefont {A.}~\bibnamefont {Ernst}}, \bibinfo
  {author} {\bibfnamefont {T.-H.}\ \bibnamefont {Chuang}}, \bibinfo {author}
  {\bibfnamefont {Y.-J.}\ \bibnamefont {Chen}}, \bibinfo {author}
  {\bibfnamefont {Y.}~\bibnamefont {Meng}},\ and\ \bibinfo {author}
  {\bibfnamefont {J.}~\bibnamefont {Kirschner}},\ }\href
  {https://doi.org/10.1103/PhysRevB.88.020404} {\bibfield  {journal} {\bibinfo
  {journal} {Phys. Rev. B}\ }\textbf {\bibinfo {volume} {88}},\ \bibinfo
  {pages} {020404} (\bibinfo {year} {2013})}\BibitemShut {NoStop}%
\bibitem [{\citenamefont {Chakraborty}\ \emph {et~al.}(2015)\citenamefont
  {Chakraborty}, \citenamefont {Wenk},\ and\ \citenamefont
  {Schliemann}}]{chakraborty2015lifetimes}%
  \BibitemOpen
  \bibfield  {author} {\bibinfo {author} {\bibfnamefont {A.}~\bibnamefont
  {Chakraborty}}, \bibinfo {author} {\bibfnamefont {P.}~\bibnamefont {Wenk}},\
  and\ \bibinfo {author} {\bibfnamefont {J.}~\bibnamefont {Schliemann}},\
  }\href {https://doi.org/10.1140/epjb/e2015-60033-6} {\bibfield  {journal}
  {\bibinfo  {journal} {Eur. Phys. J. B}\ }\textbf {\bibinfo {volume} {88}},\
  \bibinfo {pages} {1} (\bibinfo {year} {2015})}\BibitemShut {NoStop}%
\bibitem [{\citenamefont {Zhang}\ \emph {et~al.}(2012)\citenamefont {Zhang},
  \citenamefont {Chuang}, \citenamefont {Zakeri}, \citenamefont {Kirschner}
  \emph {et~al.}}]{zhang2012relaxation}%
  \BibitemOpen
  \bibfield  {author} {\bibinfo {author} {\bibfnamefont {Y.}~\bibnamefont
  {Zhang}}, \bibinfo {author} {\bibfnamefont {T.-H.}\ \bibnamefont {Chuang}},
  \bibinfo {author} {\bibfnamefont {K.}~\bibnamefont {Zakeri}}, \bibinfo
  {author} {\bibfnamefont {J.}~\bibnamefont {Kirschner}}, \emph {et~al.},\
  }\href {https://doi.org/10.1103/PhysRevLett.109.087203} {\bibfield  {journal}
  {\bibinfo  {journal} {Phys. Rev. Lett.}\ }\textbf {\bibinfo {volume} {109}},\
  \bibinfo {pages} {087203} (\bibinfo {year} {2012})}\BibitemShut {NoStop}%
\bibitem [{\citenamefont {Mentink}\ \emph {et~al.}(2010)\citenamefont
  {Mentink}, \citenamefont {Tretyakov}, \citenamefont {Fasolino}, \citenamefont
  {Katsnelson},\ and\ \citenamefont {Rasing}}]{mentink2010stable}%
  \BibitemOpen
  \bibfield  {author} {\bibinfo {author} {\bibfnamefont {J.}~\bibnamefont
  {Mentink}}, \bibinfo {author} {\bibfnamefont {M.}~\bibnamefont {Tretyakov}},
  \bibinfo {author} {\bibfnamefont {A.}~\bibnamefont {Fasolino}}, \bibinfo
  {author} {\bibfnamefont {M.}~\bibnamefont {Katsnelson}},\ and\ \bibinfo
  {author} {\bibfnamefont {T.}~\bibnamefont {Rasing}},\ }\href
  {https://doi.org/10.1088/0953-8984/22/17/176001} {\bibfield  {journal}
  {\bibinfo  {journal} {J. Phys. Condens. Matter}\ }\textbf {\bibinfo {volume}
  {22}},\ \bibinfo {pages} {176001} (\bibinfo {year} {2010})}\BibitemShut
  {NoStop}%
\bibitem [{\citenamefont {Brataas}\ \emph {et~al.}(2011)\citenamefont
  {Brataas}, \citenamefont {Tserkovnyak},\ and\ \citenamefont
  {Bauer}}]{brataas2011mag}%
  \BibitemOpen
  \bibfield  {author} {\bibinfo {author} {\bibfnamefont {A.}~\bibnamefont
  {Brataas}}, \bibinfo {author} {\bibfnamefont {Y.}~\bibnamefont
  {Tserkovnyak}},\ and\ \bibinfo {author} {\bibfnamefont {G.~E.}\ \bibnamefont
  {Bauer}},\ }\href {https://doi.org/10.1103/PhysRevB.84.054416} {\bibfield
  {journal} {\bibinfo  {journal} {Phys. Rev. B}\ }\textbf {\bibinfo {volume}
  {84}},\ \bibinfo {pages} {054416} (\bibinfo {year} {2011})}\BibitemShut
  {NoStop}%
\bibitem [{\citenamefont {Vittoria}\ \emph {et~al.}(2010)\citenamefont
  {Vittoria}, \citenamefont {Yoon},\ and\ \citenamefont
  {Widom}}]{vittoria2010relaxation}%
  \BibitemOpen
  \bibfield  {author} {\bibinfo {author} {\bibfnamefont {C.}~\bibnamefont
  {Vittoria}}, \bibinfo {author} {\bibfnamefont {S.}~\bibnamefont {Yoon}},\
  and\ \bibinfo {author} {\bibfnamefont {A.}~\bibnamefont {Widom}},\ }\href
  {https://doi.org/10.1103/PhysRevB.81.014412} {\bibfield  {journal} {\bibinfo
  {journal} {Phys. Rev. B}\ }\textbf {\bibinfo {volume} {81}},\ \bibinfo
  {pages} {014412} (\bibinfo {year} {2010})}\BibitemShut {NoStop}%
\bibitem [{\citenamefont {Rossi}\ \emph {et~al.}(2005)\citenamefont {Rossi},
  \citenamefont {Heinonen},\ and\ \citenamefont {MacDonald}}]{Rossi2005}%
  \BibitemOpen
  \bibfield  {author} {\bibinfo {author} {\bibfnamefont {E.}~\bibnamefont
  {Rossi}}, \bibinfo {author} {\bibfnamefont {O.~G.}\ \bibnamefont
  {Heinonen}},\ and\ \bibinfo {author} {\bibfnamefont {A.~H.}\ \bibnamefont
  {MacDonald}},\ }\href {https://doi.org/10.1103/PhysRevB.72.174412} {\bibfield
   {journal} {\bibinfo  {journal} {Phys. Rev. B}\ }\textbf {\bibinfo {volume}
  {72}},\ \bibinfo {pages} {174412} (\bibinfo {year} {2005})}\BibitemShut
  {NoStop}%
\bibitem [{\citenamefont {R\"uckriegel}\ and\ \citenamefont
  {Kopietz}(2015)}]{Rueckriegel2015}%
  \BibitemOpen
  \bibfield  {author} {\bibinfo {author} {\bibfnamefont {A.}~\bibnamefont
  {R\"uckriegel}}\ and\ \bibinfo {author} {\bibfnamefont {P.}~\bibnamefont
  {Kopietz}},\ }\href {https://doi.org/10.1103/PhysRevLett.115.157203}
  {\bibfield  {journal} {\bibinfo  {journal} {Phys. Rev. Lett.}\ }\textbf
  {\bibinfo {volume} {115}},\ \bibinfo {pages} {157203} (\bibinfo {year}
  {2015})}\BibitemShut {NoStop}%
\bibitem [{\citenamefont {Thonig}\ \emph {et~al.}(2015)\citenamefont {Thonig},
  \citenamefont {Henk},\ and\ \citenamefont {Eriksson}}]{thonig2015gilbert}%
  \BibitemOpen
  \bibfield  {author} {\bibinfo {author} {\bibfnamefont {D.}~\bibnamefont
  {Thonig}}, \bibinfo {author} {\bibfnamefont {J.}~\bibnamefont {Henk}},\ and\
  \bibinfo {author} {\bibfnamefont {O.}~\bibnamefont {Eriksson}},\ }\href
  {https://doi.org/10.1103/PhysRevB.92.104403} {\bibfield  {journal} {\bibinfo
  {journal} {Phys. Rev. B}\ }\textbf {\bibinfo {volume} {92}},\ \bibinfo
  {pages} {104403} (\bibinfo {year} {2015})}\BibitemShut {NoStop}%
\bibitem [{Upp()}]{UppASD}%
  \BibitemOpen
  \href@noop {} {\bibinfo {title} {Uppsala atomistic spin dynamics (uppasd)
  code available under gnu general public license}},\ \bibinfo {howpublished}
  {\url{http://physics.uu.se/ uppasd and
  http://github.com/UppASD/UppASD}}\BibitemShut {NoStop}%
\bibitem [{\citenamefont {Bergman}\ \emph {et~al.}(2010)\citenamefont
  {Bergman}, \citenamefont {Taroni}, \citenamefont {Bergqvist}, \citenamefont
  {Hellsvik}, \citenamefont {Hj{\"o}rvarsson},\ and\ \citenamefont
  {Eriksson}}]{bergman2010magnon}%
  \BibitemOpen
  \bibfield  {author} {\bibinfo {author} {\bibfnamefont {A.}~\bibnamefont
  {Bergman}}, \bibinfo {author} {\bibfnamefont {A.}~\bibnamefont {Taroni}},
  \bibinfo {author} {\bibfnamefont {L.}~\bibnamefont {Bergqvist}}, \bibinfo
  {author} {\bibfnamefont {J.}~\bibnamefont {Hellsvik}}, \bibinfo {author}
  {\bibfnamefont {B.}~\bibnamefont {Hj{\"o}rvarsson}},\ and\ \bibinfo {author}
  {\bibfnamefont {O.}~\bibnamefont {Eriksson}},\ }\href
  {https://doi.org/10.1103/PhysRevB.81.144416} {\bibfield  {journal} {\bibinfo
  {journal} {Phys. Rev. B}\ }\textbf {\bibinfo {volume} {81}},\ \bibinfo
  {pages} {144416} (\bibinfo {year} {2010})}\BibitemShut {NoStop}%
\bibitem [{\citenamefont {Mourigal}\ \emph {et~al.}(2010)\citenamefont
  {Mourigal}, \citenamefont {Zhitomirsky},\ and\ \citenamefont
  {Chernyshev}}]{mourigal2010field}%
  \BibitemOpen
  \bibfield  {author} {\bibinfo {author} {\bibfnamefont {M.}~\bibnamefont
  {Mourigal}}, \bibinfo {author} {\bibfnamefont {M.~E.}\ \bibnamefont
  {Zhitomirsky}},\ and\ \bibinfo {author} {\bibfnamefont {A.~L.}\ \bibnamefont
  {Chernyshev}},\ }\href {https://doi.org/10.1103/PhysRevB.82.144402}
  {\bibfield  {journal} {\bibinfo  {journal} {Phys. Rev. B}\ }\textbf {\bibinfo
  {volume} {82}},\ \bibinfo {pages} {144402} (\bibinfo {year}
  {2010})}\BibitemShut {NoStop}%
\bibitem [{\citenamefont {K{\"u}bler}(2017)}]{kubler2017theory}%
  \BibitemOpen
  \bibfield  {author} {\bibinfo {author} {\bibfnamefont {J.}~\bibnamefont
  {K{\"u}bler}},\ }\href@noop {} {\emph {\bibinfo {title} {Theory of itinerant
  electron magnetism}}},\ Vol.\ \bibinfo {volume} {106}\ (\bibinfo  {publisher}
  {Oxford University Press},\ \bibinfo {year} {2017})\BibitemShut {NoStop}%
\bibitem [{\citenamefont {Halilov}\ \emph {et~al.}(1998)\citenamefont
  {Halilov}, \citenamefont {Eschrig}, \citenamefont {Perlov},\ and\
  \citenamefont {Oppeneer}}]{halilov1998adiabatic}%
  \BibitemOpen
  \bibfield  {author} {\bibinfo {author} {\bibfnamefont {S.}~\bibnamefont
  {Halilov}}, \bibinfo {author} {\bibfnamefont {H.}~\bibnamefont {Eschrig}},
  \bibinfo {author} {\bibfnamefont {A.~Y.}\ \bibnamefont {Perlov}},\ and\
  \bibinfo {author} {\bibfnamefont {P.}~\bibnamefont {Oppeneer}},\ }\href
  {https://doi.org/10.1103/PhysRevB.58.293} {\bibfield  {journal} {\bibinfo
  {journal} {Phys. Rev. B}\ }\textbf {\bibinfo {volume} {58}},\ \bibinfo
  {pages} {293} (\bibinfo {year} {1998})}\BibitemShut {NoStop}%
\bibitem [{\citenamefont {Skadsem}\ \emph {et~al.}(2007)\citenamefont
  {Skadsem}, \citenamefont {Tserkovnyak}, \citenamefont {Brataas},\ and\
  \citenamefont {Bauer}}]{Skadsem2007}%
  \BibitemOpen
  \bibfield  {author} {\bibinfo {author} {\bibfnamefont {H.~J.}\ \bibnamefont
  {Skadsem}}, \bibinfo {author} {\bibfnamefont {Y.}~\bibnamefont
  {Tserkovnyak}}, \bibinfo {author} {\bibfnamefont {A.}~\bibnamefont
  {Brataas}},\ and\ \bibinfo {author} {\bibfnamefont {G.~E.~W.}\ \bibnamefont
  {Bauer}},\ }\href {https://doi.org/10.1103/PhysRevB.75.094416} {\bibfield
  {journal} {\bibinfo  {journal} {Phys. Rev. B}\ }\textbf {\bibinfo {volume}
  {75}},\ \bibinfo {pages} {094416} (\bibinfo {year} {2007})}\BibitemShut
  {NoStop}%
\bibitem [{\citenamefont {Mankovsky}\ \emph {et~al.}(2018)\citenamefont
  {Mankovsky}, \citenamefont {Wimmer},\ and\ \citenamefont
  {Ebert}}]{Mankovsky2018}%
  \BibitemOpen
  \bibfield  {author} {\bibinfo {author} {\bibfnamefont {S.}~\bibnamefont
  {Mankovsky}}, \bibinfo {author} {\bibfnamefont {S.}~\bibnamefont {Wimmer}},\
  and\ \bibinfo {author} {\bibfnamefont {H.}~\bibnamefont {Ebert}},\ }\href
  {https://doi.org/10.1103/PhysRevB.98.104406} {\bibfield  {journal} {\bibinfo
  {journal} {Phys. Rev. B}\ }\textbf {\bibinfo {volume} {98}},\ \bibinfo
  {pages} {104406} (\bibinfo {year} {2018})}\BibitemShut {NoStop}%
\bibitem [{\citenamefont {Marshall}\ and\ \citenamefont
  {Lowde}(1968)}]{Marshall1968}%
  \BibitemOpen
  \bibfield  {author} {\bibinfo {author} {\bibfnamefont {W.}~\bibnamefont
  {Marshall}}\ and\ \bibinfo {author} {\bibfnamefont {R.~D.}\ \bibnamefont
  {Lowde}},\ }\href {https://doi.org/10.1088/0034-4885/31/2/305} {\bibfield
  {journal} {\bibinfo  {journal} {Rep. Prog. Phys.}\ }\textbf {\bibinfo
  {volume} {31}},\ \bibinfo {pages} {705} (\bibinfo {year} {1968})}\BibitemShut
  {NoStop}%
\bibitem [{\citenamefont {Peduto}\ \emph {et~al.}(1991)\citenamefont {Peduto},
  \citenamefont {Frota-Pessa},\ and\ \citenamefont {Methfessel}}]{Peduto1991}%
  \BibitemOpen
  \bibfield  {author} {\bibinfo {author} {\bibfnamefont {P.~R.}\ \bibnamefont
  {Peduto}}, \bibinfo {author} {\bibfnamefont {S.}~\bibnamefont
  {Frota-Pessa}},\ and\ \bibinfo {author} {\bibfnamefont {M.~S.}\ \bibnamefont
  {Methfessel}},\ }\href {https://doi.org/10.1103/PhysRevB.44.13283} {\bibfield
   {journal} {\bibinfo  {journal} {Phys. Rev. B}\ }\textbf {\bibinfo {volume}
  {44}},\ \bibinfo {pages} {13283} (\bibinfo {year} {1991})}\BibitemShut
  {NoStop}%
\bibitem [{\citenamefont {Frota-Pess\^{o}a}(1992)}]{FrotaPessoa1992}%
  \BibitemOpen
  \bibfield  {author} {\bibinfo {author} {\bibfnamefont {S.}~\bibnamefont
  {Frota-Pess\^{o}a}},\ }\href {https://doi.org/10.1103/PhysRevB.46.14570}
  {\bibfield  {journal} {\bibinfo  {journal} {Phys. Rev. B}\ }\textbf {\bibinfo
  {volume} {46}},\ \bibinfo {pages} {14570} (\bibinfo {year}
  {1992})}\BibitemShut {NoStop}%
\bibitem [{\citenamefont {Haydock}(1980)}]{Haydock1980}%
  \BibitemOpen
  \bibfield  {author} {\bibinfo {author} {\bibfnamefont {R.}~\bibnamefont
  {Haydock}},\ }in\ \href {https://doi.org/10.1016/S0081-1947(08)60505-6}
  {\emph {\bibinfo {booktitle} {Solid state physics}}},\ Vol.~\bibinfo {volume}
  {35}\ (\bibinfo  {publisher} {Elsevier},\ \bibinfo {year} {1980})\ pp.\
  \bibinfo {pages} {215--294}\BibitemShut {NoStop}%
\bibitem [{\citenamefont {Beer}\ and\ \citenamefont
  {Pettifor}(1984)}]{Beer1984}%
  \BibitemOpen
  \bibfield  {author} {\bibinfo {author} {\bibfnamefont {N.}~\bibnamefont
  {Beer}}\ and\ \bibinfo {author} {\bibfnamefont {D.}~\bibnamefont
  {Pettifor}},\ }in\ \href@noop {} {\emph {\bibinfo {booktitle} {The Electronic
  Structure of Complex Systems}}}\ (\bibinfo  {publisher} {Springer},\ \bibinfo
  {year} {1984})\ pp.\ \bibinfo {pages} {769--777}\BibitemShut {NoStop}%
\bibitem [{\citenamefont {Andersen}(1975)}]{Andersen1975}%
  \BibitemOpen
  \bibfield  {author} {\bibinfo {author} {\bibfnamefont {O.~K.}\ \bibnamefont
  {Andersen}},\ }\href {https://doi.org/10.1103/PhysRevB.12.3060} {\bibfield
  {journal} {\bibinfo  {journal} {Phys. Rev. B}\ }\textbf {\bibinfo {volume}
  {12}},\ \bibinfo {pages} {3060} (\bibinfo {year} {1975})}\BibitemShut
  {NoStop}%
\bibitem [{\citenamefont {Frota-Pess\^{o}a}\ \emph {et~al.}(2000)\citenamefont
  {Frota-Pess\^{o}a}, \citenamefont {Muniz},\ and\ \citenamefont
  {Kudrnovsk\'y}}]{FrotaPessoa2000}%
  \BibitemOpen
  \bibfield  {author} {\bibinfo {author} {\bibfnamefont {S.}~\bibnamefont
  {Frota-Pess\^{o}a}}, \bibinfo {author} {\bibfnamefont {R.~B.}\ \bibnamefont
  {Muniz}},\ and\ \bibinfo {author} {\bibfnamefont {J.}~\bibnamefont
  {Kudrnovsk\'y}},\ }\href {https://doi.org/10.1103/PhysRevB.62.5293}
  {\bibfield  {journal} {\bibinfo  {journal} {Phys. Rev. B}\ }\textbf {\bibinfo
  {volume} {62}},\ \bibinfo {pages} {5293} (\bibinfo {year}
  {2000})}\BibitemShut {NoStop}%
\bibitem [{\citenamefont {Frota-Pess\^oa}(2004)}]{FrotaPessoa2004}%
  \BibitemOpen
  \bibfield  {author} {\bibinfo {author} {\bibfnamefont {S.}~\bibnamefont
  {Frota-Pess\^oa}},\ }\href {https://doi.org/10.1103/PhysRevB.69.104401}
  {\bibfield  {journal} {\bibinfo  {journal} {Phys. Rev. B}\ }\textbf {\bibinfo
  {volume} {69}},\ \bibinfo {pages} {104401} (\bibinfo {year}
  {2004})}\BibitemShut {NoStop}%
\bibitem [{\citenamefont {Von~Barth}\ and\ \citenamefont
  {Hedin}(1972)}]{Barth1972}%
  \BibitemOpen
  \bibfield  {author} {\bibinfo {author} {\bibfnamefont {U.}~\bibnamefont
  {Von~Barth}}\ and\ \bibinfo {author} {\bibfnamefont {L.}~\bibnamefont
  {Hedin}},\ }\href {https://doi.org/10.1088/0022-3719/5/13/012} {\bibfield
  {journal} {\bibinfo  {journal} {J. of Phys. C: Solid State Physics}\ }\textbf
  {\bibinfo {volume} {5}},\ \bibinfo {pages} {1629} (\bibinfo {year}
  {1972})}\BibitemShut {NoStop}%
\bibitem [{\citenamefont {Perdew}\ \emph {et~al.}(1996)\citenamefont {Perdew},
  \citenamefont {Burke},\ and\ \citenamefont {Ernzerhof}}]{Perdew1996}%
  \BibitemOpen
  \bibfield  {author} {\bibinfo {author} {\bibfnamefont {J.~P.}\ \bibnamefont
  {Perdew}}, \bibinfo {author} {\bibfnamefont {K.}~\bibnamefont {Burke}},\ and\
  \bibinfo {author} {\bibfnamefont {M.}~\bibnamefont {Ernzerhof}},\ }\href
  {https://doi.org/10.1103/PhysRevLett.77.3865} {\bibfield  {journal} {\bibinfo
   {journal} {Phys. Rev. Lett.}\ }\textbf {\bibinfo {volume} {77}},\ \bibinfo
  {pages} {3865} (\bibinfo {year} {1996})}\BibitemShut {NoStop}%
\bibitem [{\citenamefont {Eriksson}\ \emph {et~al.}(1990)\citenamefont
  {Eriksson}, \citenamefont {Johansson}, \citenamefont {Albers}, \citenamefont
  {Boring},\ and\ \citenamefont {Brooks}}]{Eriksson1990}%
  \BibitemOpen
  \bibfield  {author} {\bibinfo {author} {\bibfnamefont {O.}~\bibnamefont
  {Eriksson}}, \bibinfo {author} {\bibfnamefont {B.}~\bibnamefont {Johansson}},
  \bibinfo {author} {\bibfnamefont {R.~C.}\ \bibnamefont {Albers}}, \bibinfo
  {author} {\bibfnamefont {A.~M.}\ \bibnamefont {Boring}},\ and\ \bibinfo
  {author} {\bibfnamefont {M.~S.~S.}\ \bibnamefont {Brooks}},\ }\href
  {https://doi.org/10.1103/PhysRevB.42.2707} {\bibfield  {journal} {\bibinfo
  {journal} {Phys. Rev. B}\ }\textbf {\bibinfo {volume} {42}},\ \bibinfo
  {pages} {2707} (\bibinfo {year} {1990})}\BibitemShut {NoStop}%
\bibitem [{\citenamefont {Ohnuma}\ \emph {et~al.}(2002)\citenamefont {Ohnuma},
  \citenamefont {Enoki}, \citenamefont {Ikeda}, \citenamefont {Kainuma},
  \citenamefont {Ohtani}, \citenamefont {Sundman},\ and\ \citenamefont
  {Ishida}}]{Ohnuma2002}%
  \BibitemOpen
  \bibfield  {author} {\bibinfo {author} {\bibfnamefont {I.}~\bibnamefont
  {Ohnuma}}, \bibinfo {author} {\bibfnamefont {H.}~\bibnamefont {Enoki}},
  \bibinfo {author} {\bibfnamefont {O.}~\bibnamefont {Ikeda}}, \bibinfo
  {author} {\bibfnamefont {R.}~\bibnamefont {Kainuma}}, \bibinfo {author}
  {\bibfnamefont {H.}~\bibnamefont {Ohtani}}, \bibinfo {author} {\bibfnamefont
  {B.}~\bibnamefont {Sundman}},\ and\ \bibinfo {author} {\bibfnamefont
  {K.}~\bibnamefont {Ishida}},\ }\href
  {https://doi.org/10.1016/S1359-6454(01)00337-8} {\bibfield  {journal}
  {\bibinfo  {journal} {Acta Mater.}\ }\textbf {\bibinfo {volume} {50}},\
  \bibinfo {pages} {379} (\bibinfo {year} {2002})}\BibitemShut {NoStop}%
\bibitem [{\citenamefont {Billins}\ and\ \citenamefont
  {Gray}(1972)}]{American1972}%
  \BibitemOpen
  \bibinfo {editor} {\bibfnamefont {B.~H.}\ \bibnamefont {Billins}}\ and\
  \bibinfo {editor} {\bibfnamefont {D.~E.}\ \bibnamefont {Gray}},\ eds.,\
  \href@noop {} {\emph {\bibinfo {title} {American Institute of Physics
  Handbook (3rd Ed)}}}\ (\bibinfo  {publisher} {McGraw-Hill},\ \bibinfo {year}
  {1972})\BibitemShut {NoStop}%
\bibitem [{\citenamefont {Wijn}(1991)}]{Wijn1991}%
  \BibitemOpen
  \bibfield  {author} {\bibinfo {author} {\bibfnamefont {H.~P.}\ \bibnamefont
  {Wijn}},\ }\href@noop {} {\emph {\bibinfo {title} {Magnetic properties of
  metals: d-elements, alloys and compounds}}}\ (\bibinfo  {publisher} {Springer
  Science \& Business Media},\ \bibinfo {year} {1991})\BibitemShut {NoStop}%
\bibitem [{\citenamefont {Burkert}\ \emph {et~al.}(2004)\citenamefont
  {Burkert}, \citenamefont {Nordstr\"om}, \citenamefont {Eriksson},\ and\
  \citenamefont {Heinonen}}]{Burkert2004}%
  \BibitemOpen
  \bibfield  {author} {\bibinfo {author} {\bibfnamefont {T.}~\bibnamefont
  {Burkert}}, \bibinfo {author} {\bibfnamefont {L.}~\bibnamefont
  {Nordstr\"om}}, \bibinfo {author} {\bibfnamefont {O.}~\bibnamefont
  {Eriksson}},\ and\ \bibinfo {author} {\bibfnamefont {O.}~\bibnamefont
  {Heinonen}},\ }\href {https://doi.org/10.1103/PhysRevLett.93.027203}
  {\bibfield  {journal} {\bibinfo  {journal} {Phys. Rev. Lett.}\ }\textbf
  {\bibinfo {volume} {93}},\ \bibinfo {pages} {027203} (\bibinfo {year}
  {2004})}\BibitemShut {NoStop}%
\bibitem [{\citenamefont {Ma\ifmmode~\check{s}\else \v{s}\fi{}\'{\i}n}\ \emph
  {et~al.}(2013)\citenamefont {Ma\ifmmode~\check{s}\else \v{s}\fi{}\'{\i}n},
  \citenamefont {Bergqvist}, \citenamefont {Kudrnovsk\'y}, \citenamefont
  {Kotrla},\ and\ \citenamefont {Drchal}}]{Masin2013}%
  \BibitemOpen
  \bibfield  {author} {\bibinfo {author} {\bibfnamefont {M.}~\bibnamefont
  {Ma\ifmmode~\check{s}\else \v{s}\fi{}\'{\i}n}}, \bibinfo {author}
  {\bibfnamefont {L.}~\bibnamefont {Bergqvist}}, \bibinfo {author}
  {\bibfnamefont {J.}~\bibnamefont {Kudrnovsk\'y}}, \bibinfo {author}
  {\bibfnamefont {M.}~\bibnamefont {Kotrla}},\ and\ \bibinfo {author}
  {\bibfnamefont {V.}~\bibnamefont {Drchal}},\ }\href
  {https://doi.org/10.1103/PhysRevB.87.075452} {\bibfield  {journal} {\bibinfo
  {journal} {Phys. Rev. B}\ }\textbf {\bibinfo {volume} {87}},\ \bibinfo
  {pages} {075452} (\bibinfo {year} {2013})}\BibitemShut {NoStop}%
\bibitem [{\citenamefont {S\"oderlind}\ \emph {et~al.}(1992)\citenamefont
  {S\"oderlind}, \citenamefont {Eriksson}, \citenamefont {Johansson},
  \citenamefont {Albers},\ and\ \citenamefont {Boring}}]{Soderlind1992}%
  \BibitemOpen
  \bibfield  {author} {\bibinfo {author} {\bibfnamefont {P.}~\bibnamefont
  {S\"oderlind}}, \bibinfo {author} {\bibfnamefont {O.}~\bibnamefont
  {Eriksson}}, \bibinfo {author} {\bibfnamefont {B.}~\bibnamefont {Johansson}},
  \bibinfo {author} {\bibfnamefont {R.~C.}\ \bibnamefont {Albers}},\ and\
  \bibinfo {author} {\bibfnamefont {A.~M.}\ \bibnamefont {Boring}},\ }\href
  {https://doi.org/10.1103/PhysRevB.45.12911} {\bibfield  {journal} {\bibinfo
  {journal} {Phys. Rev. B}\ }\textbf {\bibinfo {volume} {45}},\ \bibinfo
  {pages} {12911} (\bibinfo {year} {1992})}\BibitemShut {NoStop}%
\bibitem [{\citenamefont {Bergman}\ and\ \citenamefont
  {Eriksson}(2006)}]{Bergman2006}%
  \BibitemOpen
  \bibfield  {author} {\bibinfo {author} {\bibfnamefont {A.}~\bibnamefont
  {Bergman}}\ and\ \bibinfo {author} {\bibfnamefont {O.}~\bibnamefont
  {Eriksson}},\ }\href {https://doi.org/10.1103/PhysRevB.74.104422} {\bibfield
  {journal} {\bibinfo  {journal} {Phys. Rev. B}\ }\textbf {\bibinfo {volume}
  {74}},\ \bibinfo {pages} {104422} (\bibinfo {year} {2006})}\BibitemShut
  {NoStop}%
\bibitem [{\citenamefont {D\'{\i}az-Ortiz}\ \emph {et~al.}(2006)\citenamefont
  {D\'{\i}az-Ortiz}, \citenamefont {Drautz}, \citenamefont {F\"ahnle},
  \citenamefont {Dosch},\ and\ \citenamefont {Sanchez}}]{Fahnle2006}%
  \BibitemOpen
  \bibfield  {author} {\bibinfo {author} {\bibfnamefont {A.}~\bibnamefont
  {D\'{\i}az-Ortiz}}, \bibinfo {author} {\bibfnamefont {R.}~\bibnamefont
  {Drautz}}, \bibinfo {author} {\bibfnamefont {M.}~\bibnamefont {F\"ahnle}},
  \bibinfo {author} {\bibfnamefont {H.}~\bibnamefont {Dosch}},\ and\ \bibinfo
  {author} {\bibfnamefont {J.~M.}\ \bibnamefont {Sanchez}},\ }\href
  {https://doi.org/10.1103/PhysRevB.73.224208} {\bibfield  {journal} {\bibinfo
  {journal} {Phys. Rev. B}\ }\textbf {\bibinfo {volume} {73}},\ \bibinfo
  {pages} {224208} (\bibinfo {year} {2006})}\BibitemShut {NoStop}%
\bibitem [{\citenamefont {Trinastic}\ \emph {et~al.}(2013)\citenamefont
  {Trinastic}, \citenamefont {Wang},\ and\ \citenamefont
  {Cheng}}]{Trinastic2013}%
  \BibitemOpen
  \bibfield  {author} {\bibinfo {author} {\bibfnamefont {J.~P.}\ \bibnamefont
  {Trinastic}}, \bibinfo {author} {\bibfnamefont {Y.}~\bibnamefont {Wang}},\
  and\ \bibinfo {author} {\bibfnamefont {H.-P.}\ \bibnamefont {Cheng}},\ }\href
  {https://doi.org/10.1103/PhysRevB.88.104408} {\bibfield  {journal} {\bibinfo
  {journal} {Phys. Rev. B}\ }\textbf {\bibinfo {volume} {88}},\ \bibinfo
  {pages} {104408} (\bibinfo {year} {2013})}\BibitemShut {NoStop}%
\bibitem [{\citenamefont {Seemann}\ \emph {et~al.}(2011)\citenamefont
  {Seemann}, \citenamefont {Freimuth}, \citenamefont {Zhang}, \citenamefont
  {Bl\"ugel}, \citenamefont {Mokrousov}, \citenamefont {B\"urgler},\ and\
  \citenamefont {Schneider}}]{Seemann2011}%
  \BibitemOpen
  \bibfield  {author} {\bibinfo {author} {\bibfnamefont {K.~M.}\ \bibnamefont
  {Seemann}}, \bibinfo {author} {\bibfnamefont {F.}~\bibnamefont {Freimuth}},
  \bibinfo {author} {\bibfnamefont {H.}~\bibnamefont {Zhang}}, \bibinfo
  {author} {\bibfnamefont {S.}~\bibnamefont {Bl\"ugel}}, \bibinfo {author}
  {\bibfnamefont {Y.}~\bibnamefont {Mokrousov}}, \bibinfo {author}
  {\bibfnamefont {D.~E.}\ \bibnamefont {B\"urgler}},\ and\ \bibinfo {author}
  {\bibfnamefont {C.~M.}\ \bibnamefont {Schneider}},\ }\href
  {https://doi.org/10.1103/PhysRevLett.107.086603} {\bibfield  {journal}
  {\bibinfo  {journal} {Phys. Rev. Lett.}\ }\textbf {\bibinfo {volume} {107}},\
  \bibinfo {pages} {086603} (\bibinfo {year} {2011})}\BibitemShut {NoStop}%
\bibitem [{\citenamefont {Lourembam}\ \emph {et~al.}(2021)\citenamefont
  {Lourembam}, \citenamefont {Khoo}, \citenamefont {Qiu}, \citenamefont {Xie},
  \citenamefont {Wong}, \citenamefont {Yap},\ and\ \citenamefont
  {Lim}}]{Lourembam2021}%
  \BibitemOpen
  \bibfield  {author} {\bibinfo {author} {\bibfnamefont {J.}~\bibnamefont
  {Lourembam}}, \bibinfo {author} {\bibfnamefont {K.~H.}\ \bibnamefont {Khoo}},
  \bibinfo {author} {\bibfnamefont {J.}~\bibnamefont {Qiu}}, \bibinfo {author}
  {\bibfnamefont {H.}~\bibnamefont {Xie}}, \bibinfo {author} {\bibfnamefont
  {S.~K.}\ \bibnamefont {Wong}}, \bibinfo {author} {\bibfnamefont {Q.~J.}\
  \bibnamefont {Yap}},\ and\ \bibinfo {author} {\bibfnamefont {S.~T.}\
  \bibnamefont {Lim}},\ }\href {https://doi.org/10.1002/aelm.202100351}
  {\bibfield  {journal} {\bibinfo  {journal} {Adv. Electron. Mater.}\ }\textbf
  {\bibinfo {volume} {7}},\ \bibinfo {pages} {2100351} (\bibinfo {year}
  {2021})}\BibitemShut {NoStop}%
\bibitem [{\citenamefont {Miranda}\ \emph {et~al.}(2021)\citenamefont
  {Miranda}, \citenamefont {Klautau}, \citenamefont {Bergman}, \citenamefont
  {Thonig}, \citenamefont {Petrilli},\ and\ \citenamefont
  {Eriksson}}]{Miranda2021}%
  \BibitemOpen
  \bibfield  {author} {\bibinfo {author} {\bibfnamefont {I.~P.}\ \bibnamefont
  {Miranda}}, \bibinfo {author} {\bibfnamefont {A.~B.}\ \bibnamefont
  {Klautau}}, \bibinfo {author} {\bibfnamefont {A.}~\bibnamefont {Bergman}},
  \bibinfo {author} {\bibfnamefont {D.}~\bibnamefont {Thonig}}, \bibinfo
  {author} {\bibfnamefont {H.~M.}\ \bibnamefont {Petrilli}},\ and\ \bibinfo
  {author} {\bibfnamefont {O.}~\bibnamefont {Eriksson}},\ }\href
  {https://doi.org/10.1103/PhysRevB.103.L220405} {\bibfield  {journal}
  {\bibinfo  {journal} {Phys. Rev. B}\ }\textbf {\bibinfo {volume} {103}},\
  \bibinfo {pages} {L220405} (\bibinfo {year} {2021})}\BibitemShut {NoStop}%
\bibitem [{\citenamefont {Shaw}\ \emph {et~al.}(2021)\citenamefont {Shaw},
  \citenamefont {Knut}, \citenamefont {Armstrong}, \citenamefont {Bhandary},
  \citenamefont {Kvashnin}, \citenamefont {Thonig}, \citenamefont
  {Delczeg-Czirjak}, \citenamefont {Karis}, \citenamefont {Silva},
  \citenamefont {Weschke}, \citenamefont {Nembach}, \citenamefont {Eriksson},\
  and\ \citenamefont {Arena}}]{Shaw2021}%
  \BibitemOpen
  \bibfield  {author} {\bibinfo {author} {\bibfnamefont {J.~M.}\ \bibnamefont
  {Shaw}}, \bibinfo {author} {\bibfnamefont {R.}~\bibnamefont {Knut}}, \bibinfo
  {author} {\bibfnamefont {A.}~\bibnamefont {Armstrong}}, \bibinfo {author}
  {\bibfnamefont {S.}~\bibnamefont {Bhandary}}, \bibinfo {author}
  {\bibfnamefont {Y.}~\bibnamefont {Kvashnin}}, \bibinfo {author}
  {\bibfnamefont {D.}~\bibnamefont {Thonig}}, \bibinfo {author} {\bibfnamefont
  {E.~K.}\ \bibnamefont {Delczeg-Czirjak}}, \bibinfo {author} {\bibfnamefont
  {O.}~\bibnamefont {Karis}}, \bibinfo {author} {\bibfnamefont {T.~J.}\
  \bibnamefont {Silva}}, \bibinfo {author} {\bibfnamefont {E.}~\bibnamefont
  {Weschke}}, \bibinfo {author} {\bibfnamefont {H.~T.}\ \bibnamefont
  {Nembach}}, \bibinfo {author} {\bibfnamefont {O.}~\bibnamefont {Eriksson}},\
  and\ \bibinfo {author} {\bibfnamefont {D.~A.}\ \bibnamefont {Arena}},\ }\href
  {https://doi.org/10.1103/PhysRevLett.127.207201} {\bibfield  {journal}
  {\bibinfo  {journal} {Phys. Rev. Lett.}\ }\textbf {\bibinfo {volume} {127}},\
  \bibinfo {pages} {207201} (\bibinfo {year} {2021})}\BibitemShut {NoStop}%
\bibitem [{\citenamefont {Liechtenstein}\ \emph {et~al.}(1987)\citenamefont
  {Liechtenstein}, \citenamefont {Katsnelson}, \citenamefont {Antropov},\ and\
  \citenamefont {Gubanov}}]{Liechtenstein1987}%
  \BibitemOpen
  \bibfield  {author} {\bibinfo {author} {\bibfnamefont {A.~I.}\ \bibnamefont
  {Liechtenstein}}, \bibinfo {author} {\bibfnamefont {M.}~\bibnamefont
  {Katsnelson}}, \bibinfo {author} {\bibfnamefont {V.}~\bibnamefont
  {Antropov}},\ and\ \bibinfo {author} {\bibfnamefont {V.}~\bibnamefont
  {Gubanov}},\ }\href {https://doi.org/10.1016/0304-8853(87)90721-9} {\bibfield
   {journal} {\bibinfo  {journal} {J. Magn. Magn. Mater.}\ }\textbf {\bibinfo
  {volume} {67}},\ \bibinfo {pages} {65} (\bibinfo {year} {1987})}\BibitemShut
  {NoStop}%
\bibitem [{\citenamefont {Pajda}\ \emph {et~al.}(2001)\citenamefont {Pajda},
  \citenamefont {Kudrnovsk\'y}, \citenamefont {Turek}, \citenamefont {Drchal},\
  and\ \citenamefont {Bruno}}]{Pajda2001}%
  \BibitemOpen
  \bibfield  {author} {\bibinfo {author} {\bibfnamefont {M.}~\bibnamefont
  {Pajda}}, \bibinfo {author} {\bibfnamefont {J.}~\bibnamefont {Kudrnovsk\'y}},
  \bibinfo {author} {\bibfnamefont {I.}~\bibnamefont {Turek}}, \bibinfo
  {author} {\bibfnamefont {V.}~\bibnamefont {Drchal}},\ and\ \bibinfo {author}
  {\bibfnamefont {P.}~\bibnamefont {Bruno}},\ }\href
  {https://doi.org/10.1103/PhysRevB.64.174402} {\bibfield  {journal} {\bibinfo
  {journal} {Phys. Rev. B}\ }\textbf {\bibinfo {volume} {64}},\ \bibinfo
  {pages} {174402} (\bibinfo {year} {2001})}\BibitemShut {NoStop}%
\bibitem [{\citenamefont {Starikov}\ \emph {et~al.}(2010)\citenamefont
  {Starikov}, \citenamefont {Kelly}, \citenamefont {Brataas}, \citenamefont
  {Tserkovnyak},\ and\ \citenamefont {Bauer}}]{Starikov2010}%
  \BibitemOpen
  \bibfield  {author} {\bibinfo {author} {\bibfnamefont {A.~A.}\ \bibnamefont
  {Starikov}}, \bibinfo {author} {\bibfnamefont {P.~J.}\ \bibnamefont {Kelly}},
  \bibinfo {author} {\bibfnamefont {A.}~\bibnamefont {Brataas}}, \bibinfo
  {author} {\bibfnamefont {Y.}~\bibnamefont {Tserkovnyak}},\ and\ \bibinfo
  {author} {\bibfnamefont {G.~E.~W.}\ \bibnamefont {Bauer}},\ }\href
  {https://doi.org/10.1103/PhysRevLett.105.236601} {\bibfield  {journal}
  {\bibinfo  {journal} {Phys. Rev. Lett.}\ }\textbf {\bibinfo {volume} {105}},\
  \bibinfo {pages} {236601} (\bibinfo {year} {2010})}\BibitemShut {NoStop}%
\bibitem [{\citenamefont {Bardos}(1969)}]{Bardos1969}%
  \BibitemOpen
  \bibfield  {author} {\bibinfo {author} {\bibfnamefont {D.}~\bibnamefont
  {Bardos}},\ }\href {https://doi.org/10.1063/1.1657673} {\bibfield  {journal}
  {\bibinfo  {journal} {J. Appl. Phys.}\ }\textbf {\bibinfo {volume} {40}},\
  \bibinfo {pages} {1371} (\bibinfo {year} {1969})}\BibitemShut {NoStop}%
\bibitem [{\citenamefont {Oogane}\ \emph {et~al.}(2006)\citenamefont {Oogane},
  \citenamefont {Wakitani}, \citenamefont {Yakata}, \citenamefont {Yilgin},
  \citenamefont {Ando}, \citenamefont {Sakuma},\ and\ \citenamefont
  {Miyazaki}}]{Oogane2006}%
  \BibitemOpen
  \bibfield  {author} {\bibinfo {author} {\bibfnamefont {M.}~\bibnamefont
  {Oogane}}, \bibinfo {author} {\bibfnamefont {T.}~\bibnamefont {Wakitani}},
  \bibinfo {author} {\bibfnamefont {S.}~\bibnamefont {Yakata}}, \bibinfo
  {author} {\bibfnamefont {R.}~\bibnamefont {Yilgin}}, \bibinfo {author}
  {\bibfnamefont {Y.}~\bibnamefont {Ando}}, \bibinfo {author} {\bibfnamefont
  {A.}~\bibnamefont {Sakuma}},\ and\ \bibinfo {author} {\bibfnamefont
  {T.}~\bibnamefont {Miyazaki}},\ }\href {https://doi.org/10.1143/JJAP.45.3889}
  {\bibfield  {journal} {\bibinfo  {journal} {Jpn. J. Appl. Phys.}\ }\textbf
  {\bibinfo {volume} {45}},\ \bibinfo {pages} {3889} (\bibinfo {year}
  {2006})}\BibitemShut {NoStop}%
\bibitem [{\citenamefont {Mankovsky}\ \emph {et~al.}(2013)\citenamefont
  {Mankovsky}, \citenamefont {K\"odderitzsch}, \citenamefont {Woltersdorf},\
  and\ \citenamefont {Ebert}}]{Mankovsky2013}%
  \BibitemOpen
  \bibfield  {author} {\bibinfo {author} {\bibfnamefont {S.}~\bibnamefont
  {Mankovsky}}, \bibinfo {author} {\bibfnamefont {D.}~\bibnamefont
  {K\"odderitzsch}}, \bibinfo {author} {\bibfnamefont {G.}~\bibnamefont
  {Woltersdorf}},\ and\ \bibinfo {author} {\bibfnamefont {H.}~\bibnamefont
  {Ebert}},\ }\href {https://doi.org/10.1103/PhysRevB.87.014430} {\bibfield
  {journal} {\bibinfo  {journal} {Phys. Rev. B}\ }\textbf {\bibinfo {volume}
  {87}},\ \bibinfo {pages} {014430} (\bibinfo {year} {2013})}\BibitemShut
  {NoStop}%
\bibitem [{\citenamefont {Khodadadi}\ \emph {et~al.}(2020)\citenamefont
  {Khodadadi}, \citenamefont {Rai}, \citenamefont {Sapkota}, \citenamefont
  {Srivastava}, \citenamefont {Nepal}, \citenamefont {Lim}, \citenamefont
  {Smith}, \citenamefont {Mewes}, \citenamefont {Budhathoki}, \citenamefont
  {Hauser}, \citenamefont {Gao}, \citenamefont {Li}, \citenamefont {Viehland},
  \citenamefont {Jiang}, \citenamefont {Heremans}, \citenamefont
  {Balachandran}, \citenamefont {Mewes},\ and\ \citenamefont
  {Emori}}]{Khodadadi2020}%
  \BibitemOpen
  \bibfield  {author} {\bibinfo {author} {\bibfnamefont {B.}~\bibnamefont
  {Khodadadi}}, \bibinfo {author} {\bibfnamefont {A.}~\bibnamefont {Rai}},
  \bibinfo {author} {\bibfnamefont {A.}~\bibnamefont {Sapkota}}, \bibinfo
  {author} {\bibfnamefont {A.}~\bibnamefont {Srivastava}}, \bibinfo {author}
  {\bibfnamefont {B.}~\bibnamefont {Nepal}}, \bibinfo {author} {\bibfnamefont
  {Y.}~\bibnamefont {Lim}}, \bibinfo {author} {\bibfnamefont {D.~A.}\
  \bibnamefont {Smith}}, \bibinfo {author} {\bibfnamefont {C.}~\bibnamefont
  {Mewes}}, \bibinfo {author} {\bibfnamefont {S.}~\bibnamefont {Budhathoki}},
  \bibinfo {author} {\bibfnamefont {A.~J.}\ \bibnamefont {Hauser}}, \bibinfo
  {author} {\bibfnamefont {M.}~\bibnamefont {Gao}}, \bibinfo {author}
  {\bibfnamefont {J.-F.}\ \bibnamefont {Li}}, \bibinfo {author} {\bibfnamefont
  {D.~D.}\ \bibnamefont {Viehland}}, \bibinfo {author} {\bibfnamefont
  {Z.}~\bibnamefont {Jiang}}, \bibinfo {author} {\bibfnamefont {J.~J.}\
  \bibnamefont {Heremans}}, \bibinfo {author} {\bibfnamefont {P.~V.}\
  \bibnamefont {Balachandran}}, \bibinfo {author} {\bibfnamefont
  {T.}~\bibnamefont {Mewes}},\ and\ \bibinfo {author} {\bibfnamefont
  {S.}~\bibnamefont {Emori}},\ }\href
  {https://doi.org/10.1103/PhysRevLett.124.157201} {\bibfield  {journal}
  {\bibinfo  {journal} {Phys. Rev. Lett.}\ }\textbf {\bibinfo {volume} {124}},\
  \bibinfo {pages} {157201} (\bibinfo {year} {2020})}\BibitemShut {NoStop}%
\bibitem [{\citenamefont {Scheck}\ \emph {et~al.}(2007)\citenamefont {Scheck},
  \citenamefont {Cheng}, \citenamefont {Barsukov}, \citenamefont {Frait},\ and\
  \citenamefont {Bailey}}]{Scheck2007}%
  \BibitemOpen
  \bibfield  {author} {\bibinfo {author} {\bibfnamefont {C.}~\bibnamefont
  {Scheck}}, \bibinfo {author} {\bibfnamefont {L.}~\bibnamefont {Cheng}},
  \bibinfo {author} {\bibfnamefont {I.}~\bibnamefont {Barsukov}}, \bibinfo
  {author} {\bibfnamefont {Z.}~\bibnamefont {Frait}},\ and\ \bibinfo {author}
  {\bibfnamefont {W.~E.}\ \bibnamefont {Bailey}},\ }\href
  {https://doi.org/10.1103/PhysRevLett.98.117601} {\bibfield  {journal}
  {\bibinfo  {journal} {Phys. Rev. Lett.}\ }\textbf {\bibinfo {volume} {98}},\
  \bibinfo {pages} {117601} (\bibinfo {year} {2007})}\BibitemShut {NoStop}%
\bibitem [{\citenamefont {Bhagat}\ and\ \citenamefont
  {Lubitz}(1974)}]{Bhagat1974}%
  \BibitemOpen
  \bibfield  {author} {\bibinfo {author} {\bibfnamefont {S.~M.}\ \bibnamefont
  {Bhagat}}\ and\ \bibinfo {author} {\bibfnamefont {P.}~\bibnamefont
  {Lubitz}},\ }\href {https://doi.org/10.1103/PhysRevB.10.179} {\bibfield
  {journal} {\bibinfo  {journal} {Phys. Rev. B}\ }\textbf {\bibinfo {volume}
  {10}},\ \bibinfo {pages} {179} (\bibinfo {year} {1974})}\BibitemShut
  {NoStop}%
\bibitem [{\citenamefont {Hsu}\ and\ \citenamefont {Berger}(1978)}]{Hsu1978}%
  \BibitemOpen
  \bibfield  {author} {\bibinfo {author} {\bibfnamefont {Y.}~\bibnamefont
  {Hsu}}\ and\ \bibinfo {author} {\bibfnamefont {L.}~\bibnamefont {Berger}},\
  }\href {https://doi.org/10.1103/PhysRevB.18.4856} {\bibfield  {journal}
  {\bibinfo  {journal} {Phys. Rev. B}\ }\textbf {\bibinfo {volume} {18}},\
  \bibinfo {pages} {4856} (\bibinfo {year} {1978})}\BibitemShut {NoStop}%
\bibitem [{\citenamefont {Schoen}\ \emph {et~al.}(2017)\citenamefont {Schoen},
  \citenamefont {Lucassen}, \citenamefont {Nembach}, \citenamefont {Koopmans},
  \citenamefont {Silva}, \citenamefont {Back},\ and\ \citenamefont
  {Shaw}}]{Schoen2017}%
  \BibitemOpen
  \bibfield  {author} {\bibinfo {author} {\bibfnamefont {M.~A.~W.}\
  \bibnamefont {Schoen}}, \bibinfo {author} {\bibfnamefont {J.}~\bibnamefont
  {Lucassen}}, \bibinfo {author} {\bibfnamefont {H.~T.}\ \bibnamefont
  {Nembach}}, \bibinfo {author} {\bibfnamefont {B.}~\bibnamefont {Koopmans}},
  \bibinfo {author} {\bibfnamefont {T.~J.}\ \bibnamefont {Silva}}, \bibinfo
  {author} {\bibfnamefont {C.~H.}\ \bibnamefont {Back}},\ and\ \bibinfo
  {author} {\bibfnamefont {J.~M.}\ \bibnamefont {Shaw}},\ }\href
  {https://doi.org/10.1103/PhysRevB.95.134411} {\bibfield  {journal} {\bibinfo
  {journal} {Phys. Rev. B}\ }\textbf {\bibinfo {volume} {95}},\ \bibinfo
  {pages} {134411} (\bibinfo {year} {2017})}\BibitemShut {NoStop}%
\bibitem [{\citenamefont {Lee}\ \emph {et~al.}(2017)\citenamefont {Lee},
  \citenamefont {Brangham}, \citenamefont {Cheng}, \citenamefont {White},
  \citenamefont {Ruane}, \citenamefont {Esser}, \citenamefont {McComb},
  \citenamefont {Hammel},\ and\ \citenamefont {Yang}}]{Lee2017}%
  \BibitemOpen
  \bibfield  {author} {\bibinfo {author} {\bibfnamefont {A.~J.}\ \bibnamefont
  {Lee}}, \bibinfo {author} {\bibfnamefont {J.~T.}\ \bibnamefont {Brangham}},
  \bibinfo {author} {\bibfnamefont {Y.}~\bibnamefont {Cheng}}, \bibinfo
  {author} {\bibfnamefont {S.~P.}\ \bibnamefont {White}}, \bibinfo {author}
  {\bibfnamefont {W.~T.}\ \bibnamefont {Ruane}}, \bibinfo {author}
  {\bibfnamefont {B.~D.}\ \bibnamefont {Esser}}, \bibinfo {author}
  {\bibfnamefont {D.~W.}\ \bibnamefont {McComb}}, \bibinfo {author}
  {\bibfnamefont {P.~C.}\ \bibnamefont {Hammel}},\ and\ \bibinfo {author}
  {\bibfnamefont {F.}~\bibnamefont {Yang}},\ }\href
  {https://doi.org/10.1038/s41467-017-00332-x} {\bibfield  {journal} {\bibinfo
  {journal} {Nat. Commun.}\ }\textbf {\bibinfo {volume} {8}},\ \bibinfo {pages}
  {1} (\bibinfo {year} {2017})}\BibitemShut {NoStop}%
\bibitem [{\citenamefont {Zhao}\ \emph {et~al.}(2018)\citenamefont {Zhao},
  \citenamefont {Liu}, \citenamefont {Tang}, \citenamefont {Jiang},
  \citenamefont {Yuan},\ and\ \citenamefont {Xia}}]{Zhao2018}%
  \BibitemOpen
  \bibfield  {author} {\bibinfo {author} {\bibfnamefont {Y.}~\bibnamefont
  {Zhao}}, \bibinfo {author} {\bibfnamefont {Y.}~\bibnamefont {Liu}}, \bibinfo
  {author} {\bibfnamefont {H.}~\bibnamefont {Tang}}, \bibinfo {author}
  {\bibfnamefont {H.}~\bibnamefont {Jiang}}, \bibinfo {author} {\bibfnamefont
  {Z.}~\bibnamefont {Yuan}},\ and\ \bibinfo {author} {\bibfnamefont
  {K.}~\bibnamefont {Xia}},\ }\href
  {https://doi.org/10.1103/PhysRevB.98.174412} {\bibfield  {journal} {\bibinfo
  {journal} {Phys. Rev. B}\ }\textbf {\bibinfo {volume} {98}},\ \bibinfo
  {pages} {174412} (\bibinfo {year} {2018})}\BibitemShut {NoStop}%
\bibitem [{\citenamefont {Karipoth}\ \emph {et~al.}(2013)\citenamefont
  {Karipoth}, \citenamefont {Thirumurugan},\ and\ \citenamefont
  {Joseyphus}}]{karipoth2013synthesis}%
  \BibitemOpen
  \bibfield  {author} {\bibinfo {author} {\bibfnamefont {P.}~\bibnamefont
  {Karipoth}}, \bibinfo {author} {\bibfnamefont {A.}~\bibnamefont
  {Thirumurugan}},\ and\ \bibinfo {author} {\bibfnamefont {R.~J.}\ \bibnamefont
  {Joseyphus}},\ }\href
  {https://doi.org/https://doi.org/10.1016/j.jcis.2013.04.041} {\bibfield
  {journal} {\bibinfo  {journal} {J. Colloid Interface Sci}\ }\textbf {\bibinfo
  {volume} {404}},\ \bibinfo {pages} {49} (\bibinfo {year} {2013})}\BibitemShut
  {NoStop}%
\bibitem [{\citenamefont {Karipoth}\ \emph {et~al.}(2016)\citenamefont
  {Karipoth}, \citenamefont {Thirumurugan}, \citenamefont {Velaga},
  \citenamefont {Greneche},\ and\ \citenamefont
  {Justin~Joseyphus}}]{karipoth2016magnetic}%
  \BibitemOpen
  \bibfield  {author} {\bibinfo {author} {\bibfnamefont {P.}~\bibnamefont
  {Karipoth}}, \bibinfo {author} {\bibfnamefont {A.}~\bibnamefont
  {Thirumurugan}}, \bibinfo {author} {\bibfnamefont {S.}~\bibnamefont
  {Velaga}}, \bibinfo {author} {\bibfnamefont {J.-M.}\ \bibnamefont
  {Greneche}},\ and\ \bibinfo {author} {\bibfnamefont {R.}~\bibnamefont
  {Justin~Joseyphus}},\ }\href {https://doi.org/10.1063/1.4962637} {\bibfield
  {journal} {\bibinfo  {journal} {J. Appl. Phys.}\ }\textbf {\bibinfo {volume}
  {120}},\ \bibinfo {pages} {123906} (\bibinfo {year} {2016})}\BibitemShut
  {NoStop}%
\bibitem [{\citenamefont {Liu}\ \emph {et~al.}(1996)\citenamefont {Liu},
  \citenamefont {Steiner}, \citenamefont {Sooryakumar}, \citenamefont {Prinz},
  \citenamefont {Farrow},\ and\ \citenamefont {Harp}}]{Liu1996}%
  \BibitemOpen
  \bibfield  {author} {\bibinfo {author} {\bibfnamefont {X.}~\bibnamefont
  {Liu}}, \bibinfo {author} {\bibfnamefont {M.~M.}\ \bibnamefont {Steiner}},
  \bibinfo {author} {\bibfnamefont {R.}~\bibnamefont {Sooryakumar}}, \bibinfo
  {author} {\bibfnamefont {G.~A.}\ \bibnamefont {Prinz}}, \bibinfo {author}
  {\bibfnamefont {R.~F.~C.}\ \bibnamefont {Farrow}},\ and\ \bibinfo {author}
  {\bibfnamefont {G.}~\bibnamefont {Harp}},\ }\href
  {https://doi.org/10.1103/PhysRevB.53.12166} {\bibfield  {journal} {\bibinfo
  {journal} {Phys. Rev. B}\ }\textbf {\bibinfo {volume} {53}},\ \bibinfo
  {pages} {12166} (\bibinfo {year} {1996})}\BibitemShut {NoStop}%
\bibitem [{\citenamefont {Walowski}\ \emph {et~al.}(2008)\citenamefont
  {Walowski}, \citenamefont {Kaufmann}, \citenamefont {Lenk}, \citenamefont
  {Hamann}, \citenamefont {McCord},\ and\ \citenamefont
  {M{\"u}nzenberg}}]{Walowski2008}%
  \BibitemOpen
  \bibfield  {author} {\bibinfo {author} {\bibfnamefont {J.}~\bibnamefont
  {Walowski}}, \bibinfo {author} {\bibfnamefont {M.~D.}\ \bibnamefont
  {Kaufmann}}, \bibinfo {author} {\bibfnamefont {B.}~\bibnamefont {Lenk}},
  \bibinfo {author} {\bibfnamefont {C.}~\bibnamefont {Hamann}}, \bibinfo
  {author} {\bibfnamefont {J.}~\bibnamefont {McCord}},\ and\ \bibinfo {author}
  {\bibfnamefont {M.}~\bibnamefont {M{\"u}nzenberg}},\ }\href
  {https://doi.org/10.1088/0022-3727/41/16/164016} {\bibfield  {journal}
  {\bibinfo  {journal} {J. Phys. D: Appl. Phys.}\ }\textbf {\bibinfo {volume}
  {41}},\ \bibinfo {pages} {164016} (\bibinfo {year} {2008})}\BibitemShut
  {NoStop}%
\bibitem [{\citenamefont {Heinrich}\ \emph {et~al.}(1979)\citenamefont
  {Heinrich}, \citenamefont {Meredith},\ and\ \citenamefont
  {Cochran}}]{Heinrich1979}%
  \BibitemOpen
  \bibfield  {author} {\bibinfo {author} {\bibfnamefont {B.}~\bibnamefont
  {Heinrich}}, \bibinfo {author} {\bibfnamefont {D.}~\bibnamefont {Meredith}},\
  and\ \bibinfo {author} {\bibfnamefont {J.}~\bibnamefont {Cochran}},\ }\href
  {https://doi.org/10.1063/1.326802} {\bibfield  {journal} {\bibinfo  {journal}
  {J. Appl. Phys.}\ }\textbf {\bibinfo {volume} {50}},\ \bibinfo {pages} {7726}
  (\bibinfo {year} {1979})}\BibitemShut {NoStop}%
\bibitem [{\citenamefont {Umetsu}\ \emph {et~al.}(2012)\citenamefont {Umetsu},
  \citenamefont {Miura},\ and\ \citenamefont {Sakuma}}]{Umetsu2012}%
  \BibitemOpen
  \bibfield  {author} {\bibinfo {author} {\bibfnamefont {N.}~\bibnamefont
  {Umetsu}}, \bibinfo {author} {\bibfnamefont {D.}~\bibnamefont {Miura}},\ and\
  \bibinfo {author} {\bibfnamefont {A.}~\bibnamefont {Sakuma}},\ }\href
  {https://doi.org/10.1143/JPSJ.81.114716} {\bibfield  {journal} {\bibinfo
  {journal} {J. Phys. Soc. Jpn.}\ }\textbf {\bibinfo {volume} {81}},\ \bibinfo
  {pages} {114716} (\bibinfo {year} {2012})}\BibitemShut {NoStop}%
\bibitem [{\citenamefont {Chen}\ \emph {et~al.}(2018)\citenamefont {Chen},
  \citenamefont {Zarzuela}, \citenamefont {Zhang}, \citenamefont {Song},
  \citenamefont {Zhou}, \citenamefont {Shi}, \citenamefont {Li}, \citenamefont
  {Zhou}, \citenamefont {Jiang}, \citenamefont {Pan} \emph
  {et~al.}}]{chen2018antidamping}%
  \BibitemOpen
  \bibfield  {author} {\bibinfo {author} {\bibfnamefont {X.}~\bibnamefont
  {Chen}}, \bibinfo {author} {\bibfnamefont {R.}~\bibnamefont {Zarzuela}},
  \bibinfo {author} {\bibfnamefont {J.}~\bibnamefont {Zhang}}, \bibinfo
  {author} {\bibfnamefont {C.}~\bibnamefont {Song}}, \bibinfo {author}
  {\bibfnamefont {X.}~\bibnamefont {Zhou}}, \bibinfo {author} {\bibfnamefont
  {G.}~\bibnamefont {Shi}}, \bibinfo {author} {\bibfnamefont {F.}~\bibnamefont
  {Li}}, \bibinfo {author} {\bibfnamefont {H.}~\bibnamefont {Zhou}}, \bibinfo
  {author} {\bibfnamefont {W.}~\bibnamefont {Jiang}}, \bibinfo {author}
  {\bibfnamefont {F.}~\bibnamefont {Pan}}, \emph {et~al.},\ }\href
  {https://doi.org/10.1103/PhysRevLett.120.207204} {\bibfield  {journal}
  {\bibinfo  {journal} {Phys. Rev. Lett.}\ }\textbf {\bibinfo {volume} {120}},\
  \bibinfo {pages} {207204} (\bibinfo {year} {2018})}\BibitemShut {NoStop}%
\bibitem [{\citenamefont {Mahfouzi}\ and\ \citenamefont
  {Kioussis}(2018)}]{mahfouzi2018damping}%
  \BibitemOpen
  \bibfield  {author} {\bibinfo {author} {\bibfnamefont {F.}~\bibnamefont
  {Mahfouzi}}\ and\ \bibinfo {author} {\bibfnamefont {N.}~\bibnamefont
  {Kioussis}},\ }\href {https://doi.org/10.1103/PhysRevB.98.220410} {\bibfield
  {journal} {\bibinfo  {journal} {Phys. Rev. B}\ }\textbf {\bibinfo {volume}
  {98}},\ \bibinfo {pages} {220410} (\bibinfo {year} {2018})}\BibitemShut
  {NoStop}%
\bibitem [{\citenamefont {Mook}\ and\ \citenamefont {Paul}(1985)}]{Mook1985}%
  \BibitemOpen
  \bibfield  {author} {\bibinfo {author} {\bibfnamefont {H.~A.}\ \bibnamefont
  {Mook}}\ and\ \bibinfo {author} {\bibfnamefont {D.~M.}\ \bibnamefont
  {Paul}},\ }\href {https://doi.org/10.1103/PhysRevLett.54.227} {\bibfield
  {journal} {\bibinfo  {journal} {Phys. Rev. Lett.}\ }\textbf {\bibinfo
  {volume} {54}},\ \bibinfo {pages} {227} (\bibinfo {year} {1985})}\BibitemShut
  {NoStop}%
\bibitem [{\citenamefont {Lynn}(1975)}]{Lynn1975}%
  \BibitemOpen
  \bibfield  {author} {\bibinfo {author} {\bibfnamefont {J.~W.}\ \bibnamefont
  {Lynn}},\ }\href {https://doi.org/10.1103/PhysRevB.11.2624} {\bibfield
  {journal} {\bibinfo  {journal} {Phys. Rev. B}\ }\textbf {\bibinfo {volume}
  {11}},\ \bibinfo {pages} {2624} (\bibinfo {year} {1975})}\BibitemShut
  {NoStop}%
\bibitem [{\citenamefont {Loong}\ \emph {et~al.}(1984)\citenamefont {Loong},
  \citenamefont {Carpenter}, \citenamefont {Lynn}, \citenamefont {Robinson},\
  and\ \citenamefont {Mook}}]{Loong1984}%
  \BibitemOpen
  \bibfield  {author} {\bibinfo {author} {\bibfnamefont {C.-K.}\ \bibnamefont
  {Loong}}, \bibinfo {author} {\bibfnamefont {J.}~\bibnamefont {Carpenter}},
  \bibinfo {author} {\bibfnamefont {J.}~\bibnamefont {Lynn}}, \bibinfo {author}
  {\bibfnamefont {R.}~\bibnamefont {Robinson}},\ and\ \bibinfo {author}
  {\bibfnamefont {H.}~\bibnamefont {Mook}},\ }\href@noop {} {\bibfield
  {journal} {\bibinfo  {journal} {Journal of applied physics}\ }\textbf
  {\bibinfo {volume} {55}},\ \bibinfo {pages} {1895} (\bibinfo {year}
  {1984})}\BibitemShut {NoStop}%
\bibitem [{\citenamefont {Balashov}(2009)}]{Balashov2009}%
  \BibitemOpen
  \bibfield  {author} {\bibinfo {author} {\bibfnamefont {T.}~\bibnamefont
  {Balashov}},\ }\emph {\bibinfo {title} {Inelastic scanning tunneling
  spectroscopy: magnetic excitations on the nanoscale}},\ \href@noop {} {Ph.D.
  thesis},\ \bibinfo  {school} {Karlsruher Institut für Technologie} (\bibinfo
  {year} {2009})\BibitemShut {NoStop}%
\bibitem [{\citenamefont {Katsnelson}\ and\ \citenamefont
  {Lichtenstein}(2004)}]{katsnelson2004magnetic}%
  \BibitemOpen
  \bibfield  {author} {\bibinfo {author} {\bibfnamefont {M.}~\bibnamefont
  {Katsnelson}}\ and\ \bibinfo {author} {\bibfnamefont {A.}~\bibnamefont
  {Lichtenstein}},\ }\href {https://doi.org/10.1088/0953-8984/16/41/023}
  {\bibfield  {journal} {\bibinfo  {journal} {J. Phys. Condens. Matter}\
  }\textbf {\bibinfo {volume} {16}},\ \bibinfo {pages} {7439} (\bibinfo {year}
  {2004})}\BibitemShut {NoStop}%
\bibitem [{\citenamefont {Barati}\ \emph {et~al.}(2014)\citenamefont {Barati},
  \citenamefont {Cinal}, \citenamefont {Edwards},\ and\ \citenamefont
  {Umerski}}]{Barati2014}%
  \BibitemOpen
  \bibfield  {author} {\bibinfo {author} {\bibfnamefont {E.}~\bibnamefont
  {Barati}}, \bibinfo {author} {\bibfnamefont {M.}~\bibnamefont {Cinal}},
  \bibinfo {author} {\bibfnamefont {D.~M.}\ \bibnamefont {Edwards}},\ and\
  \bibinfo {author} {\bibfnamefont {A.}~\bibnamefont {Umerski}},\ }\href
  {https://doi.org/10.1103/PhysRevB.90.014420} {\bibfield  {journal} {\bibinfo
  {journal} {Phys. Rev. B}\ }\textbf {\bibinfo {volume} {90}},\ \bibinfo
  {pages} {014420} (\bibinfo {year} {2014})}\BibitemShut {NoStop}%
\bibitem [{\citenamefont {Wu}\ \emph {et~al.}(2018)\citenamefont {Wu},
  \citenamefont {Liu},\ and\ \citenamefont {Luo}}]{wu2018magnon}%
  \BibitemOpen
  \bibfield  {author} {\bibinfo {author} {\bibfnamefont {X.}~\bibnamefont
  {Wu}}, \bibinfo {author} {\bibfnamefont {Z.}~\bibnamefont {Liu}},\ and\
  \bibinfo {author} {\bibfnamefont {T.}~\bibnamefont {Luo}},\ }\href
  {https://doi.org/10.1063/1.5020611} {\bibfield  {journal} {\bibinfo
  {journal} {J. Appl. Phys.}\ }\textbf {\bibinfo {volume} {123}},\ \bibinfo
  {pages} {085109} (\bibinfo {year} {2018})}\BibitemShut {NoStop}%
\bibitem [{\citenamefont {Depondt}\ and\ \citenamefont
  {Mertens}(2009)}]{depondt2009spin}%
  \BibitemOpen
  \bibfield  {author} {\bibinfo {author} {\bibfnamefont {P.}~\bibnamefont
  {Depondt}}\ and\ \bibinfo {author} {\bibfnamefont {F.}~\bibnamefont
  {Mertens}},\ }\href {https://doi.org/10.1088/0953-8984/21/33/336005}
  {\bibfield  {journal} {\bibinfo  {journal} {J. Phys. Condens. Matter}\
  }\textbf {\bibinfo {volume} {21}},\ \bibinfo {pages} {336005} (\bibinfo
  {year} {2009})}\BibitemShut {NoStop}%
\bibitem [{\citenamefont {Hickey}\ and\ \citenamefont
  {Moodera}(2009)}]{Hickey2009}%
  \BibitemOpen
  \bibfield  {author} {\bibinfo {author} {\bibfnamefont {M.~C.}\ \bibnamefont
  {Hickey}}\ and\ \bibinfo {author} {\bibfnamefont {J.~S.}\ \bibnamefont
  {Moodera}},\ }\href {https://doi.org/10.1103/PhysRevLett.102.137601}
  {\bibfield  {journal} {\bibinfo  {journal} {Phys. Rev. Lett.}\ }\textbf
  {\bibinfo {volume} {102}},\ \bibinfo {pages} {137601} (\bibinfo {year}
  {2009})}\BibitemShut {NoStop}%
\bibitem [{\citenamefont {van Seters}\ \emph {et~al.}(2022)\citenamefont {van
  Seters}, \citenamefont {Ludwig}, \citenamefont {Yuan},\ and\ \citenamefont
  {Duine}}]{Seters2022}%
  \BibitemOpen
  \bibfield  {author} {\bibinfo {author} {\bibfnamefont {D.}~\bibnamefont {van
  Seters}}, \bibinfo {author} {\bibfnamefont {T.}~\bibnamefont {Ludwig}},
  \bibinfo {author} {\bibfnamefont {H.~Y.}\ \bibnamefont {Yuan}},\ and\
  \bibinfo {author} {\bibfnamefont {R.~A.}\ \bibnamefont {Duine}},\ }\href@noop
  {} {\bibinfo {title} {Dissipation-free modes in dissipative systems}}
  (\bibinfo {year} {2022}),\ \Eprint {https://arxiv.org/abs/arXiv:2206.07471}
  {arXiv:2206.07471} \BibitemShut {NoStop}%
\bibitem [{\citenamefont {Velick\'y}(1969)}]{Velick1969}%
  \BibitemOpen
  \bibfield  {author} {\bibinfo {author} {\bibfnamefont {B.}~\bibnamefont
  {Velick\'y}},\ }\href {https://doi.org/10.1103/PhysRev.184.614} {\bibfield
  {journal} {\bibinfo  {journal} {Phys. Rev.}\ }\textbf {\bibinfo {volume}
  {184}},\ \bibinfo {pages} {614} (\bibinfo {year} {1969})}\BibitemShut
  {NoStop}%
\bibitem [{\citenamefont {Chimata}\ \emph {et~al.}(2017)\citenamefont
  {Chimata}, \citenamefont {Delczeg-Czirjak}, \citenamefont {Szilva},
  \citenamefont {Cardias}, \citenamefont {Kvashnin}, \citenamefont {Pereiro},
  \citenamefont {Mankovsky}, \citenamefont {Ebert}, \citenamefont {Thonig},
  \citenamefont {Sanyal}, \citenamefont {Klautau},\ and\ \citenamefont
  {Eriksson}}]{Chimata2017}%
  \BibitemOpen
  \bibfield  {author} {\bibinfo {author} {\bibfnamefont {R.}~\bibnamefont
  {Chimata}}, \bibinfo {author} {\bibfnamefont {E.~K.}\ \bibnamefont
  {Delczeg-Czirjak}}, \bibinfo {author} {\bibfnamefont {A.}~\bibnamefont
  {Szilva}}, \bibinfo {author} {\bibfnamefont {R.}~\bibnamefont {Cardias}},
  \bibinfo {author} {\bibfnamefont {Y.~O.}\ \bibnamefont {Kvashnin}}, \bibinfo
  {author} {\bibfnamefont {M.}~\bibnamefont {Pereiro}}, \bibinfo {author}
  {\bibfnamefont {S.}~\bibnamefont {Mankovsky}}, \bibinfo {author}
  {\bibfnamefont {H.}~\bibnamefont {Ebert}}, \bibinfo {author} {\bibfnamefont
  {D.}~\bibnamefont {Thonig}}, \bibinfo {author} {\bibfnamefont
  {B.}~\bibnamefont {Sanyal}}, \bibinfo {author} {\bibfnamefont {A.~B.}\
  \bibnamefont {Klautau}},\ and\ \bibinfo {author} {\bibfnamefont
  {O.}~\bibnamefont {Eriksson}},\ }\href
  {https://doi.org/10.1103/PhysRevB.95.214417} {\bibfield  {journal} {\bibinfo
  {journal} {Phys. Rev. B}\ }\textbf {\bibinfo {volume} {95}},\ \bibinfo
  {pages} {214417} (\bibinfo {year} {2017})}\BibitemShut {NoStop}%
\bibitem [{\citenamefont {Turek}\ \emph {et~al.}(2015)\citenamefont {Turek},
  \citenamefont {Kudrnovsk\'y},\ and\ \citenamefont {Drchal}}]{Turek2015}%
  \BibitemOpen
  \bibfield  {author} {\bibinfo {author} {\bibfnamefont {I.}~\bibnamefont
  {Turek}}, \bibinfo {author} {\bibfnamefont {J.}~\bibnamefont
  {Kudrnovsk\'y}},\ and\ \bibinfo {author} {\bibfnamefont {V.}~\bibnamefont
  {Drchal}},\ }\href {https://doi.org/10.1103/PhysRevB.92.214407} {\bibfield
  {journal} {\bibinfo  {journal} {Phys. Rev. B}\ }\textbf {\bibinfo {volume}
  {92}},\ \bibinfo {pages} {214407} (\bibinfo {year} {2015})}\BibitemShut
  {NoStop}%
\bibitem [{\citenamefont {Oogane}\ \emph {et~al.}(2015)\citenamefont {Oogane},
  \citenamefont {Kubota}, \citenamefont {Naganuma},\ and\ \citenamefont
  {Ando}}]{Oogane2015}%
  \BibitemOpen
  \bibfield  {author} {\bibinfo {author} {\bibfnamefont {M.}~\bibnamefont
  {Oogane}}, \bibinfo {author} {\bibfnamefont {T.}~\bibnamefont {Kubota}},
  \bibinfo {author} {\bibfnamefont {H.}~\bibnamefont {Naganuma}},\ and\
  \bibinfo {author} {\bibfnamefont {Y.}~\bibnamefont {Ando}},\ }\href
  {https://doi.org/10.1088/0022-3727/48/16/164012} {\bibfield  {journal}
  {\bibinfo  {journal} {J. Phys. D: Appl. Phys.}\ }\textbf {\bibinfo {volume}
  {48}},\ \bibinfo {pages} {164012} (\bibinfo {year} {2015})}\BibitemShut
  {NoStop}%
\bibitem [{\citenamefont {Guimaraes}\ \emph {et~al.}(2019)\citenamefont
  {Guimaraes}, \citenamefont {Suckert}, \citenamefont {Chico}, \citenamefont
  {Bouaziz}, \citenamefont {dos Santos~Dias},\ and\ \citenamefont
  {Lounis}}]{Guimaraes2019}%
  \BibitemOpen
  \bibfield  {author} {\bibinfo {author} {\bibfnamefont {F.~S.}\ \bibnamefont
  {Guimaraes}}, \bibinfo {author} {\bibfnamefont {J.~R.}\ \bibnamefont
  {Suckert}}, \bibinfo {author} {\bibfnamefont {J.}~\bibnamefont {Chico}},
  \bibinfo {author} {\bibfnamefont {J.}~\bibnamefont {Bouaziz}}, \bibinfo
  {author} {\bibfnamefont {M.}~\bibnamefont {dos Santos~Dias}},\ and\ \bibinfo
  {author} {\bibfnamefont {S.}~\bibnamefont {Lounis}},\ }\href
  {https://doi.org/10.1088/1361-648X/ab1239} {\bibfield  {journal} {\bibinfo
  {journal} {J. Phys. Condens. Matter}\ }\textbf {\bibinfo {volume} {31}},\
  \bibinfo {pages} {255802} (\bibinfo {year} {2019})}\BibitemShut {NoStop}%
\bibitem [{\citenamefont {Yuan}\ \emph {et~al.}(2014)\citenamefont {Yuan},
  \citenamefont {Hals}, \citenamefont {Liu}, \citenamefont {Starikov},
  \citenamefont {Brataas},\ and\ \citenamefont {Kelly}}]{Yuan2014}%
  \BibitemOpen
  \bibfield  {author} {\bibinfo {author} {\bibfnamefont {Z.}~\bibnamefont
  {Yuan}}, \bibinfo {author} {\bibfnamefont {K.~M.~D.}\ \bibnamefont {Hals}},
  \bibinfo {author} {\bibfnamefont {Y.}~\bibnamefont {Liu}}, \bibinfo {author}
  {\bibfnamefont {A.~A.}\ \bibnamefont {Starikov}}, \bibinfo {author}
  {\bibfnamefont {A.}~\bibnamefont {Brataas}},\ and\ \bibinfo {author}
  {\bibfnamefont {P.~J.}\ \bibnamefont {Kelly}},\ }\href
  {https://doi.org/10.1103/PhysRevLett.113.266603} {\bibfield  {journal}
  {\bibinfo  {journal} {Phys. Rev. Lett.}\ }\textbf {\bibinfo {volume} {113}},\
  \bibinfo {pages} {266603} (\bibinfo {year} {2014})}\BibitemShut {NoStop}%
\bibitem [{\citenamefont {Sakuma}(2015)}]{Sakuma2015}%
  \BibitemOpen
  \bibfield  {author} {\bibinfo {author} {\bibfnamefont {A.}~\bibnamefont
  {Sakuma}},\ }\href {https://doi.org/10.1063/1.4905429} {\bibfield  {journal}
  {\bibinfo  {journal} {J. Appl. Phys.}\ }\textbf {\bibinfo {volume} {117}},\
  \bibinfo {pages} {013912} (\bibinfo {year} {2015})}\BibitemShut {NoStop}%
\end{thebibliography}%
\bibliographystyle{apsrev4-2}
\clearpage
\appendix
\section{Numerical solver}
\label{appendix:solver}
In this Appendix, the numerical method to solve Eq. \ref{eq:sllg} is described. In previous studies, several numerical approaches have been proposed to solve the local LLG equations, including HeunP method, implicit midpoint method, Depondt-Merten’s method~\cite{depondt2009spin}, semi-implicit A (SIA) and semi-implicit B (SIB) methods~\cite{mentink2010stable}. To solve this non-local LLG equation, we use the fixed-point iteration midpoint 
method. We have done convergence tests on this method and find that it preserve the energy and spin length of the system, which is demonstrated in Fig. \ref{Fig:Dimer} for the case of a dimer.
With stable outputs, the solver allows for a relatively large time step size, typically of the order of $\Delta t\sim0.1-1$ fs. 

Following the philosophy of an implicit midpoint method, the implemented algorithm can be described as follows. Let $\vec{m}_i^{t}$ be the magnetic moment of site $i$ at a given time step $t$. Then we can define the quantity $\vec{m}_{mid}$ and the time derivative of $\vec{m}_i$, respectively, as 

\begin{equation}
\begin{aligned}
\vec{m}_{mid}=\frac{\vec{m}_i^{t+1}+\vec{m}_i^t}{2}, \\
\frac{\partial{\vec{m}_i}}{\partial{t}}=\frac{\vec{m}_i^{t+1}-\vec{m}_i^{t}}{\Delta t}.
\end{aligned}
\end{equation}\\

Using this definition in Eq. \ref{eq:sllg_nonlocal}, the equation of motion of the $i$-th spin becomes:

\begin{equation}
\begin{split}
\frac{\partial{\vec{m}_i}}{\partial{t}}=\vec{m}_{mid}\times\left(-\gamma\left[\vec{B}_i(\vec{m}_{mid})+\vec{b}_{i}(t)\right]+\sum_{j}\frac{\alpha_{ij}}{m_j}\frac{\partial{\vec{m}_j}}{\partial{t}}\right).
\end{split}
\end{equation}

Thus, with a fixed-point scheme, we can do the following iteration

\begin{widetext}
\begin{equation}
\label{eq:convergence-moment-damping}
\vec{m}_i^{t+1(k+1)}=\vec{m}_i^t+\Delta t\left(\left(\frac{\vec{m}_i^{t+1(k)}+\vec{m}_i^t}{2}\right)\times\left(-\gamma\left[\vec{B}_i\left(\frac{\vec{m}_i^{t+1(k)}+\vec{m}_i^t}{2}\right)+\vec{b}_{i}(t)\right]+\sum_{j}\frac{\alpha_{ij}}{m_j}\frac{\vec{m}_j^{t+1(k)}-\vec{m}_j^{t}}{\Delta t}\right)\right).
\end{equation}
\end{widetext}

If $\vec{m}_i^{t+1(k+1)}\approx \vec{m}_i^{t+1(k)}$, the self-consistency converges. Typically, about 6 iteration steps are needed. This solver was implemented in the software package UppASD \cite{UppASD} for this work.

\section{Analytical model of anti-damping in dimers}
\label{undamped dimer}
In the dimer model, there are two spins on site 1 and site 2 denoted by $\vec{m}_1$ and $\vec{m}_2$, which are here supposed to be related to the same element -- so that, naturally, $\alpha_{11}=\alpha_{22}>0$. Also, let's consider a sufficiently low temperature so that $\vec{b}_i(t)\rightarrow0$, which is a reasonable assumption, given that damping has an intrinsic origin \cite{Hickey2009}. This simple system allows us to provide explicit expressions for the Hamiltonian, the effective magnetic fields and the damping term. From the analytical solution, it is found that the dimer spin system becomes an undamped system when local damping is equal to non-local damping, \textit{i.e.} the effective damping of the system is zero. 

Following the definition given by Eq. \ref{eq:sllg_nonlocal} in the main text, the equation of motion for spin 1 reads:
\begin{equation}\label{eq:moment-spin1}
\frac{\partial{\vec{m}_1}}{\partial{t}}=\vec{m}_1\times\left(-\gamma \vec{B}_1+\frac{\alpha_{11}}{m_1}\frac{\partial{\vec{m}_1}}{\partial{t}}+\frac{\alpha_{12}}{m_2}\frac{\partial{\vec{m}_2}}{\partial{t}}\right),
\end{equation}
\noindent and an analogous expression can be written for spin 2. For sake of simplicity, the Zeeman term is zero and the effective field only includes the contribution from Heisenberg exchange interactions. Thus, we have $\vec{B}_1=2J_{12}\vec{m}_2 $ and $\vec{B}_2=2J_{21}\vec{m}_1$. With $|\alpha_{ij}|\ll 1  $, we can take the LL form $ \frac{\partial{\vec{m}_i}}{\partial{t}}=-\gamma \vec{m}_i\times \vec{B}_i $ to approximate the time-derivative on the right-hand side of the LLG equation.
Let $ m_1=m_2$ and $\alpha_{12}=\lambda \alpha_{11}$. Since $J_{12}=J_{21} $ and $\vec{m}_1\times \vec{m}_2=-\vec{m}_2 \times \vec{m}_1 $, then we have
\begin{equation}\label{eq:moment-spin1-specialcase}
\frac{\partial{\vec{m}_1}}{\partial{t}}=-2\gamma J_{12}\vec{m}_1\times\left[\vec{m}_2+(1-\lambda)\frac{\alpha_{11}}{m_1}(\vec{m}_1\times \vec{m}_2)\right].
\end{equation}

Therefore, when $\alpha_{12}= \alpha_{21}= \alpha_{11} $ (\textit{i.e.}, $\lambda=1$), Eq. \ref{eq:moment-spin1} is reduced to:
\begin{equation}
\frac{\partial{\vec{m}_1}}{\partial{t}}=-2\gamma J_{12}\vec{m}_1\times \vec{m}_2,
\end{equation}
\noindent and the system becomes undamped. It is however straightforward that, for the opposite case of a strong negative non-local damping ($\lambda=-1$), Eq. \ref{eq:moment-spin1-specialcase} describes a common damped dynamics. A side (and related) consequence of Eq. \ref{eq:moment-spin1-specialcase}, but important for the discussion in Section \ref{sec:remagnetization}, is the fact that the effective onsite damping term $\alpha_{11}^{*}=(1-\lambda)\alpha_{11}$ becomes less relevant to the dynamics as the positive non-local damping increases ($\lambda\rightarrow1$), or, in other words, as $\alpha_{\textnormal{tot}}=(\alpha_{11}+\alpha_{12})$ strictly increases due to the non-local contribution. Exactly the same reasoning can be made for a trimer, for instance, composed by atoms with equal moments and exchange interactions ($m_1=m_2=m_3$, $J_{12}=J_{13}=J_{23}$), and same non-local dampings ($\alpha_{13}=\alpha_{12}=\lambda\alpha_{11}$).

\begin{figure}[H]
        \includegraphics[width=\columnwidth]{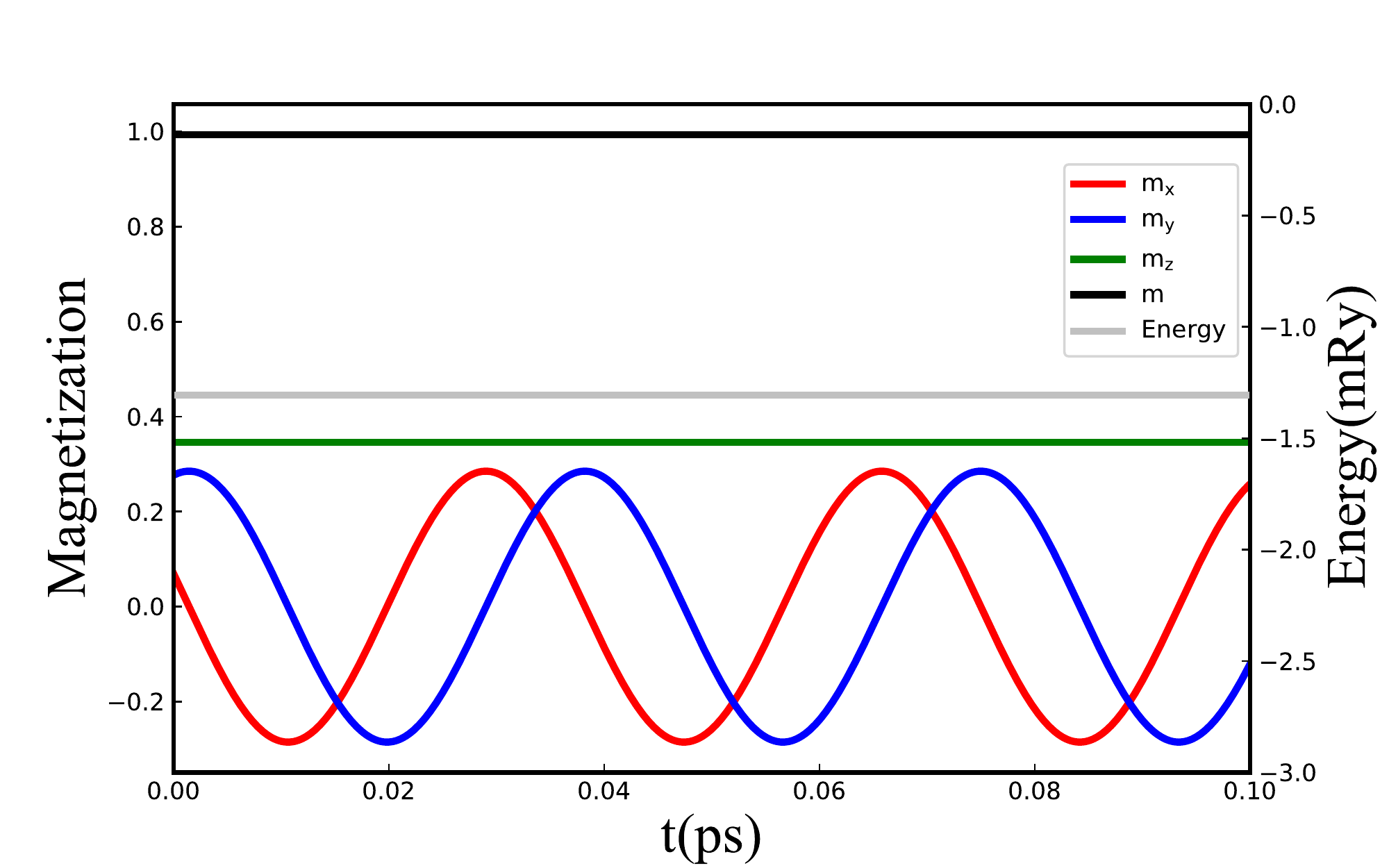} 
        \caption{Spin dynamics at $T=0$ K of an undamped dimer in which $\alpha_{12}= \alpha_{21}= \alpha_{11}$ (see text). The vector $\vec{m}_1$ is normalized and its Cartesian components are labeled in the figure as $m_x$, $m_y$ and $m_z$. The black and grey lines indicate the length of spin and energy (in mRy), respectively.} 
        \label{Fig:Dimer} 
\end{figure}

The undamped behavior can be directly observed from ASD simulations of a dimer with $\alpha_{12}=\alpha_{11}$, as shown in Fig. \ref{Fig:Dimer}. Here the magnetic moment and the exchange are taken the same of an Fe dimer, $m_1=2.23\,\mu_{B}$ and $J_{12}=1.34$ mRy. Nevertheless, obviously the overall behavior depicted in Fig. \ref{Fig:Dimer} is not dependent on the choice of $\vec{m}_1$ and $J_{12}$. The $z$ component is constant, while the $x$ and $y$ components of $ \vec{m}_1 $ oscillate in time, indicating a precessing movement.  

In a broader picture, this simple dimer case exemplifies the connection between the eigenvalues of the damping matrix $\boldsymbol{\alpha}=\left(\alpha_{ij}\right)$ and the damping behavior. 
The occurrence of such undamped dynamics has been recently discussed in Ref.~\cite{Seters2022}, where it is shown that a dissipation-free mode can occur in a system composed of two subsystems coupled to the same bath.

\section{Effective and onsite damping in the FeCo and CoNi alloys}
\label{sec:vca-comparison}
\begin{figure}[h]
        \includegraphics[width=1.2\columnwidth]{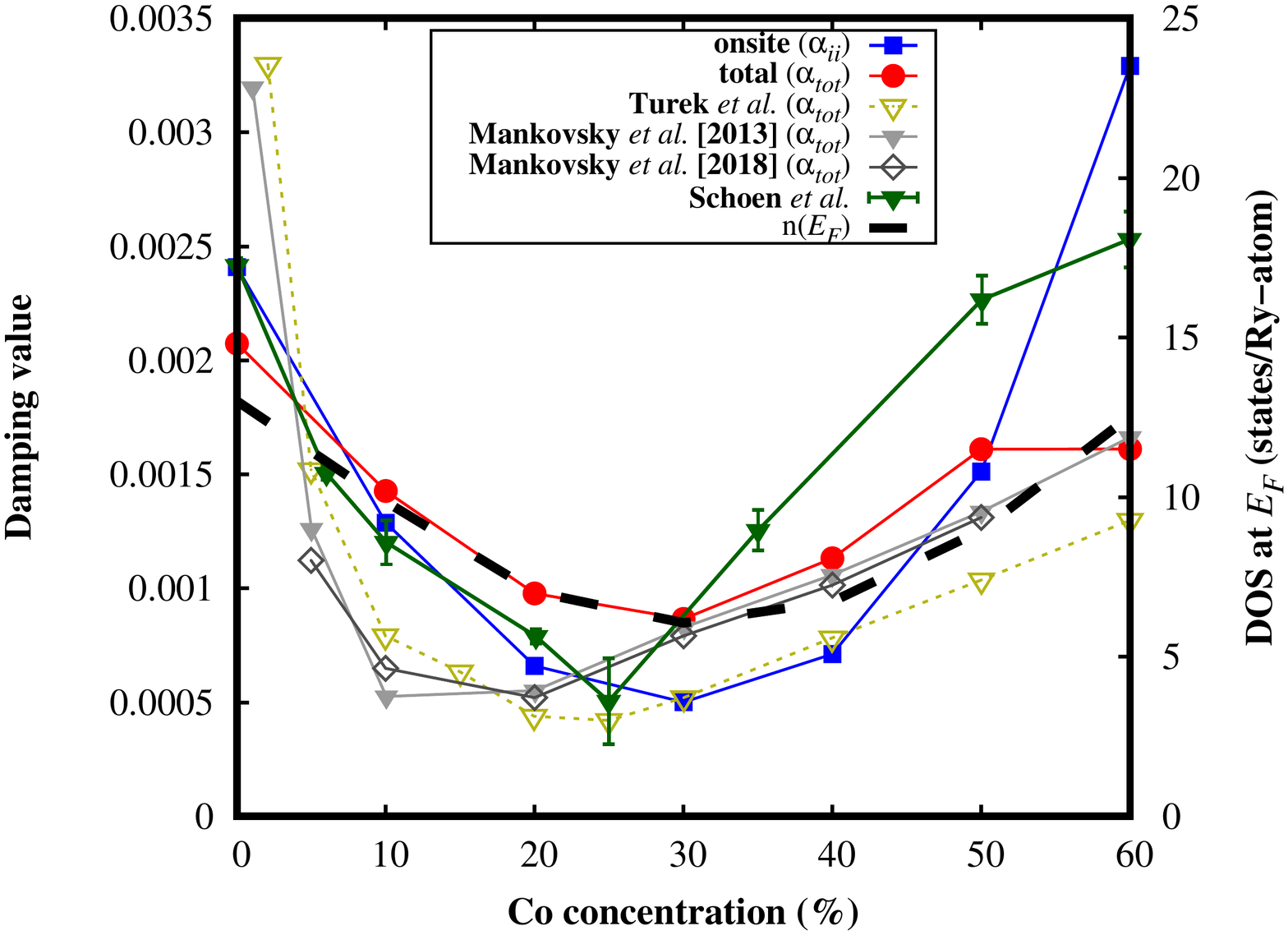} 
        \caption{(Color online) \textit{Left scale}: Computed Gilbert effective ($\alpha_{tot}$, red circles) and onsite ($\alpha_{ii}$, blue squares) damping parameters as a function of Co the concentration ($x$) for bcc Fe$_{1-x}$Co$_x$ binary alloy in the virtual-crystal approximation. The values are compared with previous theoretical results using CPA, from Ref. \cite{Mankovsky2013} (gray full triangles), Ref. \cite{Mankovsky2018} (black open rhombus), Ref. \cite{Turek2015} (yellow open triangles), and room-temperature experimental data \cite{Schoen2016}. \textit{Right scale}: The calculated density of states (DOS) at the Fermi level as a function of $x$, represented by the black dashed line.} 
        \label{fig:alloy-feco}
\end{figure}

As mentioned in Section \ref{sec:methods}, the simple VCA model allows us to account for the disorder in $3d$-transition-metal alloys in a crude but efficient way which avoids the use of large supercells with random chemical distributions. With exactly the same purpose, the coherent potential approximation (CPA) \cite{Velick1969} has also been employed to analyze damping in alloys (\textit{e.g.}, in Refs. \cite{Mankovsky2013,Chimata2017,Turek2015}), showing a very good output with respect to trends, when compared to experiments \cite{Schoen2016,Starikov2010}. In Fig. \ref{fig:alloy-feco} we show the normalized calculated local (onsite, $\alpha_{ii}$) and effective damping ($\alpha_{tot}$) parameters for the zero-temperature VCA Fe$_{1-x}$Co$_{x}$ alloy in the bcc structure, consistent with a concentration up to $x\approx 60\%$ of Co \cite{Schoen2016}. The computed values in this work (blue, representing $\alpha_{ii}$, and red points, representing $\alpha_{tot}$) are compared to previous theoretical CPA results and room-temperature experimental data. 
The trends with VCA are reproduced in a good agreement with respect to experiments and CPA calculations, showing a minimal $\alpha_{tot}$ when the Co concentration is $x\approx 30\%$. This behavior is well correlated with the local density of states (LDOS) at the Fermi level, as expected by the simplified Kamberský equation \cite{Oogane2015}, and the onsite contribution. 
Despite the good agreement found, the values we have determined are subjected to a known error of the VCA with respect to the experimental results.

This discrepancy can be partially explained by three reasons: (\textit{i}) the significant influence of local environments (local disorder and/or short-range order) to $\alpha_{tot}$ \cite{Miranda2021,li2019giant}; (\textit{ii}) the fact that the actual electronic lifetime (\textit{i.e.}, the mean time between two consecutive scattering events) is subestimated by the VCA average for randomness in the FeCo alloy, which can have a non-negligible impact in the damping parameter \cite{gilmore2007identification,Guimaraes2019}; and (\textit{iii}) the influence on damping of noncollinear spin configurations in finite temperature measurements \cite{Yuan2014,Mankovsky2018}. On top of that, it is also notorious that damping is dependent on the imaginary part of the energy (broadening) \cite{gilmore2007identification,Guimaraes2019}, $\delta$, which can be seen as an empirical quantity, and accounts for part of the differences between theory and experiments.

\begin{figure}[h]
        
        \includegraphics[width=\columnwidth]{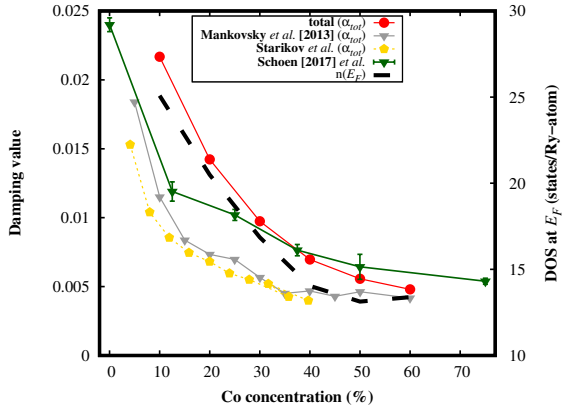} 
        \caption{(Color online) \textit{Left scale}: Computed Gilbert effective ($\alpha_{tot}$, red circles) damping parameters as a function of the Co concentration ($x$) for fcc Co$_{x}$Ni$_{1-x}$ binary alloy in the virtual-crystal approximation. The values are compared with previous theoretical results using CPA, from Ref. \cite{Mankovsky2013} (gray full triangles), Ref. \cite{Starikov2010} (gold full circles), and room-temperature experimental data \cite{Schoen2017}. \textit{Right scale}: The calculated density of states (DOS) at the Fermi level as a function of $x$, represented by the black dashed line.} 
        \label{fig:alloy-coni}
\end{figure}

In the spirit of demonstrating the effectiveness of the simple VCA to qualitatively (and also, to some extent, quantitatively) describe the properties of Gilbert damping in suitable magnetic alloys, we also show in Fig. \ref{fig:alloy-coni} the results obtained for Co$_{x}$Ni$_{1-x}$ systems. The CoNi alloys are known to form in the fcc structure for a Ni concentration range of $10\%-100\%$. Therefore, here we modeled Co$_{x}$Ni$_{1-x}$ by a big fcc cluster containing $\sim530000$ atoms in real-space with the equilibrium lattice parameter of $a=3.46\,$\AA. The number of recursion levels considered is $LL=41$. A good agreement with experimental results and previous theoretical calculations can be noticed. In particular, the qualitative comparison with theory from Refs. \cite{Mankovsky2013,Starikov2010} indicates the equivalence between the torque correlation and the spin correlation models for calculating the damping parameter, which was also investigated by Sakuma \cite{Sakuma2015}. The onsite contribution for each Co concentration, $\alpha_{ii}$, is omitted from Fig. \ref{fig:alloy-coni} due to an absolute value $2-4$ times higher than $\alpha_{tot}$, but follows the same decreasing trend. Again, the overall effective damping values are well correlated with the LDOS, and reflect the variation of the quantity $\frac{1}{m_t}$ with Co concentration (see Eq. \ref{eq:damping-dft}).

\section{Effect of further neighbors in the magnon lifetimes}
\label{sec:effect-further}
When larger cutoff radii ($R_{cut}$) of $\alpha_{ij}$ parameters are included in ASD, Eq. \ref{eq:convergence-moment-damping} takes longer times to achieve a self-consistent convergence. In practical terms, to reach a sizeable computational time for the calculation of a given system, $R_{cut}$ needs to be chosen in order to preserve the main features of the magnon properties as if $R_{cut}\rightarrow\infty$. A good quantity to rely on is the magnon lifetime $\tau_{\textbf{q}}$, as it consists of both magnon frequency and $\textbf{q}$-resolved damping (Eq. \ref{eq:magnon lifetime}). In Section \ref{sec:magnon-spectra}, we have shown the equivalence between Eq. \ref{eq:magnon lifetime} and the inverse of FWHM on the energy axis of $S(\textbf{q},\omega)$ for the ferromagnets investigated here. Thus, the comparison of two $\tau_{\textbf{q}}$ spectra for different $R_{cut}$ can be done directly and in an easier way using Eq. \ref{eq:magnon lifetime}.

\begin{figure}[htb]
        \includegraphics[width=\columnwidth]{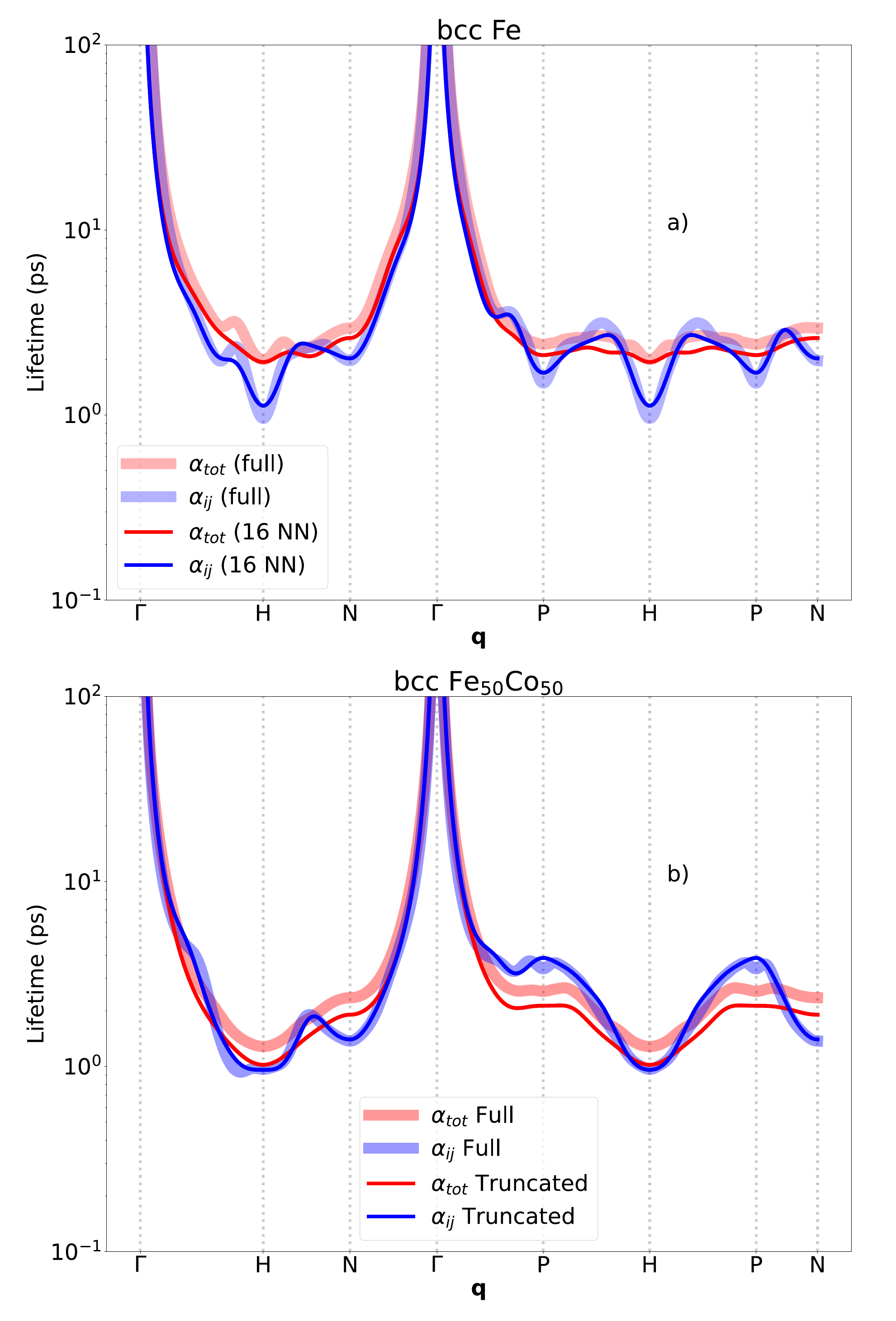} 
        \caption{(Color online) Magnon lifetimes calculated using Eq. \ref{eq:magnon lifetime} for: (a) bcc Fe; and (b) bcc Fe$_{50}$Co$_{50}$, using a reduced set of 16 NN shells (opaque lines), and the full set of 136 NN shells (transparent lines).}
        \label{fig:comparison-lifetimes-bccFe}
\end{figure}

An example is shown in Figure \ref{fig:comparison-lifetimes-bccFe} for bcc Fe and bcc Fe$_{50}$Co$_{50}$. Here we choose the first 16 NN ($R_{cut}\sim3.32a$) and compare the results with the full calculated set of 136 NN ($R_{cut}=10a$). It is noticeable that the reduced set of neighbors can capture most of the features of the $\tau_{\textbf{q}}$ spectrum for a full NN set. However, long-range influences of small magnitudes, such as extra oscillations around the point $\textbf{q}=\textbf{H}$ in Fe, can occur. In particular, these extra oscillations arise mainly due to the presence of Kohn anomalies in the magnon spectrum of Fe, already reported in previous works \cite{Pajda2001,halilov1998adiabatic}. In turn, for the case of Fe$_{50}$Co$_{50}$, the long-range $\alpha_{ij}$ reduces $\alpha_{tot}$, and causes the remagnetization times for non-local and effective dampings to be very similar (see Fig. \ref{fig:spin dynamics}). For the other ferromagnets considered in the present research, comparisons of the reduced $R_{cut}$ with analogous quality were reached.

\end{document}